# The Super-FRS GEM-TPC prototype development

## Technical report

F.Garcia[1], C.Caesar[2], T.Grahn[3,1], A.Prochazka[2], and B.Voss[2]

[1]Helsinki Institute of Physics, University of Helsinki, 00014 Helsinki, Finland

[2]GSI Helmholtzzentrum für Schwerionenforschung, Darmstadt 64291, Germany

[3]University of Jyvaskyla, Department of Physics, FI-40014 University of Jyvaskyla, Finland

June 1, 2015

**Abstract**

This document contains the pre-design of the beam diagnostics components, tracking detectors, for the Super-FRS magnetic separator of FAIR. A GEM-TPC detector has been suggested as a suitable tracking detector for the ion and fragment beams produced at the in-flight separator Super-FRS under construction at the FAIR facility. The detector concept combines two widely used approaches in gas-filled detectors, the Time Projection Chamber (TPC) and the Gas Electron Multiplication (GEM). Three detector generations (prototypes) have been tested in 2011, 2012 and 2014 with relativistic ion beams at GSI. Due to the high-resolution achromatic mode of Super-FRS, highly homogeneous transmission tracking detectors are crucial to tag the momentum of the ion/fragment beam. They must be able to provide precise information on the (horizontal and vertical) deviation from nominal beam optics, while operated with slow-extracted beam on event-by-event basis, in order to provide unambiguous identification of the fragments, together with time-of-flight and $\Delta E$ measurements. The main requirements are a maximum active area horizontally and vertically of 380 $\times$ 80 mm$^2$, a position resolution up to 1 mm, a maximum rate capability of 10 MHz and a dynamic range of about 600 fC. In total 32 tracking detectors operating in vacuum are needed along the Super-FRS beam line.

# 1 FAIR Super-FRS tracking detector concept

A GEM-TPC detector, which combines two major concepts in modern gas-filled detectors, the Time Projection Chamber (TPC) and the Gas Electron Multiplication (GEM), has been suggested as tracking detector for the ion/fragment beams produced at the in-flight separator Super-FRS under construction at the FAIR facility [1]. The present report outlines the design of the prototype of the Super-FRS GEM-TPC tracking detector. The requirements and constraints determined by the experimental conditions and their impact on the design will be discussed. The progress in the detailed design of the Super-FRS such as radiation environment, has set new requirements that have not necessarily been considered in the beginning of the project. Such constraints have to be taken into account and thus implementations of proposed solutions will be discussed.

## 1.1 Super-FRS

Super-FRS will be the most powerful in-flight magnetic separator for exotic nuclei up to relativistic energies compared to any existing device. Rare isotopes of all elements up to uranium can be produced and spatially separated within some hundred nanoseconds, thus very short-lived nuclei can be studied efficiently.

Super-FRS is a large-acceptance superconducting fragment separator with three branches serving different experimental areas including a new storage-ring complex. The magnetic system will consist of three branches connecting different experimental areas, see Fig. 1. Reaction studies with highest energies under complete kinematics, comprising of a large dipole magnet and particle detectors, will be performed at the High-Energy Branch (HEB). Unique studies of e.g. nuclear masses and ground-state properties will be performed in the ring branch consisting mainly of Collector Ring (CR). The Low-Energy Branch (LEB) is mainly dedicated to precision experiments with energy-bunched beams stopped in a gas cell or an active stopper. This branch is complementary to the ISOL (Isotope Separator On-Line) facilities [2], since all elements and short-lived isotopes can be studied.

The layout of Super-FRS consists of magnets with a maximum magnetic rigidity $B\rho_{\max} = 20$ Tm. This increase in rigidity compared with the already existing FRS facility at GSI is determined by the goal to circumvent atomic charge-changing collisions up to the heaviest projectile fragments. It is also planned to use the Super-FRS as a separator-spectrometer, in which secondary and/or tertiary targets and possible ancillary detectors will be placed at various focal planes allowing the use of the remaining section of Super-FRS as a spectrometer.

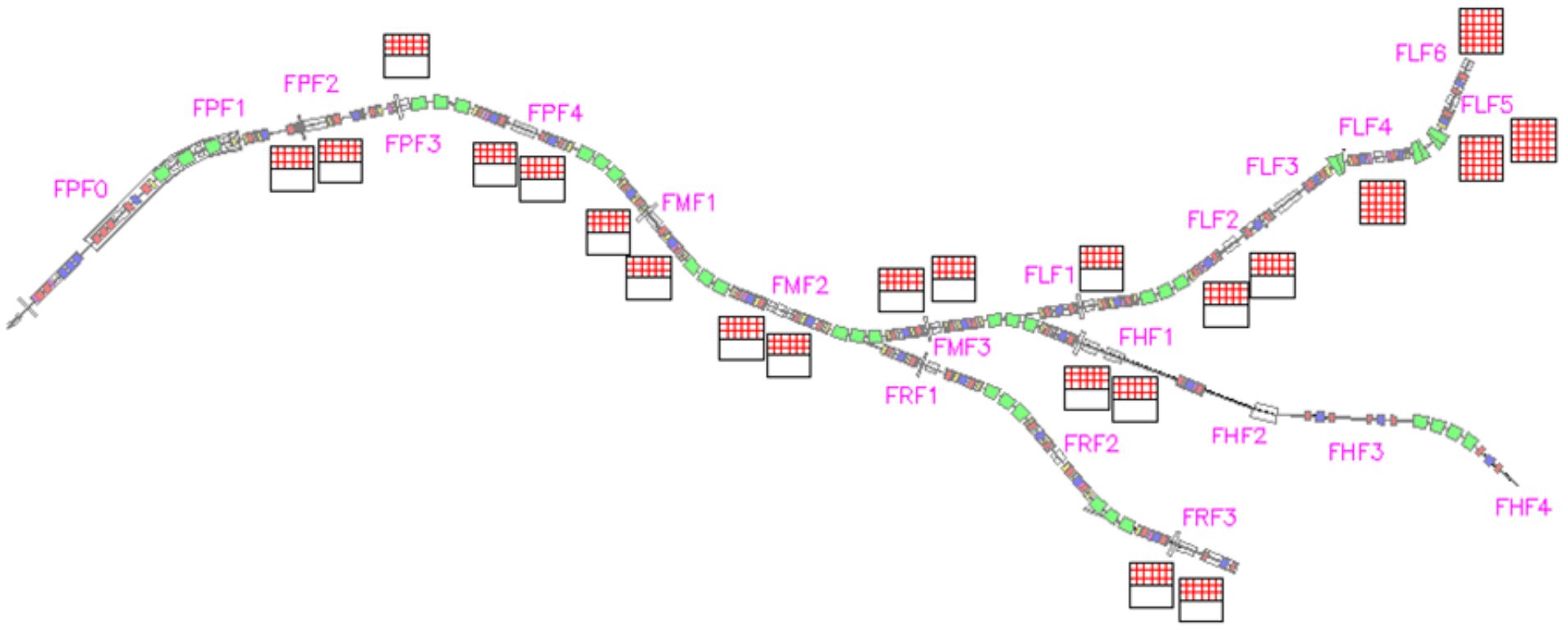

Figure 1: Layout of Super-FRS and the tracking detectors (red symbols). At all focal planes, except FLF4, FLF5 and FLF6, each tracking detector is integrated in a ladder.

Super-FRS has to efficiently separate in-flight rare isotopes produced via projectile fragmentation of all primary beams up to $^{238}$U and via fission of $^{238}$U beam. The latter reaction is a prolific source of medium-mass very neutron-rich nuclei. However, due to the relatively large amount of kinetic energy released in the fission reaction, the products populate a large phase space and thus is one reason for the need of a large acceptance for Super-FRS. This has obvious implications to the physical side of the tracking detectors.

Besides the fragment intensities, the selectivity and sensitivity are crucial parameters that strongly influence the success of an experiment with very rare nuclei. A prerequisite for a clean isotopic separation is that the fragments have to be fully ionized to avoid cross contamination from different ionic charge states. Multiple separation stages are necessary to efficiently reduce the background of such contaminants.

The main parameters of Super-FRS are given in Appendix I and a brief introduction to the physics case in Appendix II.

## 1.2 Tracking system Overview

The main task of the tracking detectors is to provide $B\rho$ of particles through track reconstruction *event-by-event* that is required for particle identification. The $B\rho$ value is related to the particle momentum. Since particle momentum is proportional to the particle position at the focal plane, the $B\rho$ value can be deduced from particle tracking.

The tracking detectors of the Super-FRS are therefore expected to measure the space coordinates of the slow-extracted ion/fragment beam, i.e. transversal (horizontal and vertical) to the beam axis, event-by-event. Operating in pair at the same station (or focal plane) they provide the horizontal and vertical angles of the ion trajectory. The designated locations of the tracking detectors along the Super-FRS are shown in Figure 1. Three areas are considered: the Pre-Separator (PS), the Main-Separator (MS) and the Energy Buncher (EB) (see Appendix I). The total number of detectors per station is listed in Table 1.

The resolving power of the spectrometer should not be compromised by the detectors, i.e. the material inserted in beam should be minimal. Interactions between the detector material and beam should not change the beam composition in $A$ or $Z$. In addition, the velocity vectors of the beam particles should remain minimally affected.

The detectors are operated in vacuum inside a beam diagnostic chamber, except at FMF2a, FHF1a and FLF6, where some units are foreseen to be mounted and working in air, depending on the particular experiment. Vacuum is foreseen to be on the order of $10^{-7}$ mbar. Each detector inside the vacuum chamber is vertically moveable along the transversal beam cross section. The detectors are mounted on the ladder whose movement is conducted by stepper motor driven linear actuators on an installation flange. All detectors have to be handled from the top. While inserted they will provide the horizontal and vertical ion position measurement event-by-event. The design of the tracking detectors installed in the PS region, i.e. at FPF2, FPF3 and FPF4, must support robot handling.

## 1.3 Parameter requirements

The tracking detectors of Super-FRS must be able to provide information on the deviation from nominal beam optics, while operated with slow-extracted beam on event-by event basis. Moreover, in the MS they are used to tag the momentum of the ion/fragment beam, in order to provide the $B\rho$ values required for unambiguous fragment identification in mass.

Due to the high-resolution achromatic mode of Super-FRS, the tracking detectors are crucial to obtain a precise momentum measurement of the produced fragments. They can be also used to apply a position correction for the ion velocity which is measured with another detector system, whenever needed. The technical design is defined by the main beam parameters (e.g. beam energy and intensity, electric charge, duty cycle, etc.). At the Super-FRS, high-energy primary and secondary fully stripped ion beams up to uranium (see Appendix I) need to be considered.

| Station | Unit | Separator region |
|---|---|---|
| FPF2 | 4 | PS |
| FPF3 | 1 | PS |
| FPF4 | 2 | PS |
| FMF1 | 2 | MS |
| FMF2 | 4 | MS |
| FMF2a | 4 | MS |
| FMF3 | 2 | MS |
| FHF1 | 2 | MS |
| FHF1a | 2 | MS |
| FLF1 | 1 | MS |
| FLF2 | 2 | MS |
| FRF3 | 2 | MS |
| FLF4 | 1 | EB |
| FLF5 | 2 | EB |
| FLF6 | 1 | EB |

Table 1: Tracking detectors at the Super-FRS stations. PS=Pre-Separator, MS=Main-Separator and EB= Energy Buncher of the Super-FRS

The Super-FRS tracking detectors should operate up to 10 MHz beam intensity close to 100% efficiency at the stations FPF4, FMF1, FMF2, FMF3, FHF1, FLF2 and FLF4. At the focal planes FPF2, FPF3, FLF1, FRF3, FLF5 and FLF6 high detection efficiency is not mandatory since event-by-event particle identification is not required or possible at those stages.

The technical specifications of the tracking detectors are listed in Tables 2 and 3.

Monte Carlo simulations were performed to estimate the horizontal and vertical aperture of the detectors along Super-FRS. The typical case used to explore the full acceptance is represented by the fission of uranium beam impinging on a target at the entrance of the separator. A primary beam of $^{238}$U at 1.5 GeV/u (up to 3x$10^{11}$ ions/spill) and a graphite production target are used to produce fragments in the Sn region at about 1 GeV/u. Simulated horizontal and vertical position distributions of $^{132}$Sn fragments are plotted in Figures 2-6. Except at FPF3, two different positions are considered for tracking, one in front of the focus and one behind the focus. In both cases the distance from the focus is about 1550 mm. In the pre-separator, where the first fragment selection takes place, the rate is still prohibitive and no tracking can be performed during experiments. Only with reduced intensity, e.g. $10^4$ ions/spill, the detectors will be inserted and used to steer the primary

| **Device** | **Parameter** | **Value** | **Unit** |
|---|---|---|---|
| Tracking detectors operated in vacuum | Maximum beam spot size | $380 \times 150$ | mm$^2$ |
| | Maximum allowed length in beam direction | see Table 3 | |
| | Position resolution | <0.5 | mm |
| | Positioning | $\pm$ 0.1 | mm |
| | Time resolution | <3 | ns |
| | Maximum rate capability | 1-3 | kHz/mm$^2$ |
| | Dynamic range | 600 (see section 10) | fC |
| | Gas | P10 or other | |
| | Minimum active area | see Table 3 | |
| Tracking detectors operated in air | Maximum beam spot size | $380 \times 80$ | mm$^2$ |
| | Position resolution | <0.5 | mm |
| | Time resolution (FWHM) | <3 | ns |
| | Maximum rate capability | 1-3 | kHz/mm$^2$ |
| | Dynamic range | 600 (see Section 10) | fC |
| | Gas | P10 or other | |
| | Minimum active area | see Table 2.3 | |

Table 2: Technical specification of the tracking detectors.

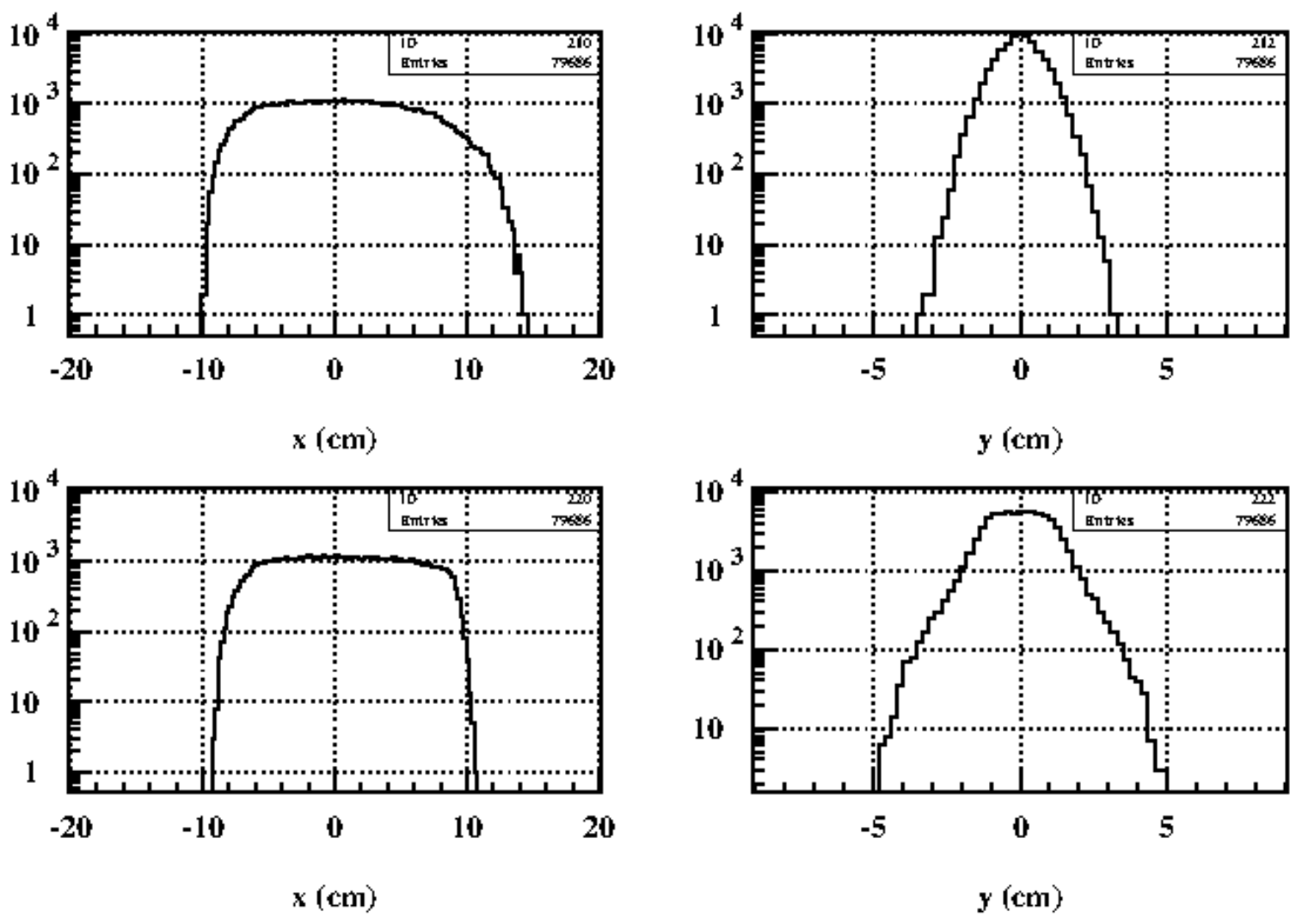

Figure 2: Horizontal and vertical distributions at FPF2 (dispersive focus) in front (upper panels) and behind (lower panels) the focus.

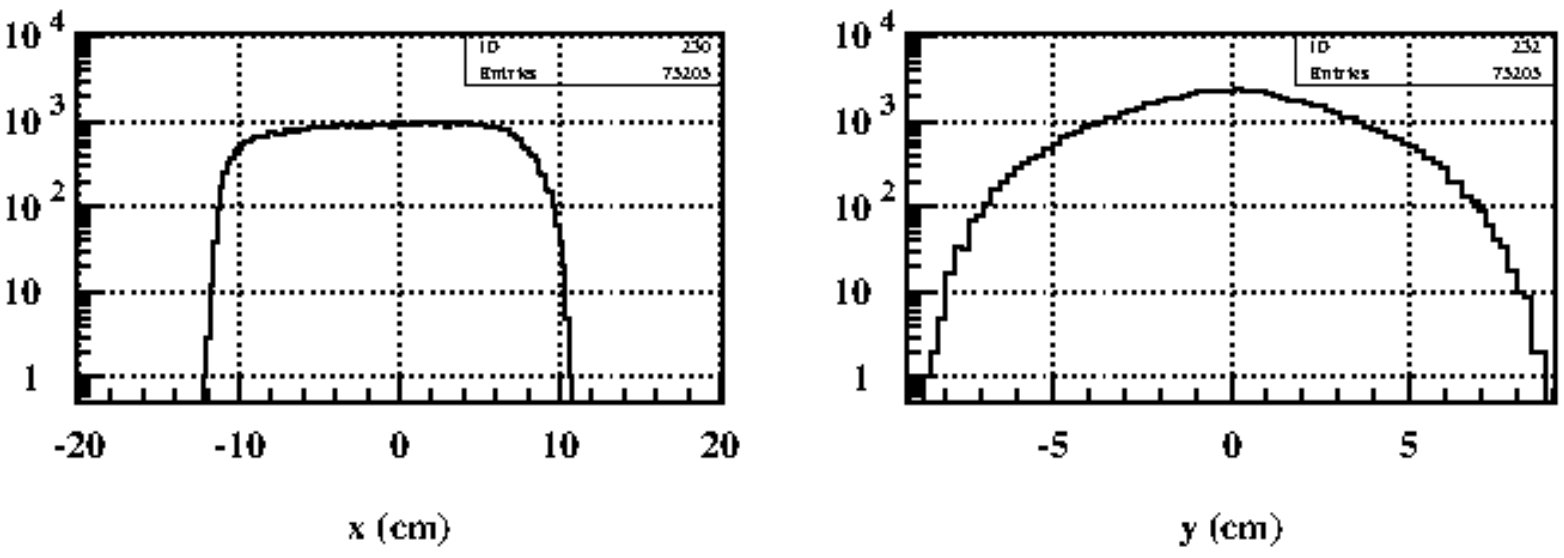

Figure 3: Horizontal and vertical distributions at FPF3.

beam.

At the FPF3, where there is no real focus, only one detector is needed for diagnostics.

At FPF4 the beam-spot size is reduced. For symmetry the same size of the beam can be considered at FHF1.

During the standard Super-FRS operation the detectors at FMF2 need to be always in place. Since FMF2 is a dispersive focus in the horizontal plane, better precision is required to reconstruct the horizontal focal position and angle. High-rate capability is needed here in experiments, which require the identification of the fragment delivered to the three different branches. Tracking detectors at at FMF1 and FMF3 are expected to work with similar particle rates.

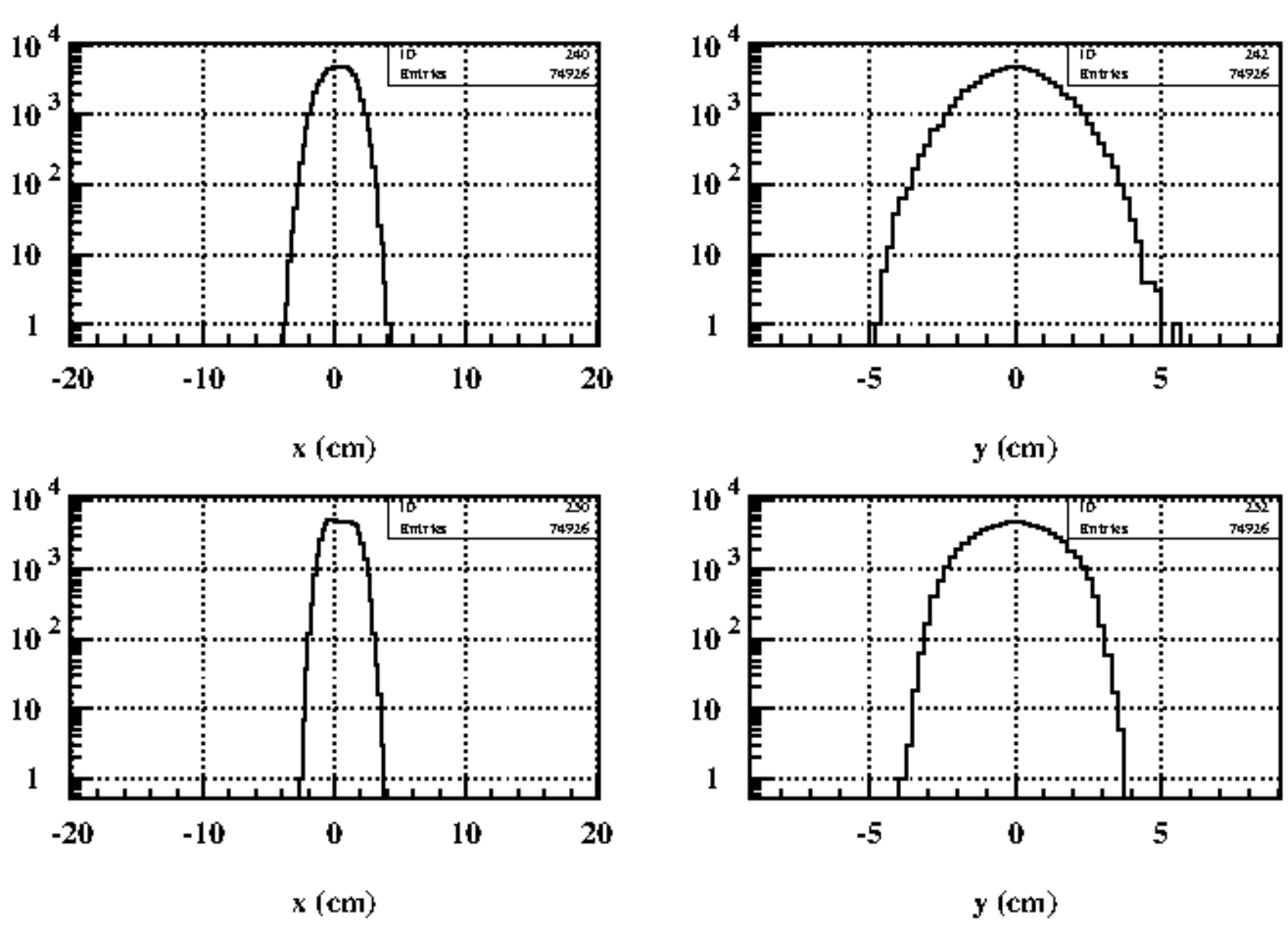

Figure 4: Horizontal and vertical distributions at FPF4 (achromatic focus) in front (upper panels) and behind (lower panels) the focus.

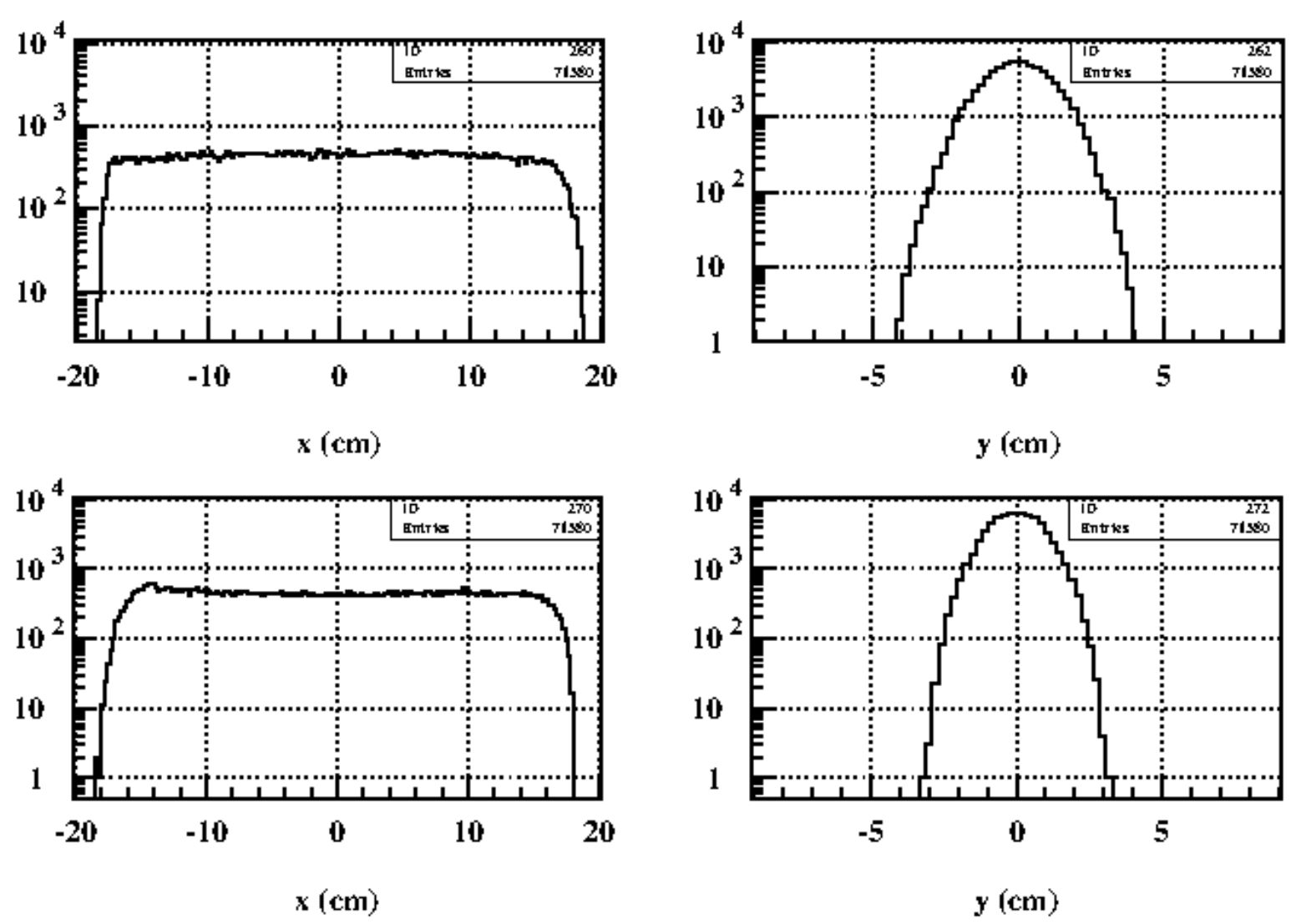

Figure 5: Horizontal and vertical distributions at FMF1/FMF3 in front (upper panels) and behind (lower panels) the focus.

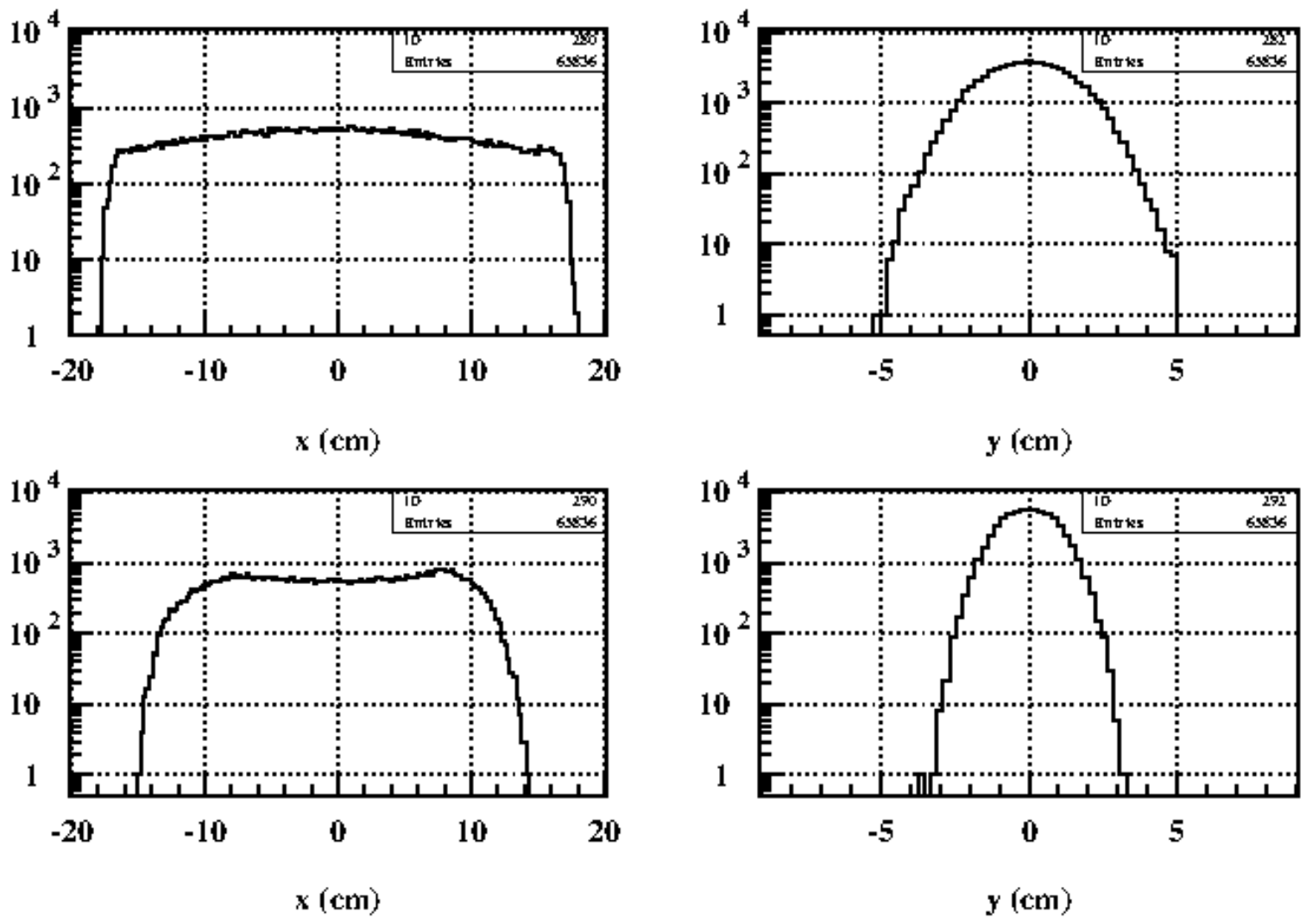

Figure 6: Horizontal and vertical distributions at FMF2 (achromatic focus) in front (upper panels) and behind (lower panels) the focus.

Starting from a total standard rate of $10^7$ ions/spill and assuming an area of $200 \times 50$ mm$^2$ (see Figure 6) with a spill duration equal to 1 s, a local rate of about 1 kHz/mm$^2$ is achieved. At FHF1 and FLF2 (or FRF3), after the second fragment separation, the maximum rate is reduced by several orders of magnitudes. On the other hand, there will be experiments where not so exotic species need to be delivered. For this reason, $10^7$ ions/spill maximum rate capability must be considered also at the end of the HEB and LEB. Due to the beam achromaticity, assuming a reduction in the spot size of a factor 2-3, the local rate increases up to 2-3 kHz/mm$^2$.

The dynamic range of the tracking detector has to be flexible. All detectors have to work with all beam species from proton to heavy ions up to uranium. Therefore, the total kinetic energy of the beam varies largely from few hundreds of MeV to hundreds of GeV.

## 1.4 The GEM-TPC detector

The field cage of the detectors is an improvement of the design of the time-projection chamber detectors currently used at the FRS at GSI for beam tracking. The present design is a novel concept that combines the TPC detector with an amplification stage based on GEM foils [3, 4]. The GEM-TPC to be established must have high electric field uniformity along the charge-carrier drift field of the TPC chamber, low GEM-foil leakage current, high amplification of each of the GEM-foil levels and a fast signal readout. In addition, the signal to noise ratio needs to be high.

| Focal plane | X (mm) | Y (mm) | Z (mm) | Separator region |
|---|---|---|---|---|
| FPF2 | 380 | 80 | 250 | PS |
| FPF3 | 380 | 150 | 250 | PS |
| FPF4 | 200 | 80 | 400 | PS |
| FMF1 | 380 | 80 | 400 | MS |
| FMF2 | 380 | 80 | 400 | MS |
| FMF2a | 380 | 80 | 400 | MS |
| FMF3 | 380 | 80 | 400 | MS |
| FHF1 | 200 | 80 | 400 | MS |
| FHF1a | 200 | 80 | 400 | MS |
| FLF1 | 380 | 80 | 250 | MS |
| FLF2 | 200 | 80 | 400 | MS |
| FRF3 | 200 | 80 | 250 | MS |
| FLF4 | 380 | 150 | 400 | EB |
| FLF5 | 380 | 150 | 250 | EB |
| FLF6 | 200 | 80 | 250 | EB |

Table 3: Minimum active area and maximum allowed length in beam direction of the tracking detectors.

Due to high radiation-level environment, detectors that will be installed for the beam diagnostics are expected to be operational as long as possible without any need for human intervention.

The optimization of the GEM-amplification stage is important for operating at high rates. To avoid decline in efficiency and in resolution, the field uniformity inside the TPC drift volume has to be optimized.

The GEM-TPC detectors for particle tracking in the main separator will be operated in twin field-cage configuration with two twin detectors at almost each focal plane. Such a configuration will unambiguously detect the particle track $(x, y, z)$ coordinates at the entrance and exit) and thus its magnetic rigidity $(B\rho)$ can be deduced. The idea behind the twin field cage configuration is to achieve 100% tracking efficiency even at highest rates (10 MHz). The concept of the twin GEM-TPC detector prototype HGB4 (hereafter HGB4 when referring to the prototype) working in air will be described in the following sections.

# 2 Detector Mechanical Structure and its Components

An overview of the geometry of HGB4 in its actual design is shown in Fig. 7. The principal idea of the design of the prototype was to keep the future user-case in mind and to study required technologies. Any changes required to adapt to different geometries at various places of operation should be possible with a minimum of work with regards to design changes and fabrication. Special care was taken to achieve a maximum of compactness and end up with a minimum of components to be disassembled/assembled during maintenance.

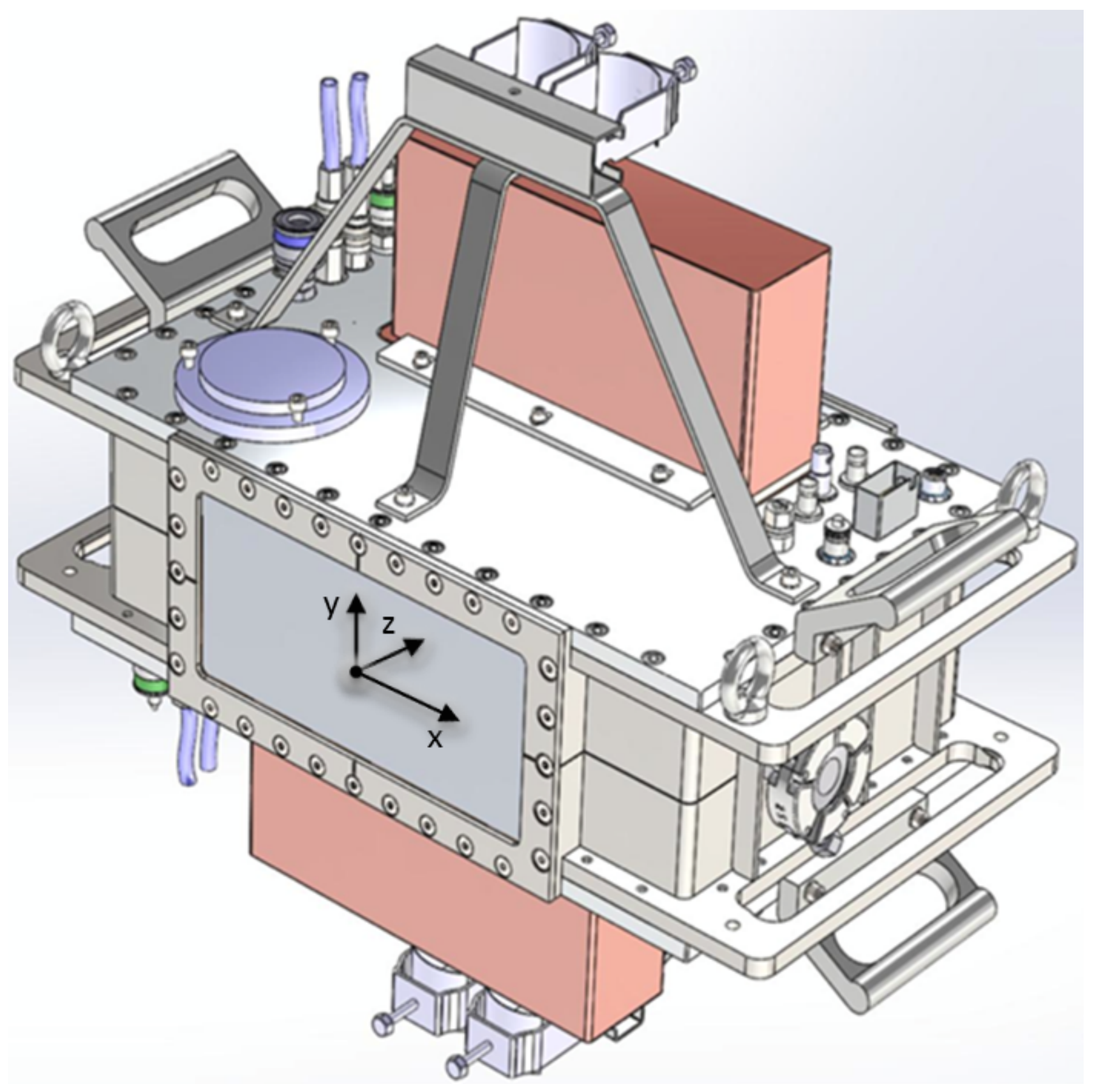

Figure 7: The HGB4 detector.

## 2.1 Introduction

The version of the GEM-TPC detector described herein is based on a twin configuration of two identical active volumes, each enclosing a drift volume of rectangular or boxed shape with a constant electrical field with up to 550 V/cm. The direction of the fields has been chosen to be in vertical axis ($Y$) perpendicular to the beam of traversing particles, their orientation is opposite within the two volumes following each other along the beams path ($Z$). The centres of the two field-cage compartments are adjusted to be on one line horizontally ($X$) and vertically ($Y$) within a precision of 0.1 mm. A sketch of this arrangement is shown in Figs. 8 and 9. Due to the intended operation in a magnetic spectrometer, great care has been taken to optimize, respectively minimize the material the beam particles have to traverse when passing the detector volume. A sketch of the various polyimide foils for windows (25 $\mu$m) and field-cages ($2 \times 7.5$ $\mu$m) are shown in Fig. 9.

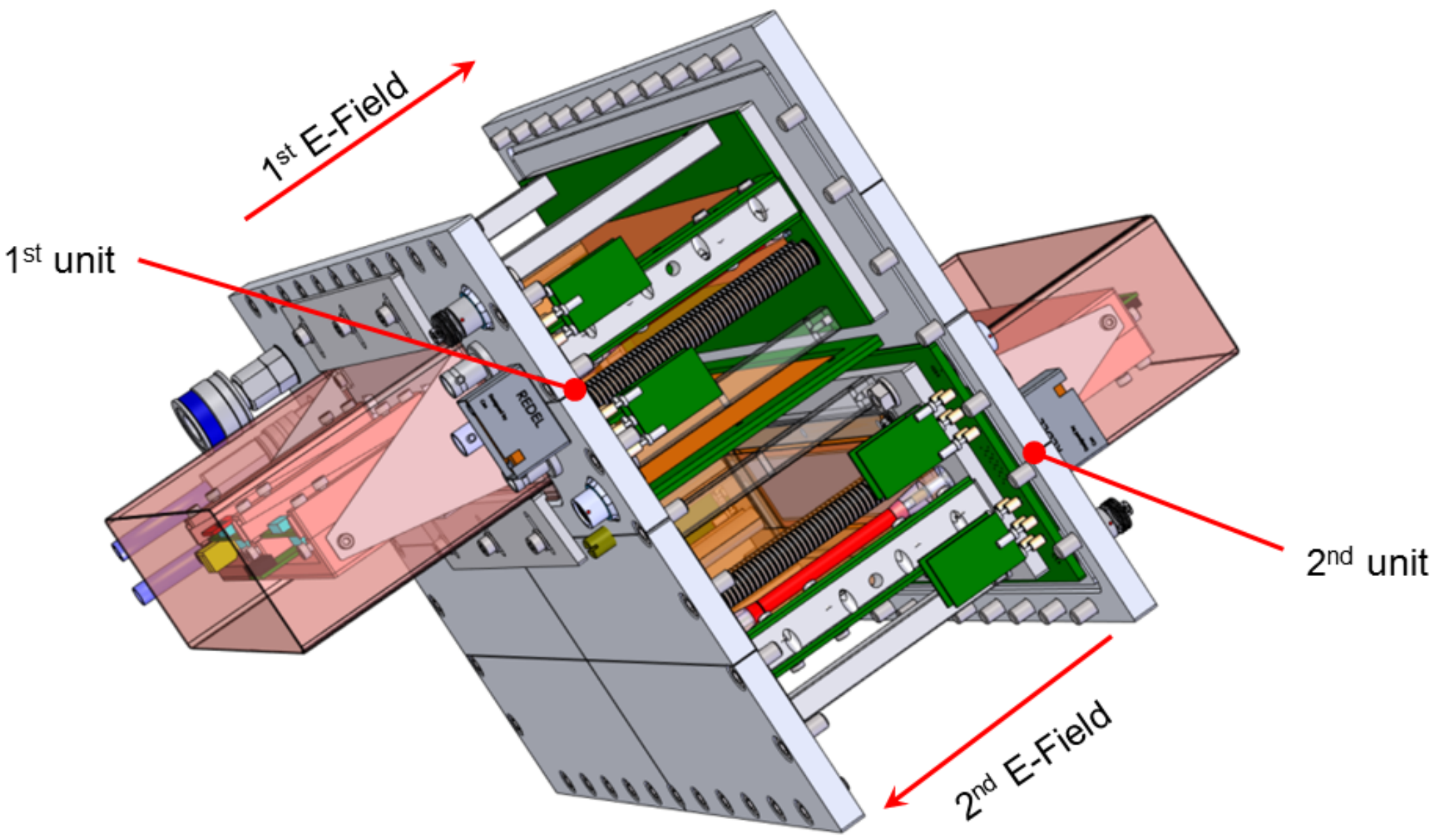

Figure 8: CAD sketch of the field-cage compartments of HGB4.

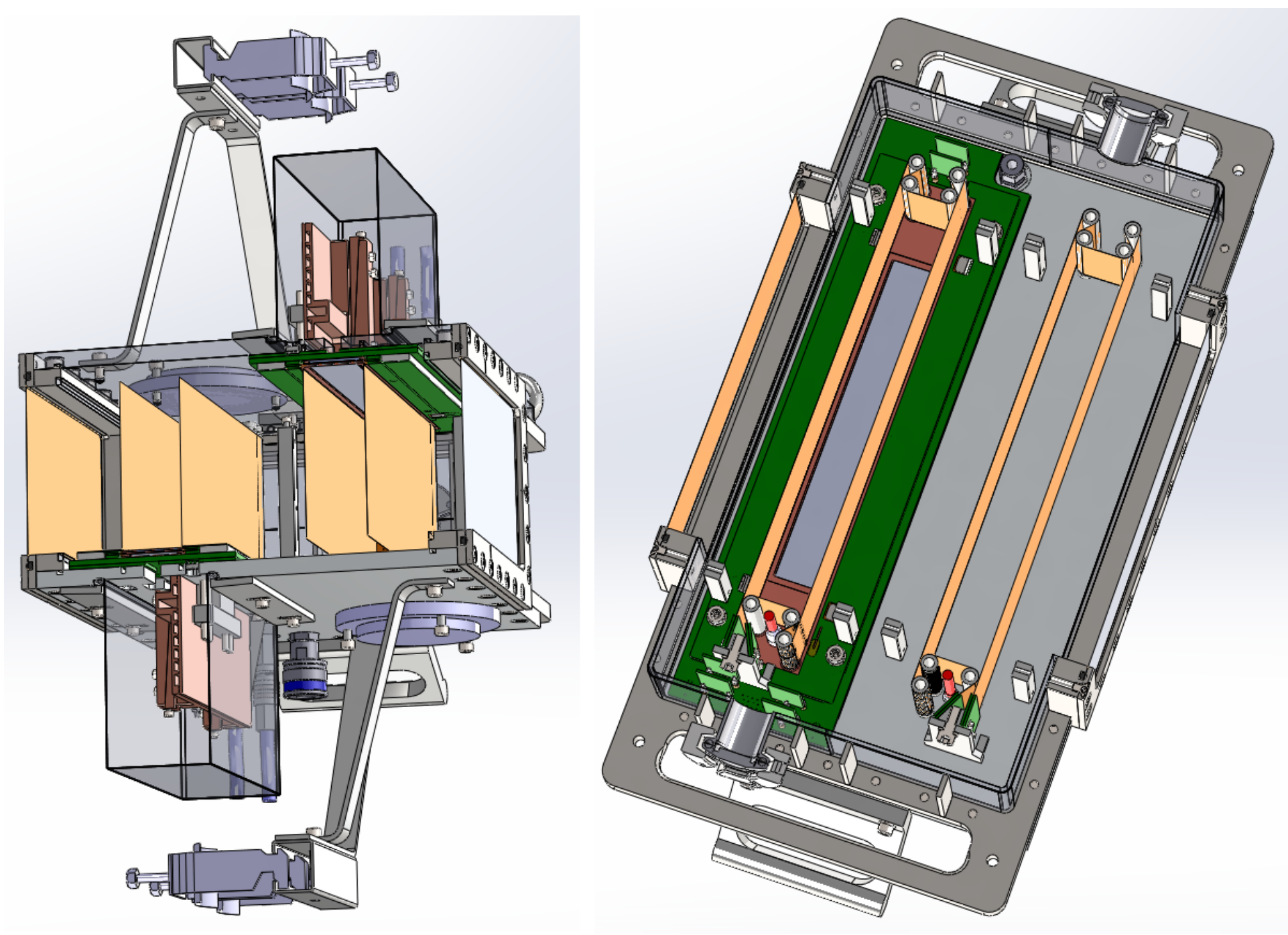

Figure 9: CAD sketch of the field-cage compartments of HGB4 demonstrating their arrangement and the various thin layers of material the beam has to pass (horizontally from left to right).

## 2.2 Housing and Vessel

The housing of the detector has been realized in two types of constructions for testing purposes:

1. one version is made from stainless steel in a welding-type design as shown in Fig. 10

2. another specimen has been machined from a full block of aluminium as shown in Fig. 11

The current designs of the housing vessel allow a precision mounting of the two compartments of HGB4 relative to each other with an accuracy of better than 0.1 mm in all directions.

Both types allow operation under the reduced pressure of e.g. 0.3 mbar up to an overpressure of approximately 1 bar. The welded-type specimen of the vessel can be either pumped down to several mbar and/or may be applied in a vacuum surrounding by the appropriate choice of foils, depending on the experimental conditions, of the entrance and exit windows ($20 \times 10 cm^2$) made of either 25$\mu$m thick Kapton or 50$\mu$m thick stainless steel.

## 2.3 Flanges

Due to principal design rules, the whole set-up of the two opposing compartments of HGB4 were set up to be essentially identical.

An overview picture of a single panel flange of HGB4 is shown in Fig. 12. The two groups of supplies (electrical and fluidal) as well as high-density connectors for the read out of the signals are clearly discriminable. Line-engravings are realized for LASER adjustment together with a bubble-level.

In order to maximize compactness, the actual flanges also serve to distribute all electrical and fluidal supplies as indicated in Fig. 13. Moreover, they provide appropriate stiffening of the pad-planes glued to them in order to achieve optimal flatness of the pad plane PCB. They are part of the screening concept against electromagnetic interferences and close the Faraday cage together with the metallic vessel of the detector.

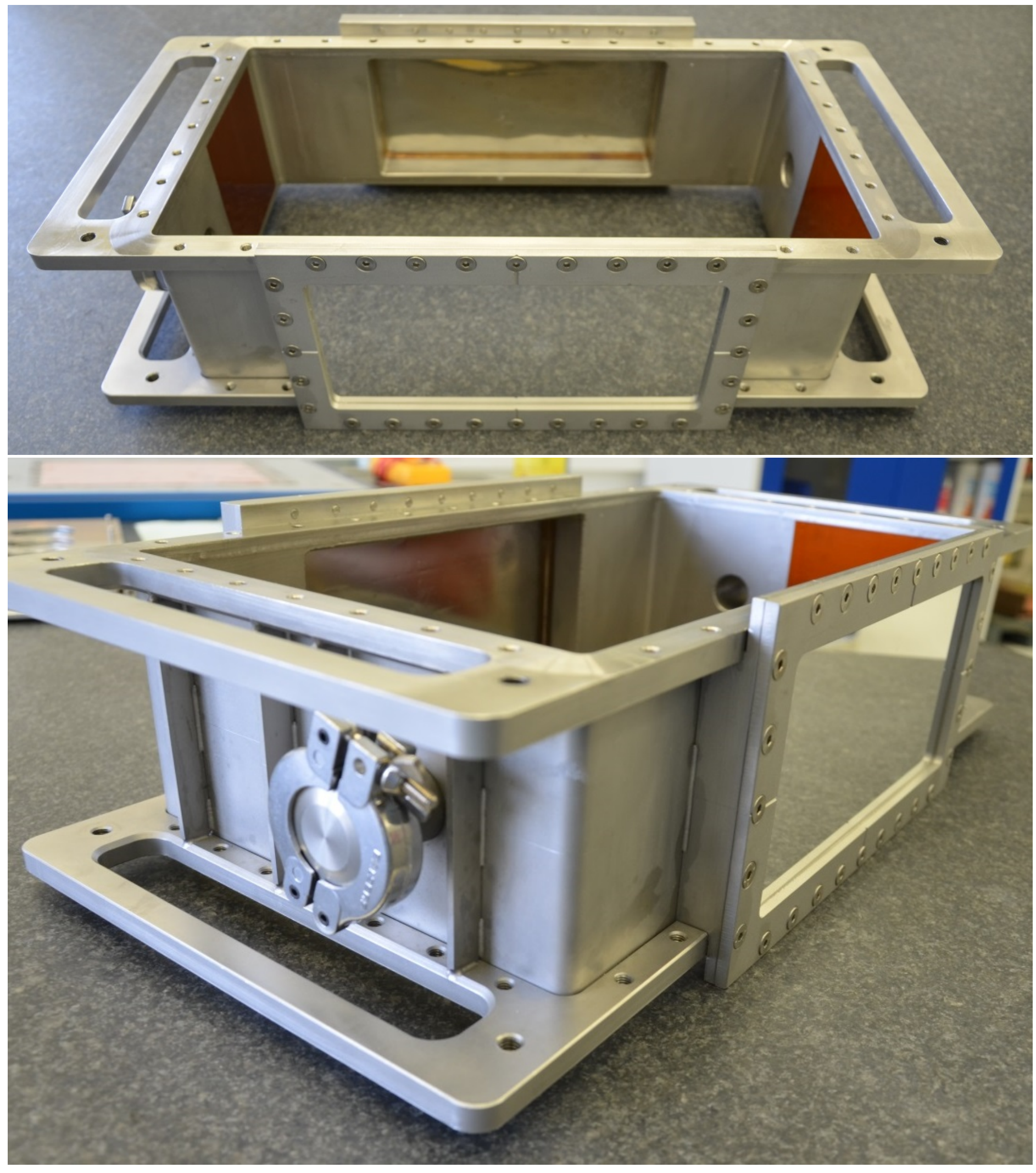

Figure 10: Housing made from stainless steel in a welding-type construction. The KF-type exhausts can be used to evacuate the vessel enforcing outgassing of the detector materials installed inside in order to achieve better gas-purity.

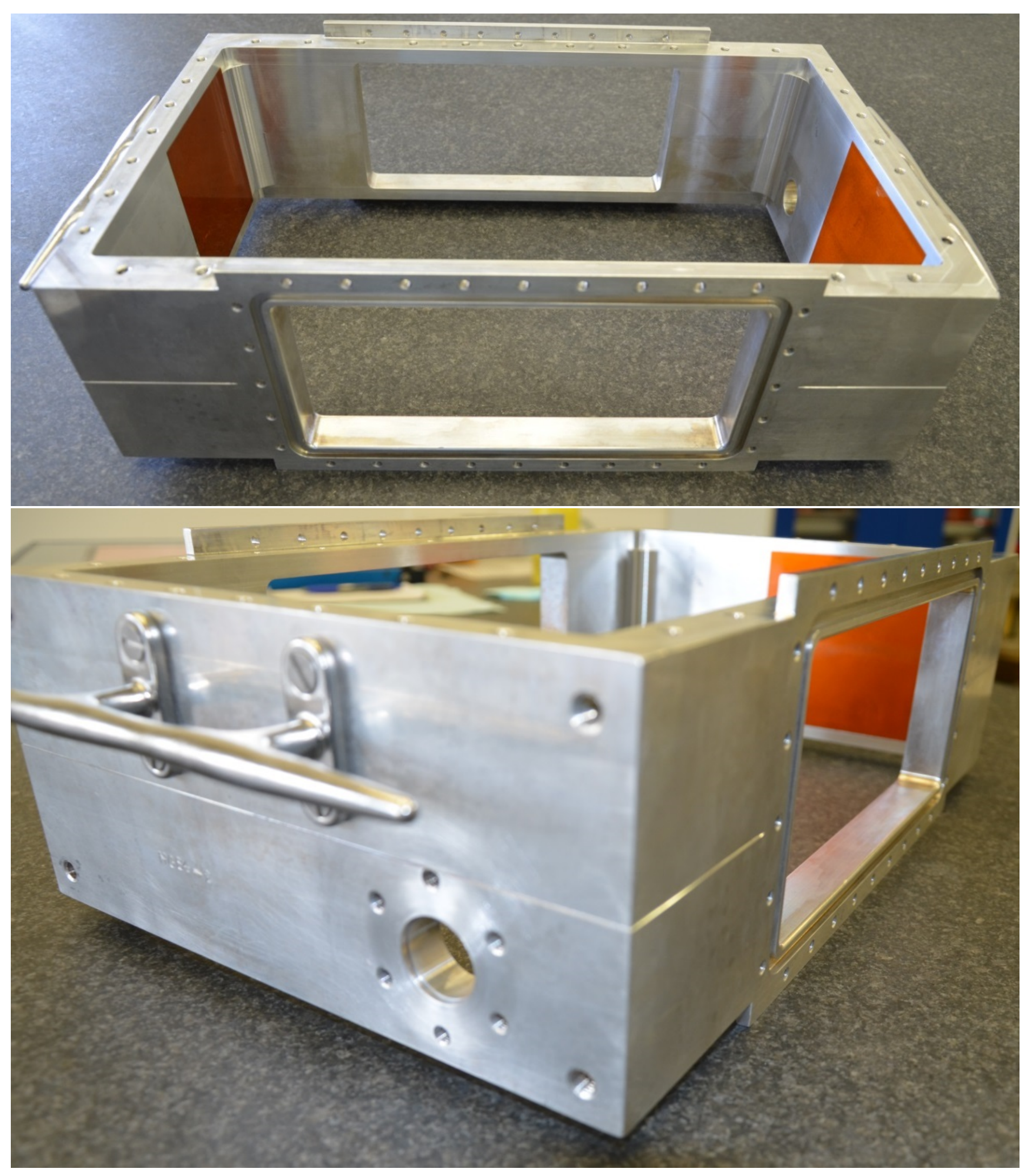

Figure 11: Housing milled-out from a single block of Aluminium.

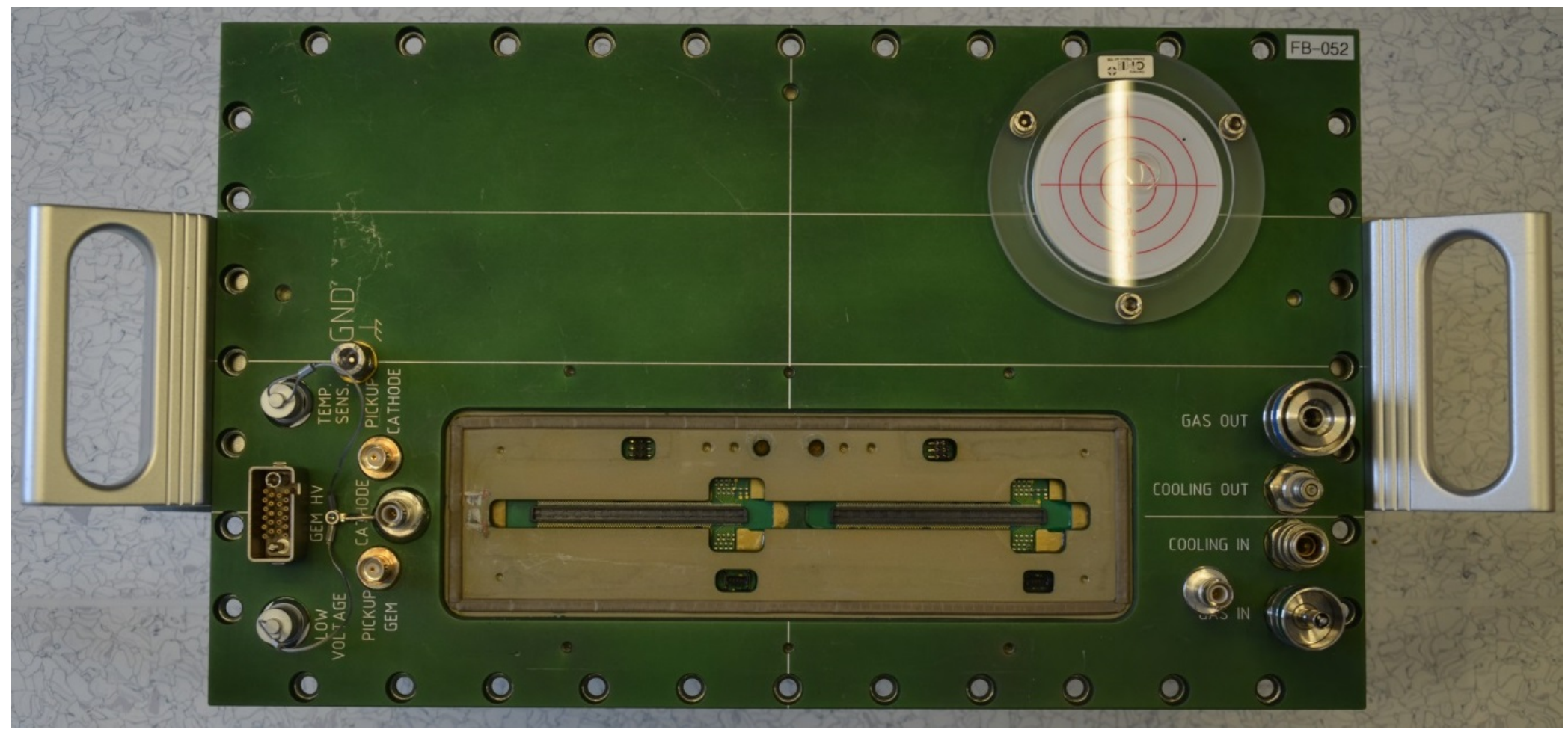

Figure 12: Overview picture of one of the back-side of the flanges mounted to HGB4 facing the inside volume of the detector. The two groups of grooves as part of the distribution of the fluidal supplies are visible as well as the screw-mounted connectors

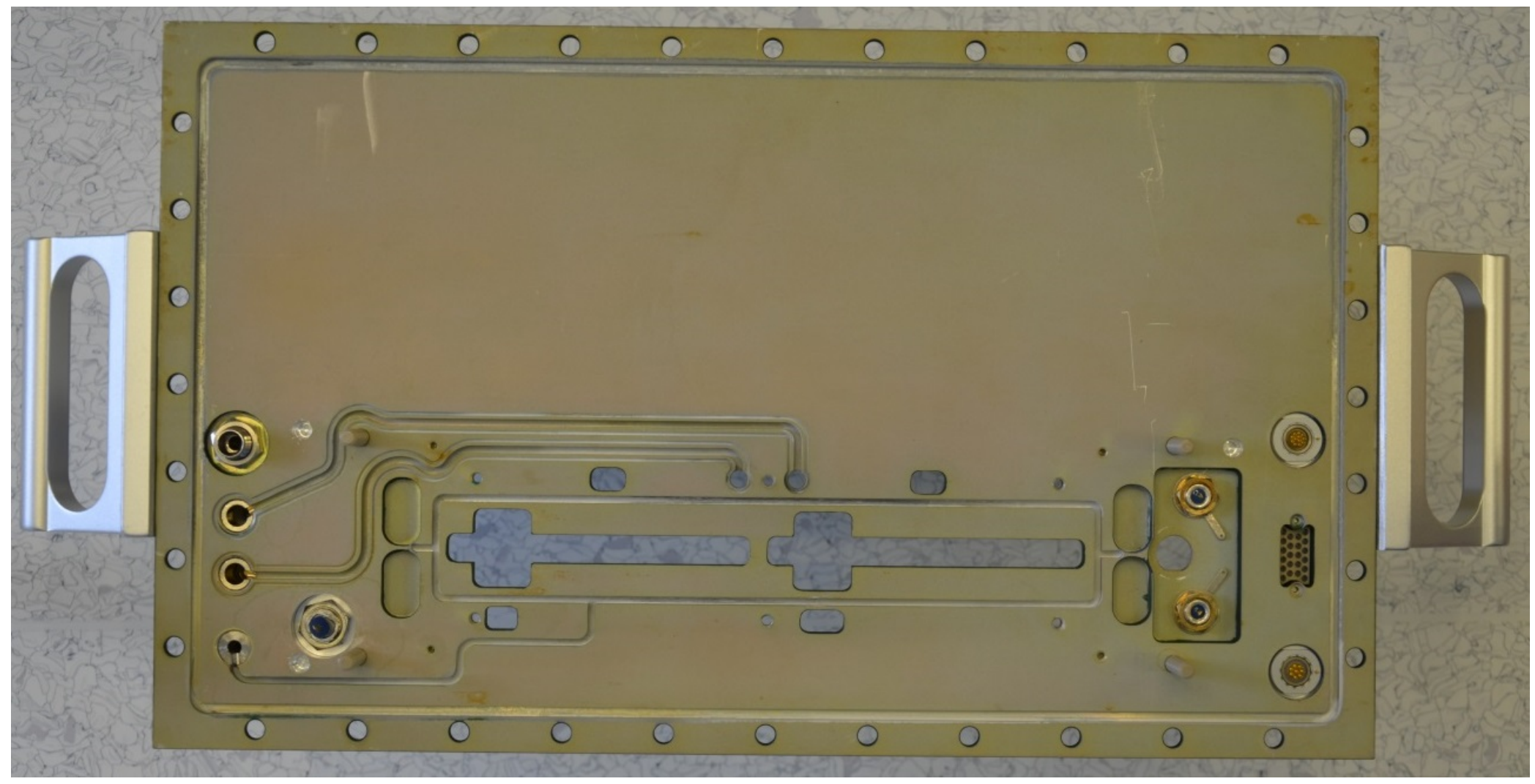

Figure 13: Overview picture of one of the flanges mounted to HGB4 facing the surrounding ambient.

## 2.4 Adjustment

Ease of operation was a principal design goal to be achieved for the pre-series detector. Thus, several design details have been realized as shown in Fig. 14 in order to facilitate adjustment.

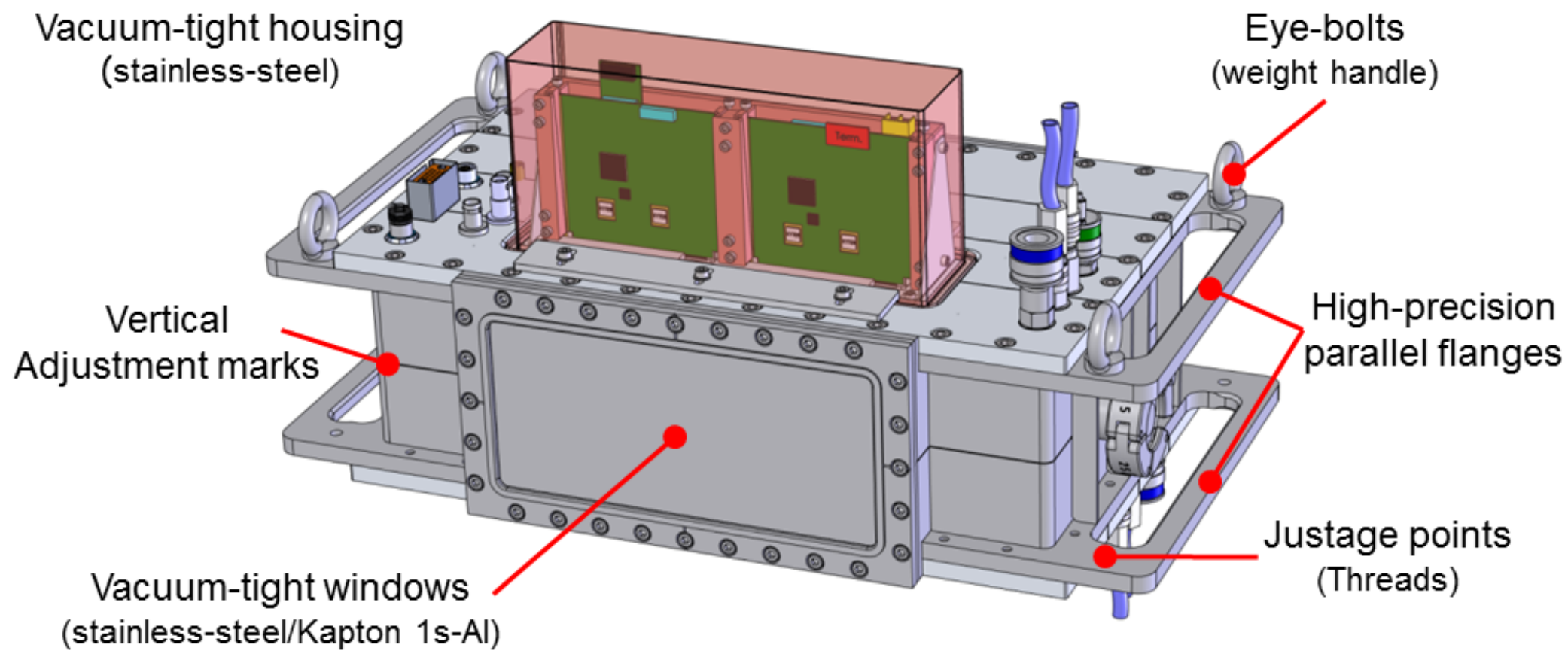

Figure 14: CAD sketch of the chamber housing emphasizing various design details which support the adjustment during mounting.

Engraved marks centred on the vessels outer surfaces (see Figs. 10 and 11) as well as corresponding engravings on the panel flanges (see Figs. 12) allow for an adjustment with the help of external Laser lines with an accuracy down to 0.1 mm, depending on the achievable width of the LASER lines.

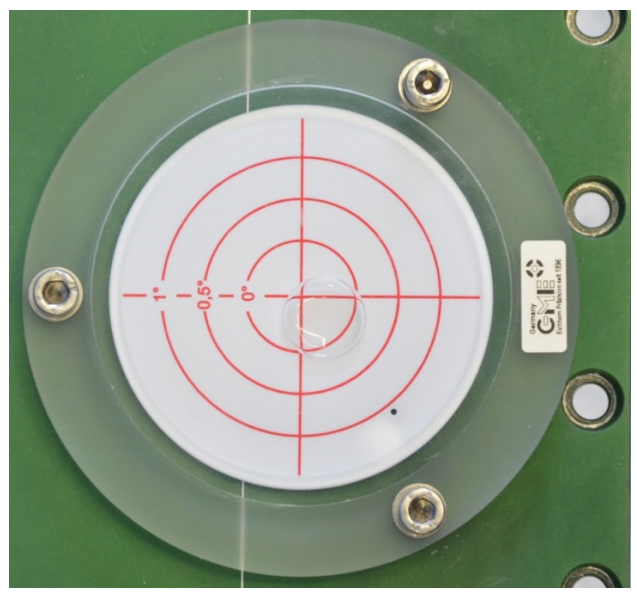

Figure 15: Picture of the bubble-level device mounted on the detector flanges allowing a horizontal adjustment of the detector.

Moreover, a bubble-level device shown in Fig. 15 is installed on the top-level flange (see Fig. 12). It helps to ease adjusted mounting of the chamber at least horizontally with an accuracy of 0.1° which is especially helpful in a crowded experimental surrounding where the LASER beams might not reach the chamber surfaces.

## 2.5 Supplies and Connectors

Two groups of supplies were grouped together to facilitate interconnection during mounting of the detector system.

1. The electricalal supplies and slow-control signals like ground (GND), low, high and bias voltages; sensor and pick-up signals (see Fig. 16).

2. The fluidal supplies such as the detector gas as well as the cooling fluid (see Fig. 17)

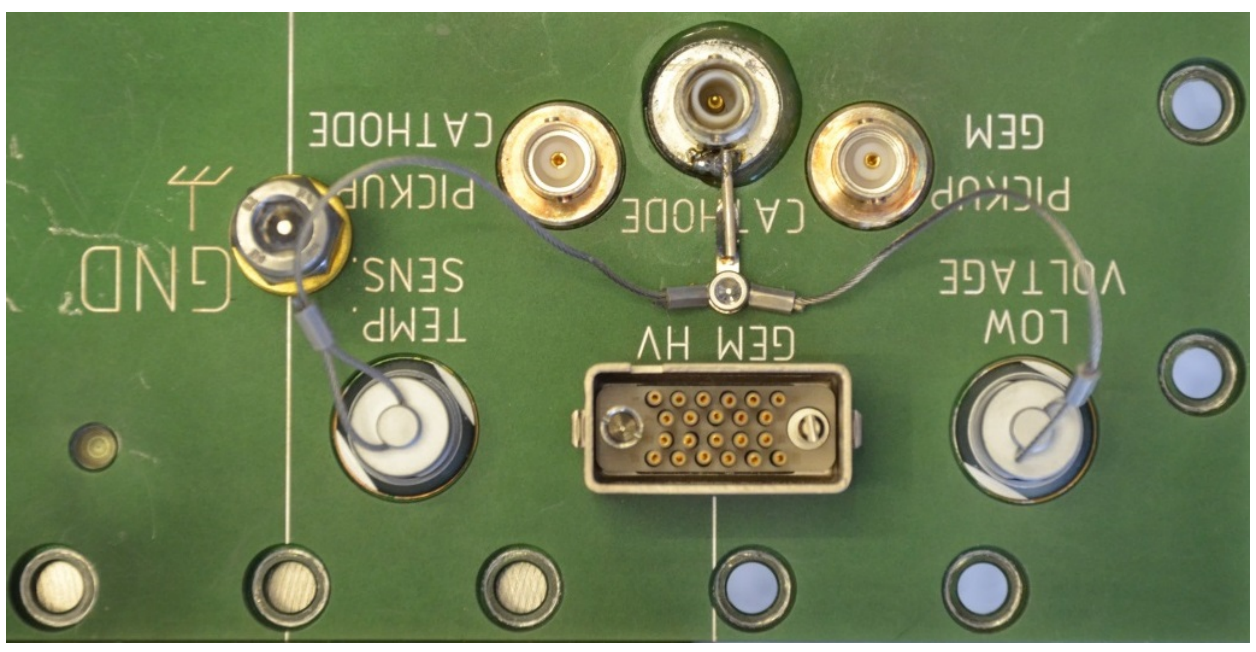

Figure 16: Picture of the electrical supplies and signals: Ground (GND), low, high, and bias voltages, sensor and pick-up signals.

In order to ease connection and to avoid errors, coded self-closing valves have been selected for all fluidal media.

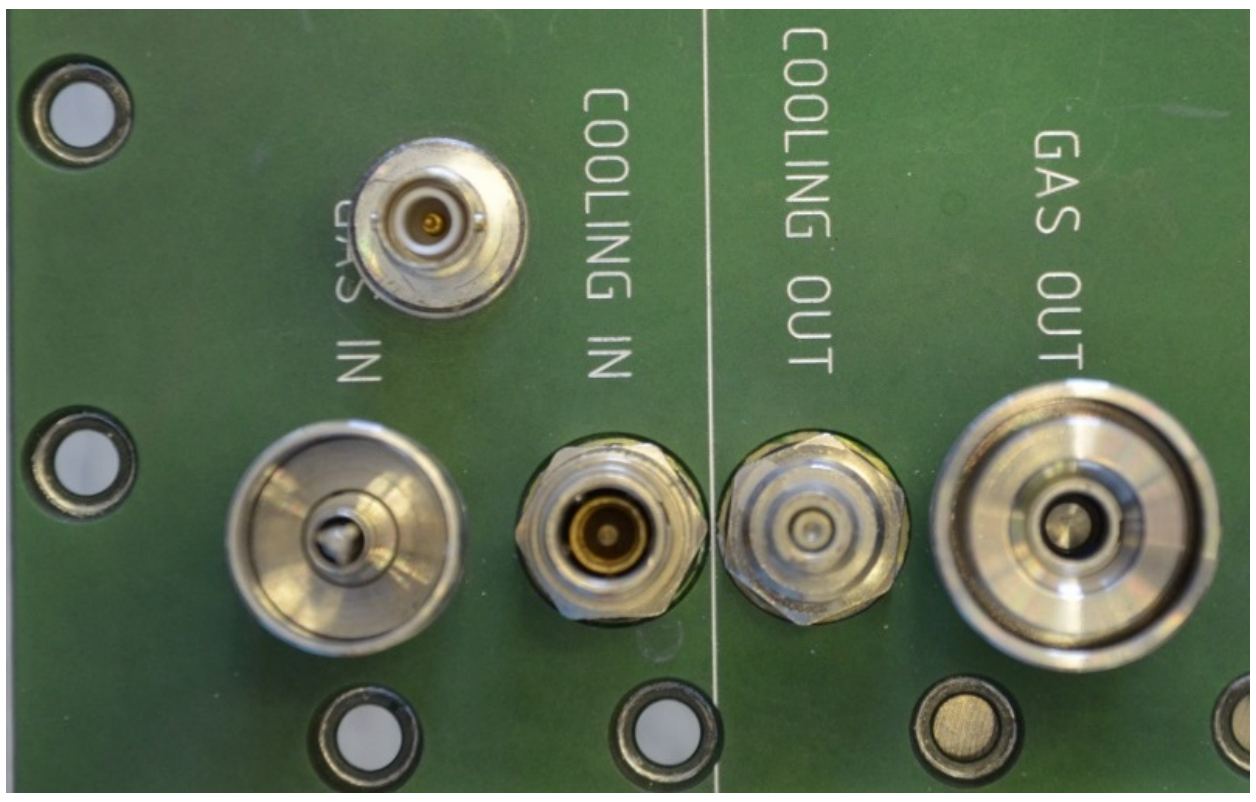

Figure 17: Picture of the fluidal supplies, the detector gas as well as the cooling fluid. Note that it has been decided only after the flanges have already been produced to provide the first strip of the field cage with autonomous biasing.

## 2.6 Cooling System

The current version of the read-out ASIC, which is the XYTER-type NX1.0 and NX1.1 [12], as well as the following detector-near digital read-out chain (ADC and FPGA) require active cooling and a temperature stabilized wit $\pm 1^{\circ}C$ accuracy. In order to facilitate the supply with the appropriate water-based cooling fluids, the distribution of the coolant was designed into the Aluminium flange (see Fig. 13) offering in- and outlet via a connectors with self-closing valves.

A picture of the actual cooler is shown in Fig. 18 without and in Fig. 19 with the cover heat-screen removed and the actual GEMEX1C [13] read-out boards installed (the interlinking GEMCON2 PCB bridge has been removed too). The read-out electronics employed is discussed in detail in Chapter 9.

## 2.7 Gas System

The gas-distribution systems are integrated into the detector panel as well as into the cathode plate (see Fig. 17). Each compartment of HGB4 is provided with an autonomous supply. Inside the vessel, fresh gas is directly supplied into the respective induction gap (the volume just above the pad plane, facing GEM-foil no.3) of each field-cage compartment via the pad-plane. The corresponding exhausts are part of the cathode support structure of each compartment. This allows for a more or less guided flow across each drift volume.

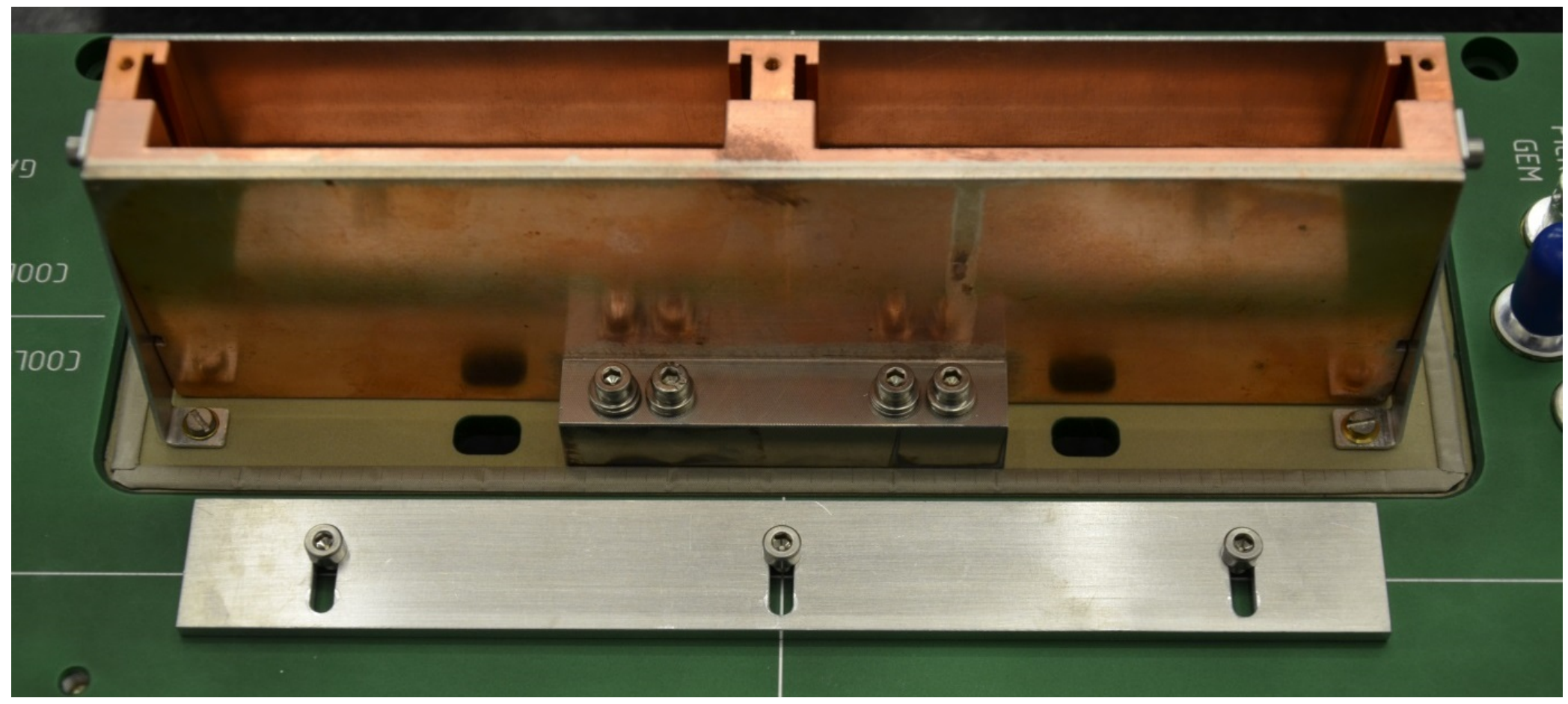

Figure 18: The active copper cooler.

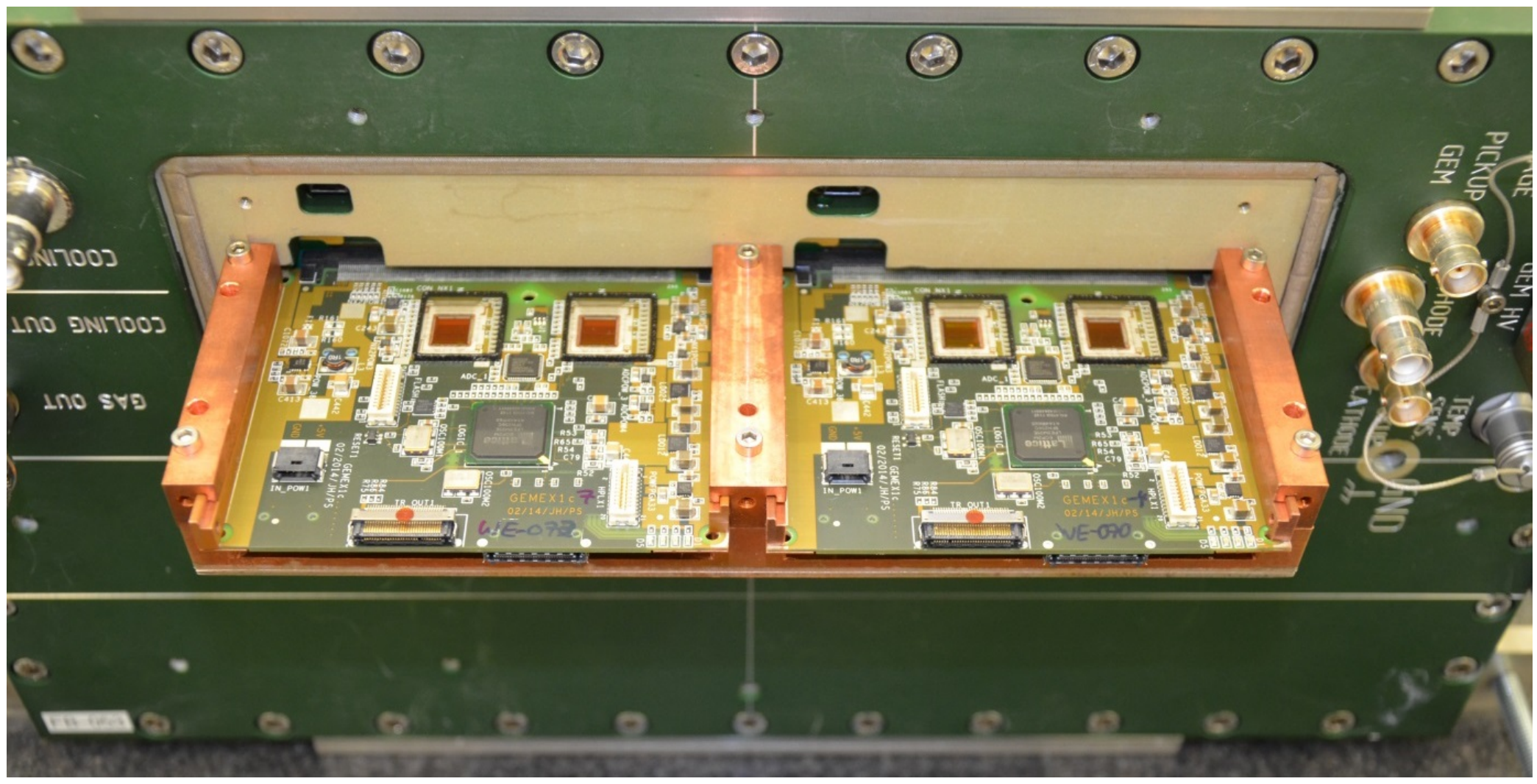

Figure 19: The active copper cooler with the cover heat-screen removed and the actual GEMEX1C read-out boards installed (the interlinking GEMCON PCB bridge has also been removed).

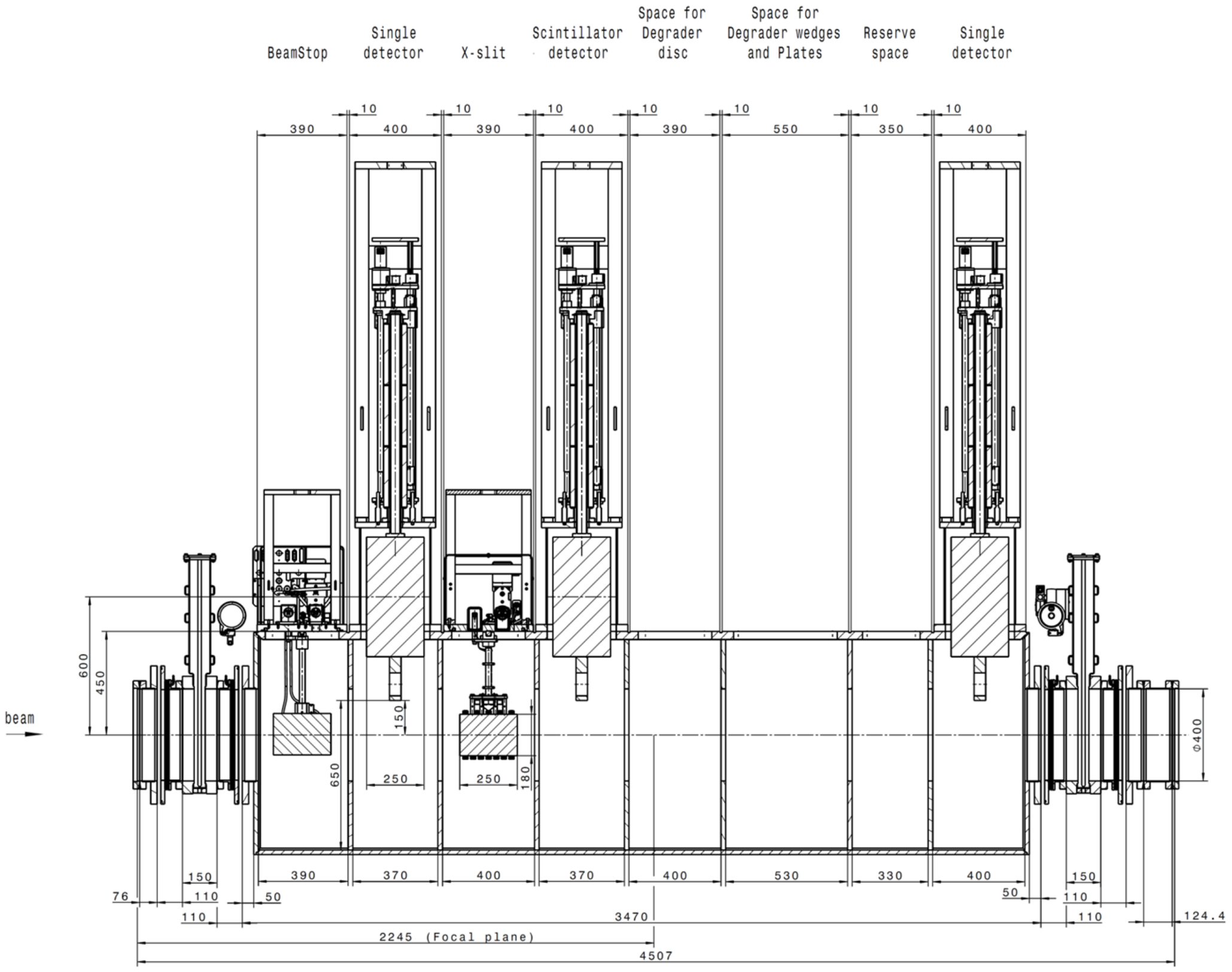

Figure 20: Super-FRS diagnostic chamber at FPF2.

# 3 Diagnostic Station and its Components

An example of Super-FRS diagnostic station (FPF2) is shown in Figure 20. The dimension of vacuum flanges in the beam direction is limited by the presence of the other beam diagnostics components. This is one of the main reason of having tracking detectors and profile monitors on the same ladder/drive. The maximum length of the vacuum flanges of the tracking detector at each station is shown in Table 4. The ladder arms and service cables are inserted into the vacuum volume through the vacuum flanges. The detailed design of these is under consideration.

## 3.1 Actuator and Ladder

The final, full-size GEM-TPC detector is attached on the ladder arm that in turn is used to move the detector vertically down to the beam line using a standard linear actuator stepper motor. In addition the ladder arm is used to guide cables, gas and cooling pipes from the detector to outside the vacuum chamber. The connectors are situated on the top flange of the detector and at the other end on the top flange of the ladder arm. The SEM-Grid beam profile detector is situated on the same ladder arm beneath the GEM-TPC detector. Since the SEM-Grid detector is optional to GEM-TPC in the PS section of Super-FRS when using fast extraction, it can be attached on to the same ladder as GEM-TPC. The schematic design of the ladder is shown in Fig. 21. The detailed design of the ladder system is under consideration.

| **Focal plane** | **Flange length (mm)** | **Separator region** |
|---|---|---|
| FPF2 | 400 | PS |
| FPF3 | 400 | PS |
| FPF4 | 550 | PS |
| FMF1 | 550 | MS |
| FMF2 | 550 | MS |
| FMF2a | - | MS |
| FMF3 | 550 | MS |
| FHF1 | 550 | MS |
| FHF1a | - | MS |
| FLF1 | 400 | MS |
| FLF2 | 550 | MS |
| FRF3 | 400 | MS |
| FLF4 | 550 | EB |
| FLF5 | 400 | EB |
| FLF6 | - | EB |

Table 4: Maximum allowed flange dimension along the beam direction.

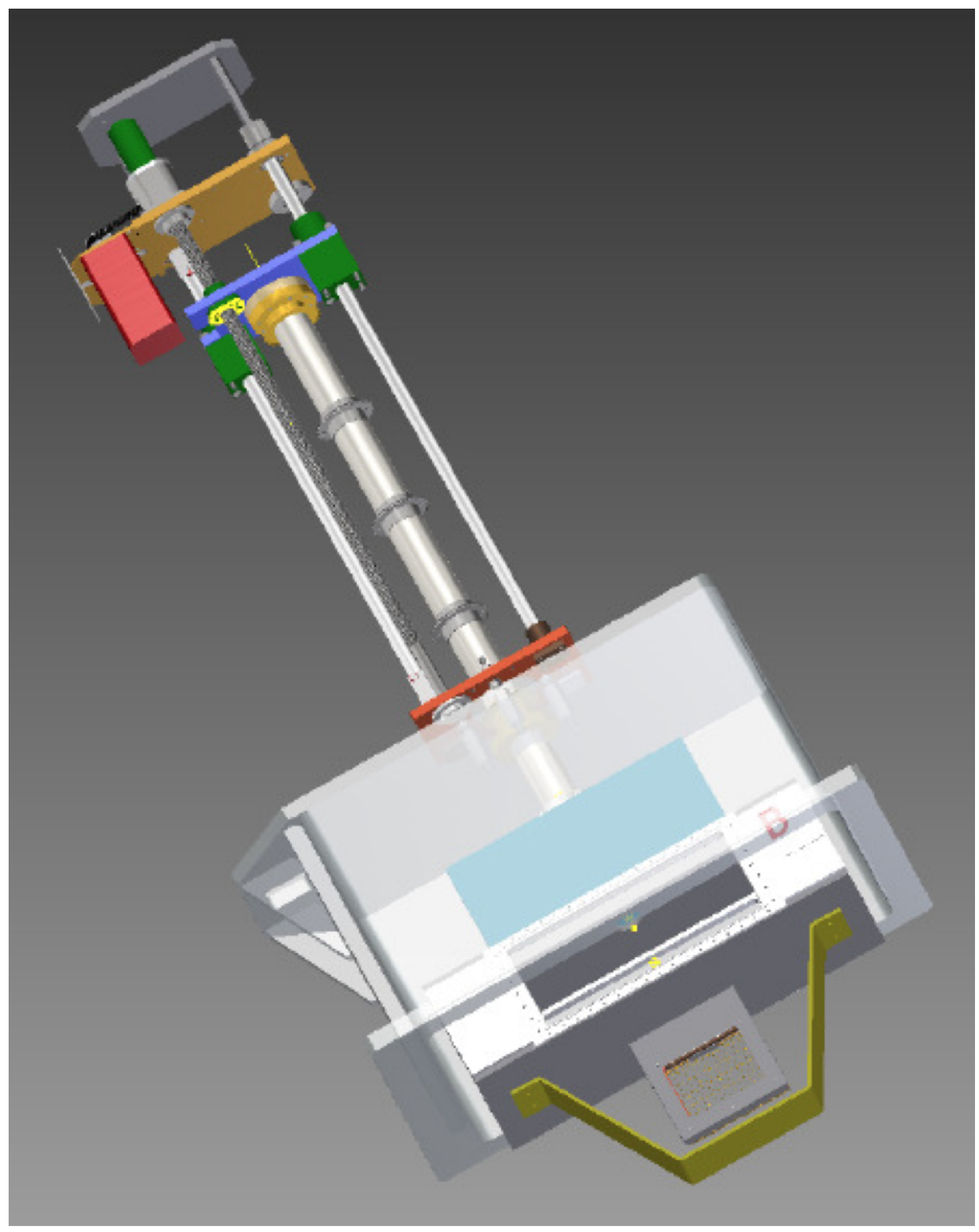

Figure 21: HGB4 attached on to the ladder arm. The SEM-Grid beam profile monitor is attached on the bottom flange of HGB4.

# 4 Gas Choice

The current choice for the gas has been mainly driven by the fact that the TPC type [23] installed at the present FRS is operating with P10 gas. For the direct comparison of the detector systems the same gas was chosen for the Super-FRS GEM-TPC detector.

The calculated [30] electron drift velocity, longitudinal and transverse diffusions for the P10 gas as a function of electric field are plotted in Figures 22, 23.

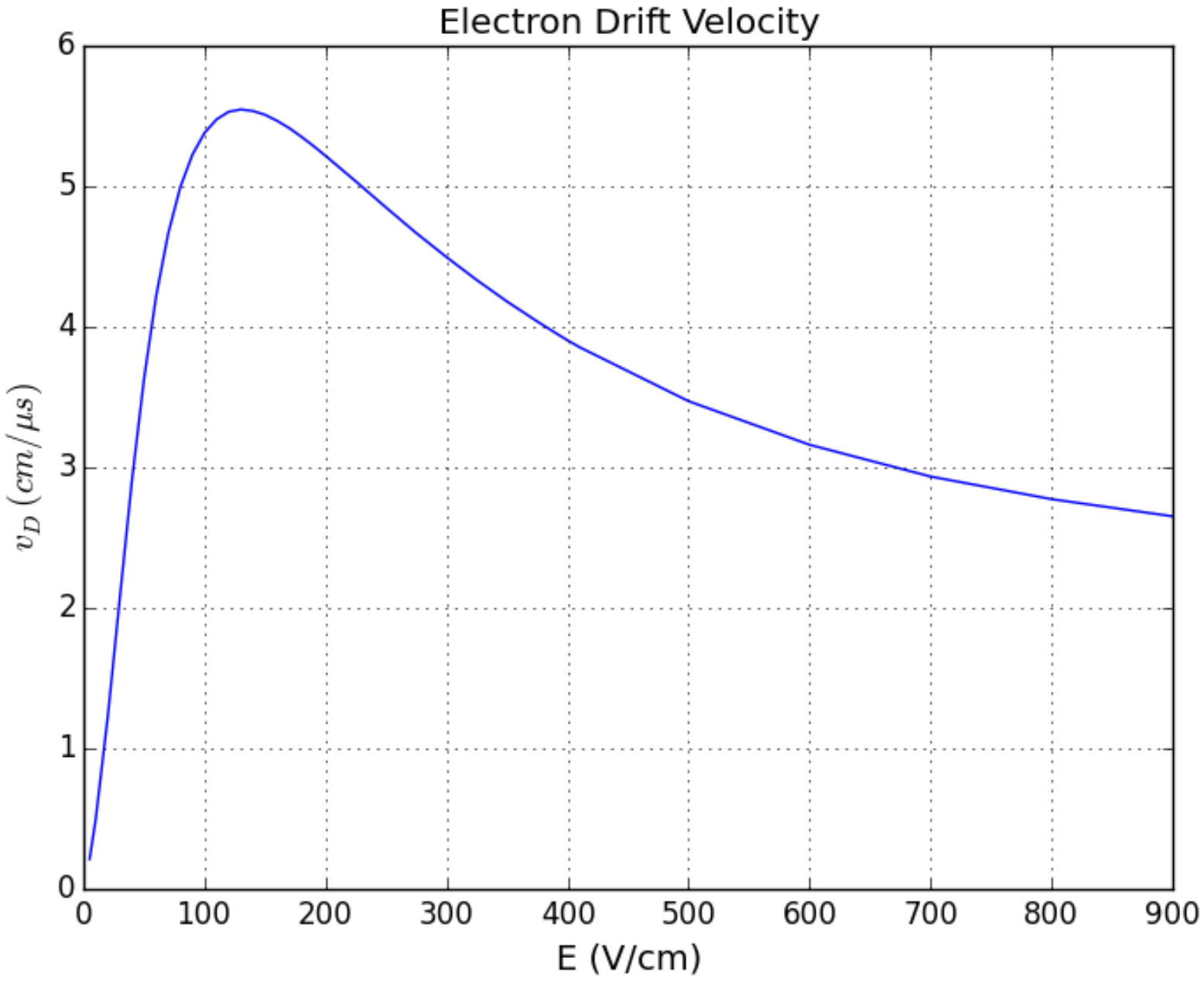

Figure 22: Electron drift velocity in a P10 gas mixture as a function of the applied electrical field.

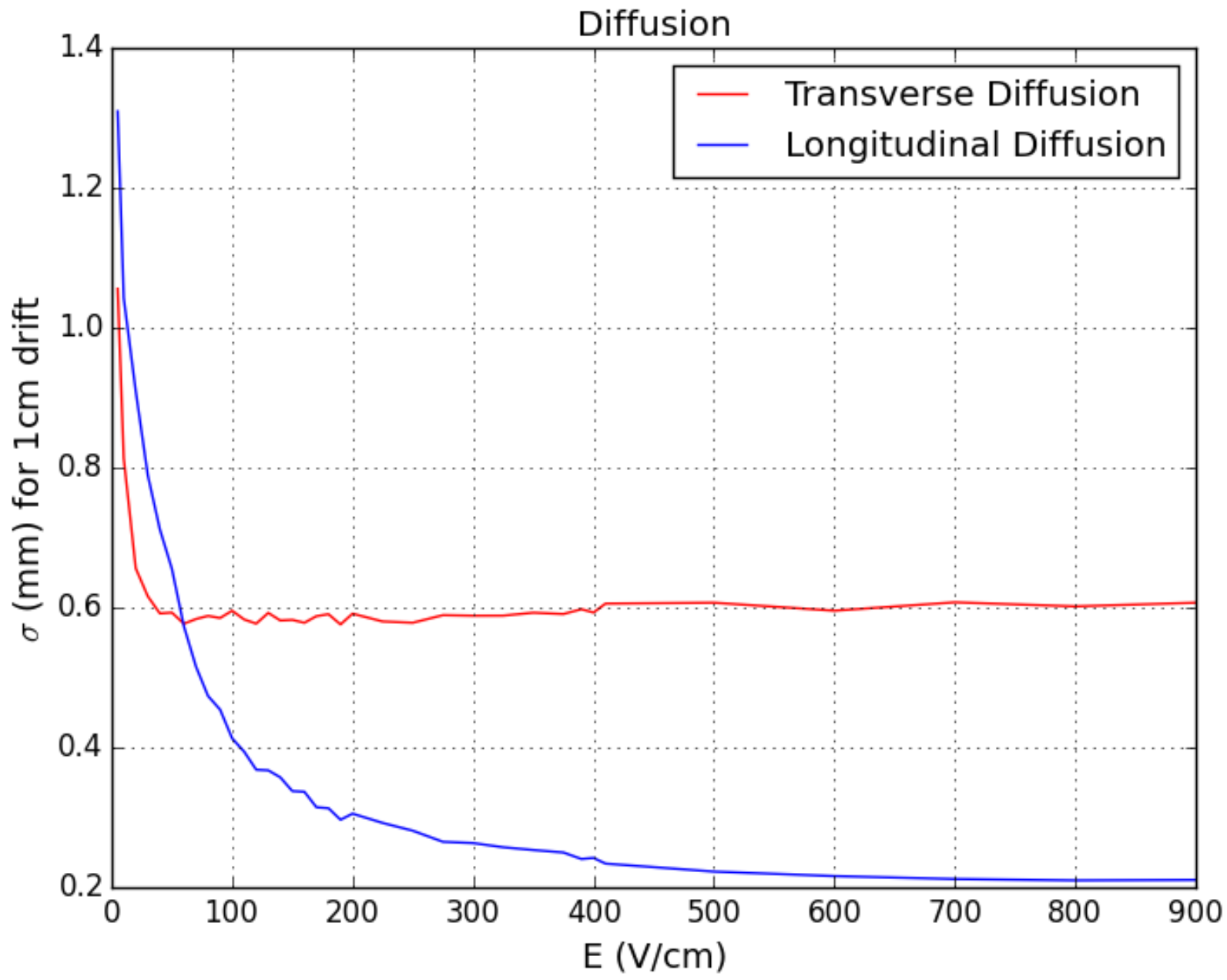

Figure 23: Longitudinal and transvere diffusion of electrons in a P10 gas mixture as a function of the applied electrical field.

# 5 Field Cage and its Quality Assurance

It is the main task of a field cage of a detector system employed in a magnetic spectrometer such as the future Super-FRS to define and enclose a volume of constant, very homogeneous electrical field without unacceptable interference with the particles traversing this so-called active drift volume. To serve that purpose, several requirements apply:

- In a high-rate environment such as present at several places in the future Super-FRS, with very heavy particles impinging, the materials in use should be radiation hard to withstand several years of operation.
- The resolving power of the spectrometer should not be decreased by altering the properties of traversing particles (beam quality), e.g. by changing its distribution in
  - isotopic composition ($Z$, $A$) thus avoiding nuclear reactions
  - charge state $Z_{\text{eff}}$ due to charge pick-up
  - the emittance (directions transversal and longitudinal to the beam) by adding straggling contributions with respect to
    - the geometry of particle distribution
    - the velocity and/or momentum of the beam particles
- All mechanical key-values (see Sec. 2) of the materials itself as well as their geometry have to be stable over long-time for a high-precision operation.
- The whole unit has to be immune against external electromagnetic interferences. Thus appropriate screening properties need to be considered.
- The electric field generated inside the active drift volume should be as homogeneous as possible in order not to spoil the geometry of the trajectories of traversing particles and/or the timing information carried by the drifting charge carriers. As a rule of thumb, calculations of the electrical potentials enclosed in a strip-line type field cage show that the inhomogeneities produced due to their discrete geometrical electrode structure decrease to a bearable level at distances from the wall which are twice the value of the structural sizes of the strip-type electrodes.

- The field-defining surfaces have to be geometrically flat since any structure in the geometry defining the local electric potential will directly spoil its homogeneity. Considering foils in general, this also has implications of their material properties with respect to the forces to be applied to keep several $\mu$m thin materials flat.

These requirements call for a very thin, flat and mechanically stable and highly durable material with low-$Z$ and low density and good homogeneity in all its properties.

One possible material, which in general meets the above-mentioned requirements, is a foil made of polyimide. A 7.5 $\mu$m thick Kapton-HN foil by DuPont with double-sided staggered/mirrored strip-wise aluminized layers forming equipotential areas have been chosen for HBG4.

The sequence of aluminized strips is impressed with continuously and stepwise changing electrical potentials provided by a chain of resistors mounted on a separate PCB. For the reason of electrical operation stability and in order to maximize the uniformity of the generated electric field, the structural sizes of the strip-type electrodes need to be minimized. This optimizes the potential difference between adjacent strips. Employing double-sided metallizations in a 'staggered' arrangement of the electrode strips further improves the field homogeneity.

As a compromise between structural sizes and technical constraints in production, a strip width of 3 mm and a pitch of 5 mm has been realized for the field-cages of the Super-FRS Twin GEM-TPC detectors.

Each field-cage assembly comprises a variety of parts that all together define and enclose a volume of constant electric field:

1. Cathode plate
2. Skimmer/ $1^{st}$-strip field terminator
3. Strip-line foil with resistor chain(s)

Several small pieces are used to mount everything together mechanically and electrically.

## 5.1 Cathode

The cathode plate is shown in Fig. 24. It is made from a 1.5 mm thick FR4 printed-circuit board mounted on a 5 mm thick Stesalit 4412/G10 backing which is providing appropriate stiffness and flatness as well as an integrated gas-distribution system.

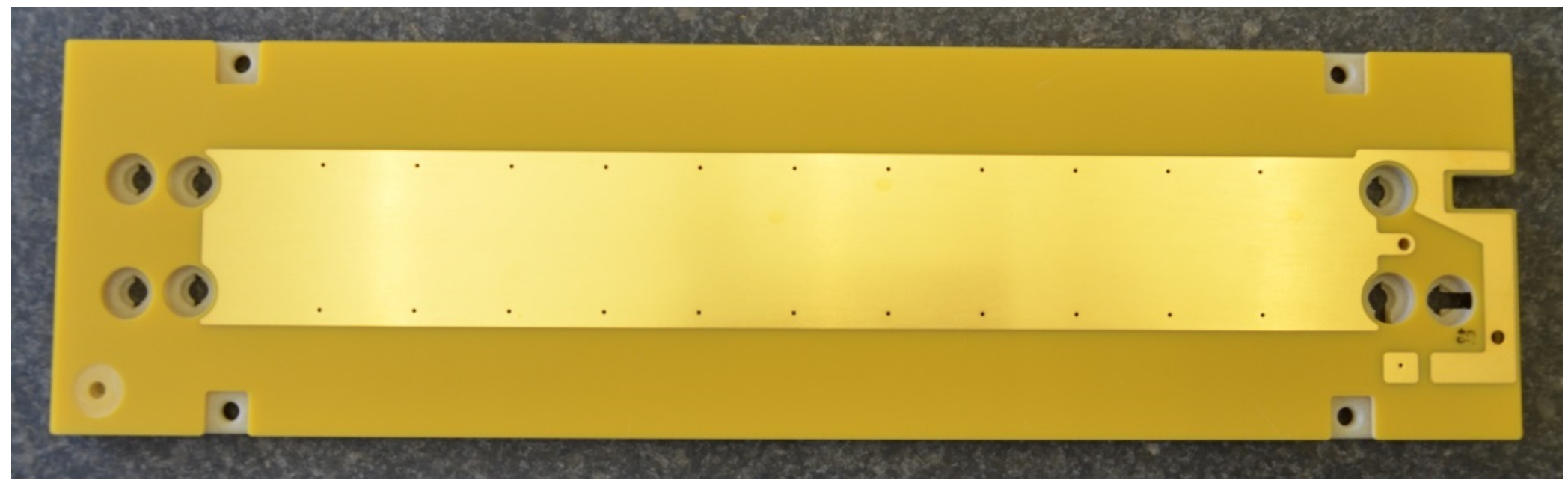

Figure 24: Picture of the cathode plate mounted on its backing.

## 5.2 Skimmer and field terminator

The field terminating plate is shown in Fig. 25. It is made from a 0.5 mm thick FR4 printed-circuit board and has been glued to a backing support made from 3.2 mm thick glass-fibre reinforced PEEK plate which is providing appropriate stiffness and flatness.

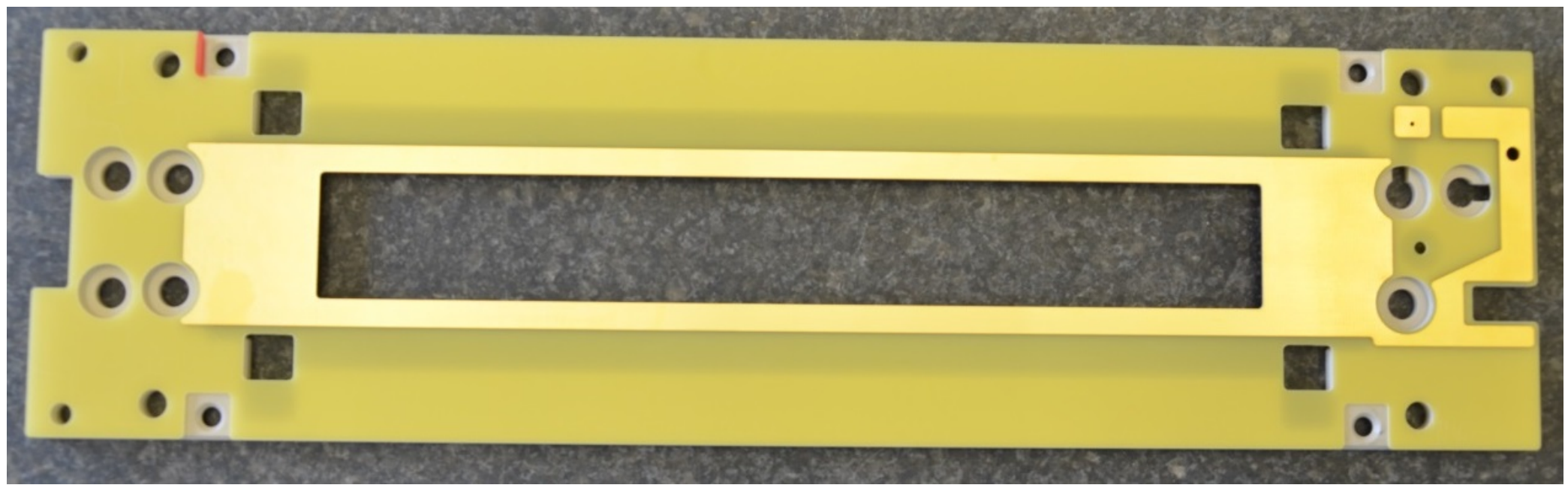

Figure 25: Picture of the field-terminating plate mounted on its backing.

This electrode is kept on the same electric potential as the first strip of the strip-line foils of the field-cage. Its absolute value can be adjusted via a separately fed supply voltage to be identical to the top-surface of the first GEM foil facing the drift volume.

## 5.3 Field-cage Strip-Line Foil

In order to minimize the effect of the step-wise electric gradient on the homogeneity of the electric field enclosed in the drift volume, the electrode structure has to be kept as small as possible. Following technical boundary conditions a strip width of 2.15±0.03 mm and a gap of 0.98±0.02 mm with a pitch of 3.12±0.05 mm have been chosen. A picture of the foil is shown in

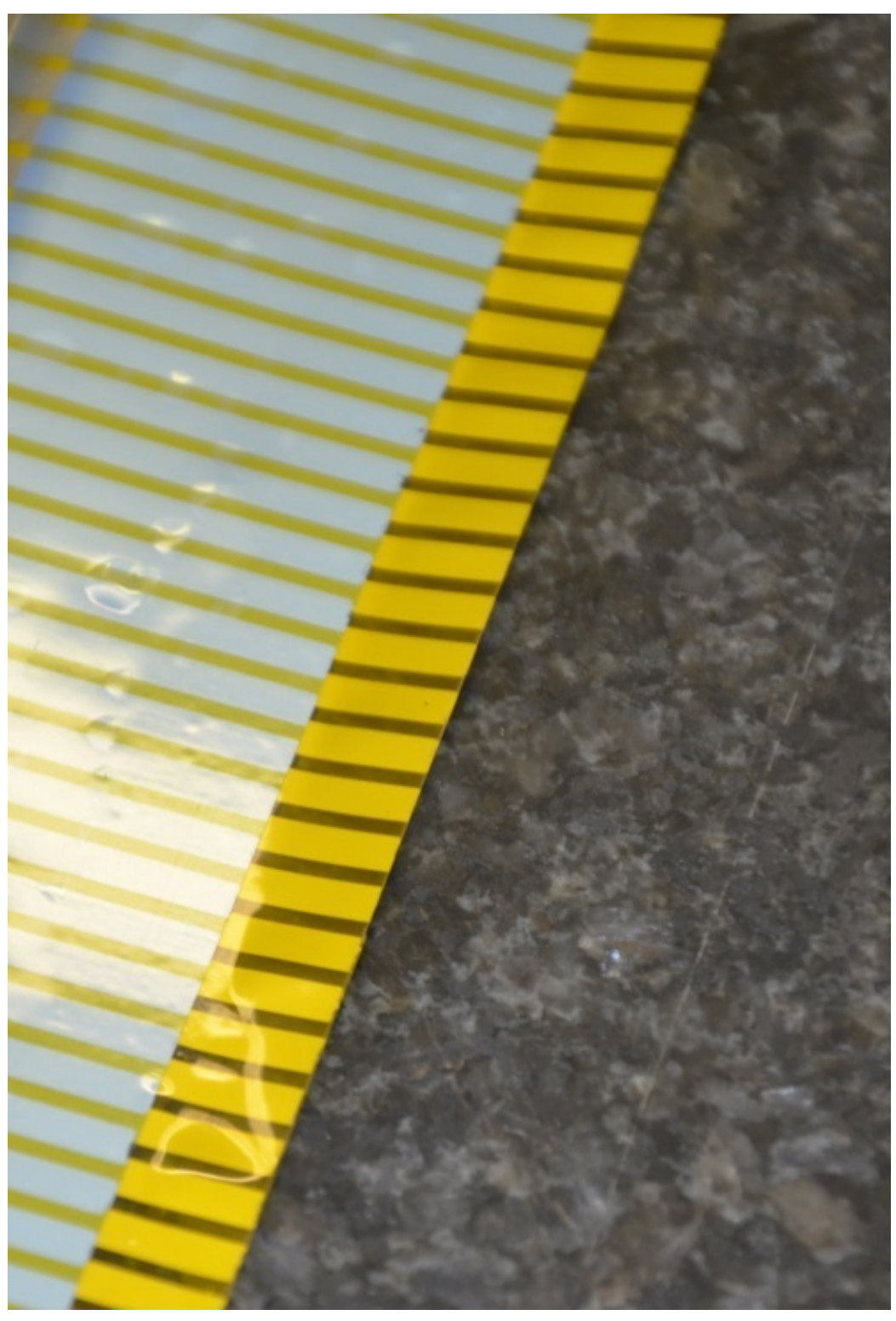

Figure 26: Picture of the field-cage strip-line foil, the staggered structure of the field defining aluminium stripes in the front as well as the mirror stripes on the back-side surface are clearly visible.

Fig. 26. The staggered structure of the field-defining aluminium strips in the front as well as the mirror strips on the back-side surface of the polyimide foil of 7.5 $\mu$m thickness are clearly visible.

Choosing aluminium of reasonable thickness (several 100 nm) as coating material for the strips preserves a reasonably low resistivity and resistance (and thus a voltage-drop) over the length of a strip which is well below the value of the respective resistors used to supply an electric potential to its surface. The latter is described in more detail in the next section.

Stretching of the foil is done using a system of bars in the edges of the active area (see Fig. 8). Allowing the walls of the field cage to be slightly larger than the copper-plated area of the GEMs ensures a nearly disturbance-free drift in the active volume. This ensures a clean, undisturbed projection of a particles track to the subsequent read-out pad-plane.

## 5.4 Field-cage voltage divider

Two identical resistor dividers operated in parallel are mounted on both ends of the strip-line type field-cage foil. Choosing this double-sided ending splits the currents across the respective dividers into half, distributing power consumption and heat generated by the flow of the respective electric currents. A picture of these dividers is shown in Fig. 27.

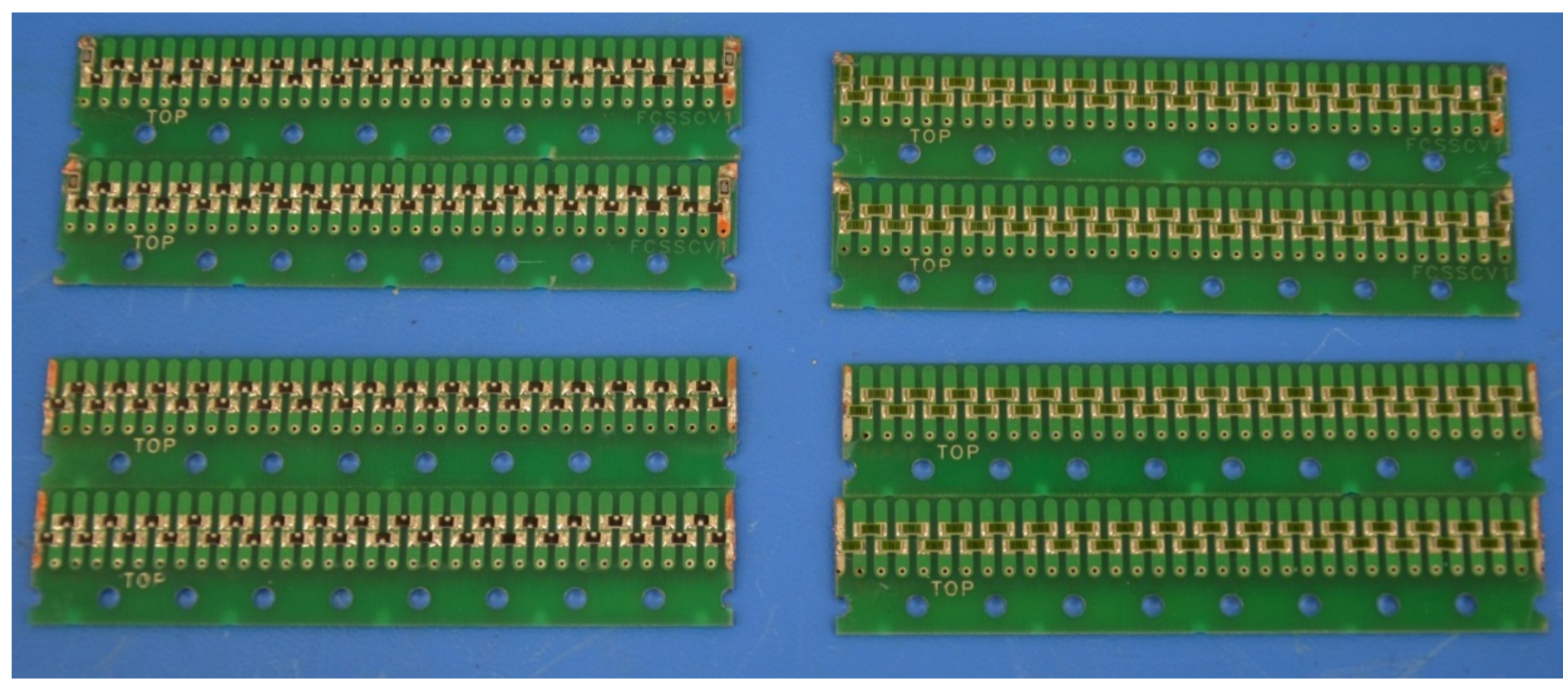

Figure 27: Picture of the resistor dividers for two identical field-cages in HGB4. The left part shows the PCB with standard resistors part-mounted, the right pair is equipped with high-precision high-voltage type SMD resistors.

Half of the resistor dividers have been equipped with $4 \times 32$ pieces real high-voltage resistors SMD 1206 of 3.0 MΩ resistivity, providing excellent tolerance (0.1%) and low TCR values (25-100 ppm/C). They sum up to a total mean resistivity across the mounted field cage of $24.79 \pm 0.01$ MΩ.

For comparison and evaluation, the other half was equipped with $4 \times 32$ pieces standard type resistors SMD 1206 of 3.3 MΩ and 0.5% tolerance (0.1% via selection) which are much cheaper. They sum up to a total mean resistivity of across the field cage of $27.72 \pm 0.01$ MΩ.

The full field-cage assembly is shown in Figs. 28 and 29. The latter also shows the double-sided metallized foil mounted in between the two field cage compartments of HGB4 configuration for screening purpose.

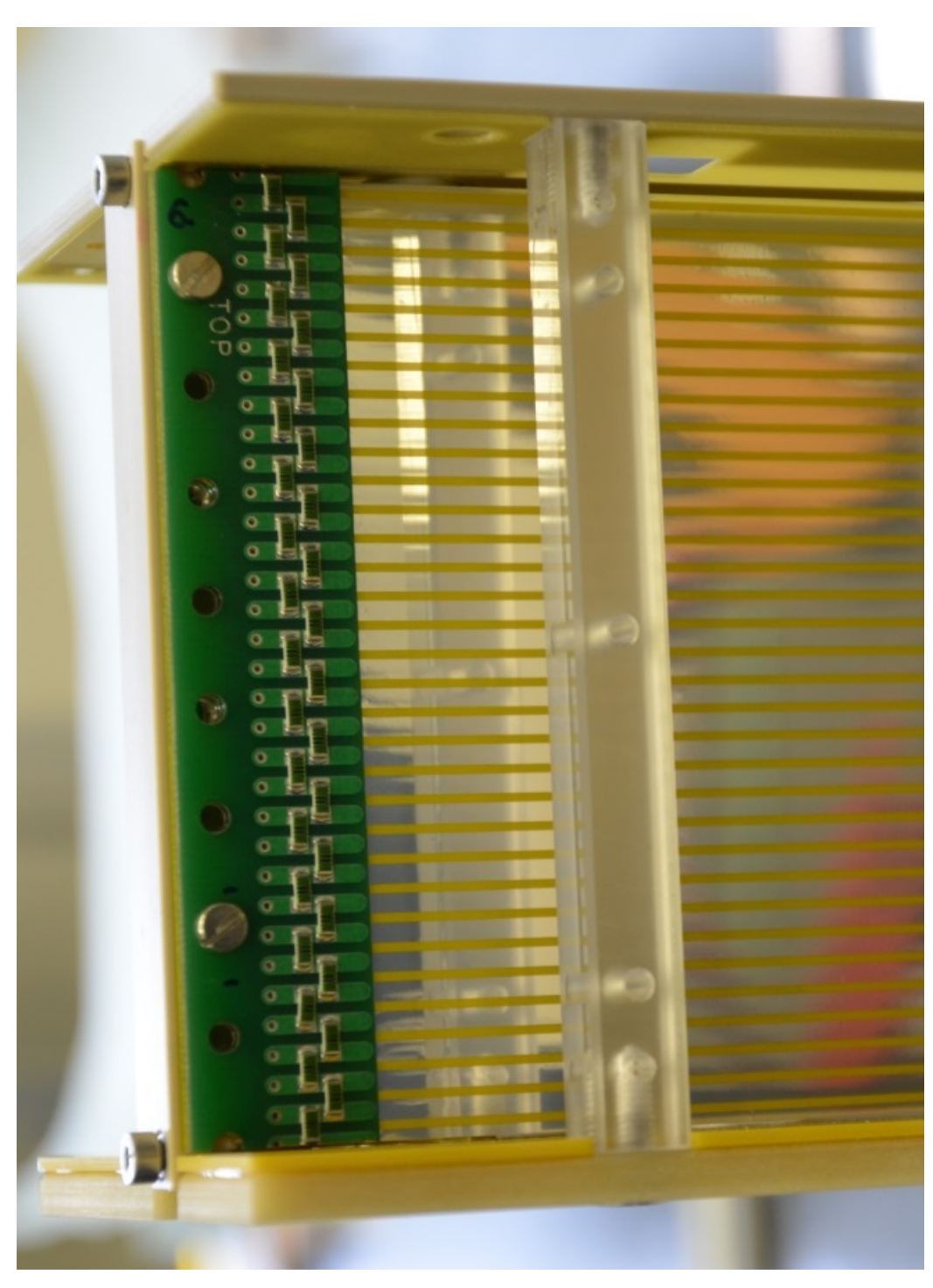

Figure 28: Picture of one of the two field-cages of HGB4.

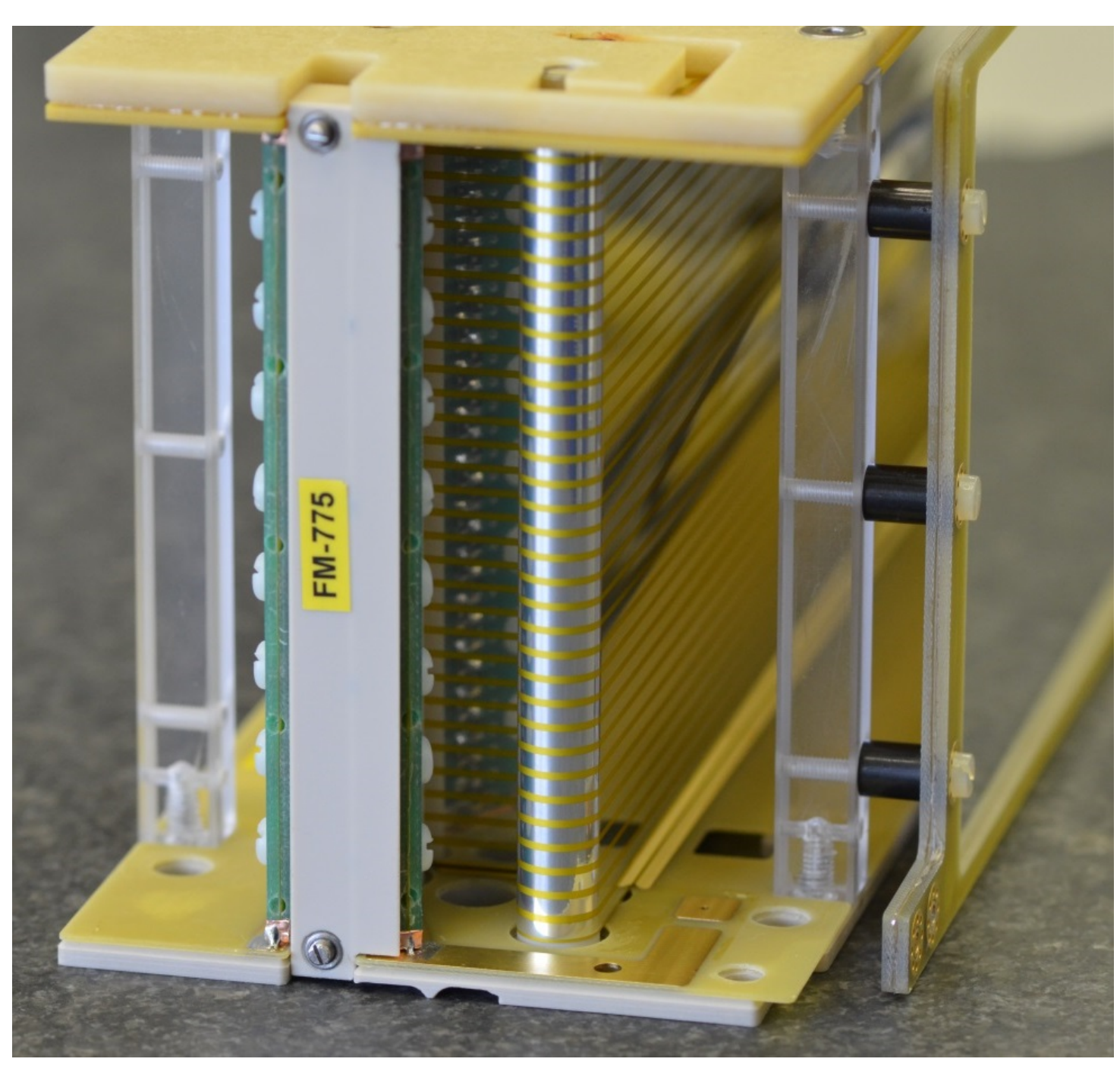

Figure 29: Picture of one of the two field-cages of HGB4. Two identical resistor dividers operated in parallel are mounted on both ends of the strip-line type field-cage foil. The double–sided metallized foil mounted in between the two field cage compartments of HGB4 for screening purpose is also visible.

## 5.5 Assembling and mounting

The whole design of the detector is highly modular and as compact as possible in order to minimize the overall space requirements in view of the later conditions to be faced at the Super-FRS. During installation only very little manual interactions are required to potentially exchange faulty parts. Except for the flaps of the framed GEM foils in the GEM-stack no electric connection has to be done explicitly and only one fluid connector (gas out) needs to be joined.

The following main elements can be identified which need to be mounted successively in order to assemble a single autonomously operating GEM-TPC as shown in Fig. 30.

1. Detector panel-flange integrating flange, pad-plane, connectors, cooler and read-out electronics
2. GEM-stack
3. HV-filters
4. Field-defining system enclosing the drift-volume consisting of cathode, strip-line field-cage foil and skimmer

Having two of those sets assembled, the two compartments can be joined with the detector housing/vessel as shown in Fig. 31. As can be seen easily, isolation distances with respect to the surrounding housing on electrical ground potential are kept to a minimum in order to minimize the overall space requirements. In the current design this is limiting the maximum operation voltage to 6 kV for a safe and spark-free operation.

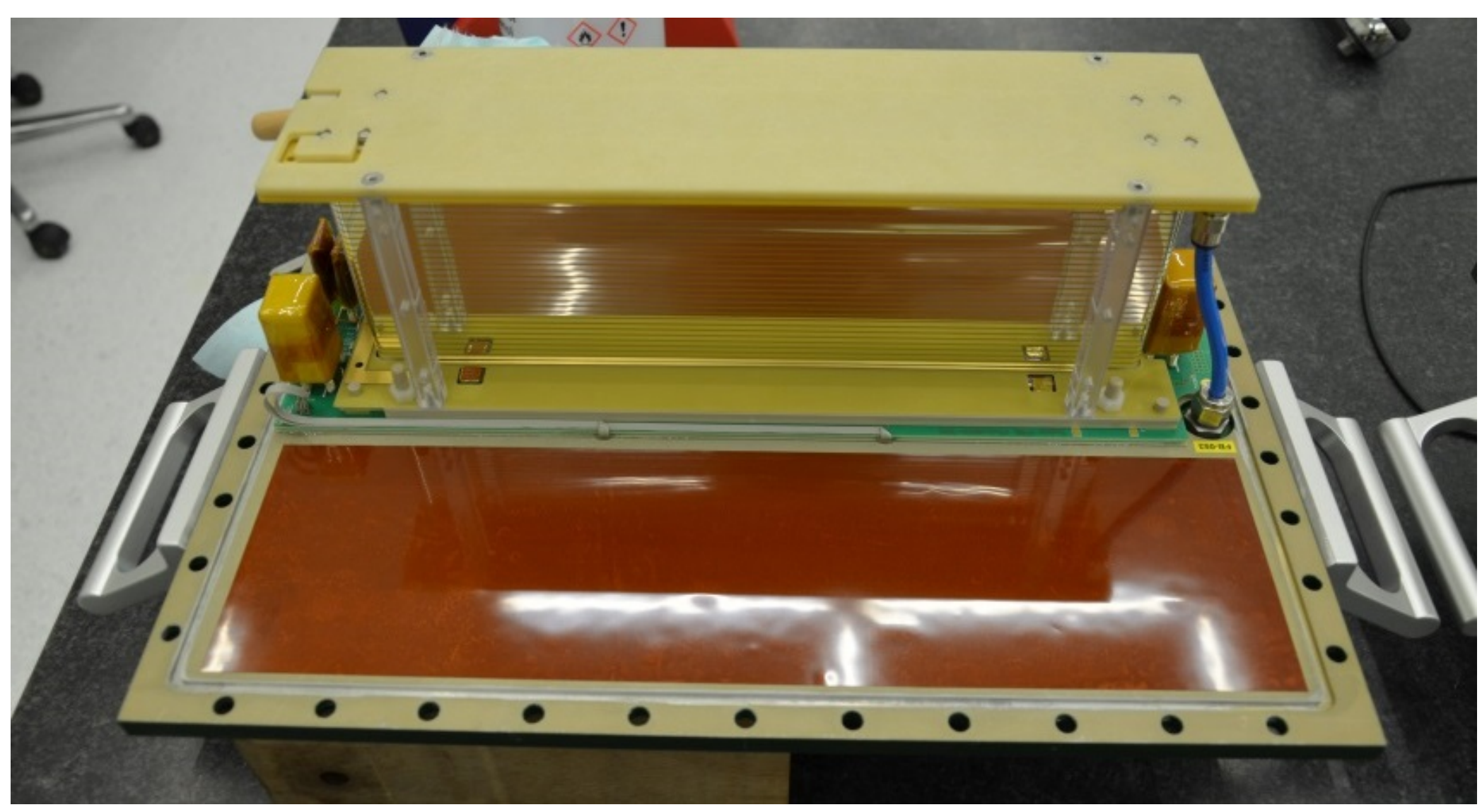

Figure 30: Picture of one half of HGB4 mounted on its panel.

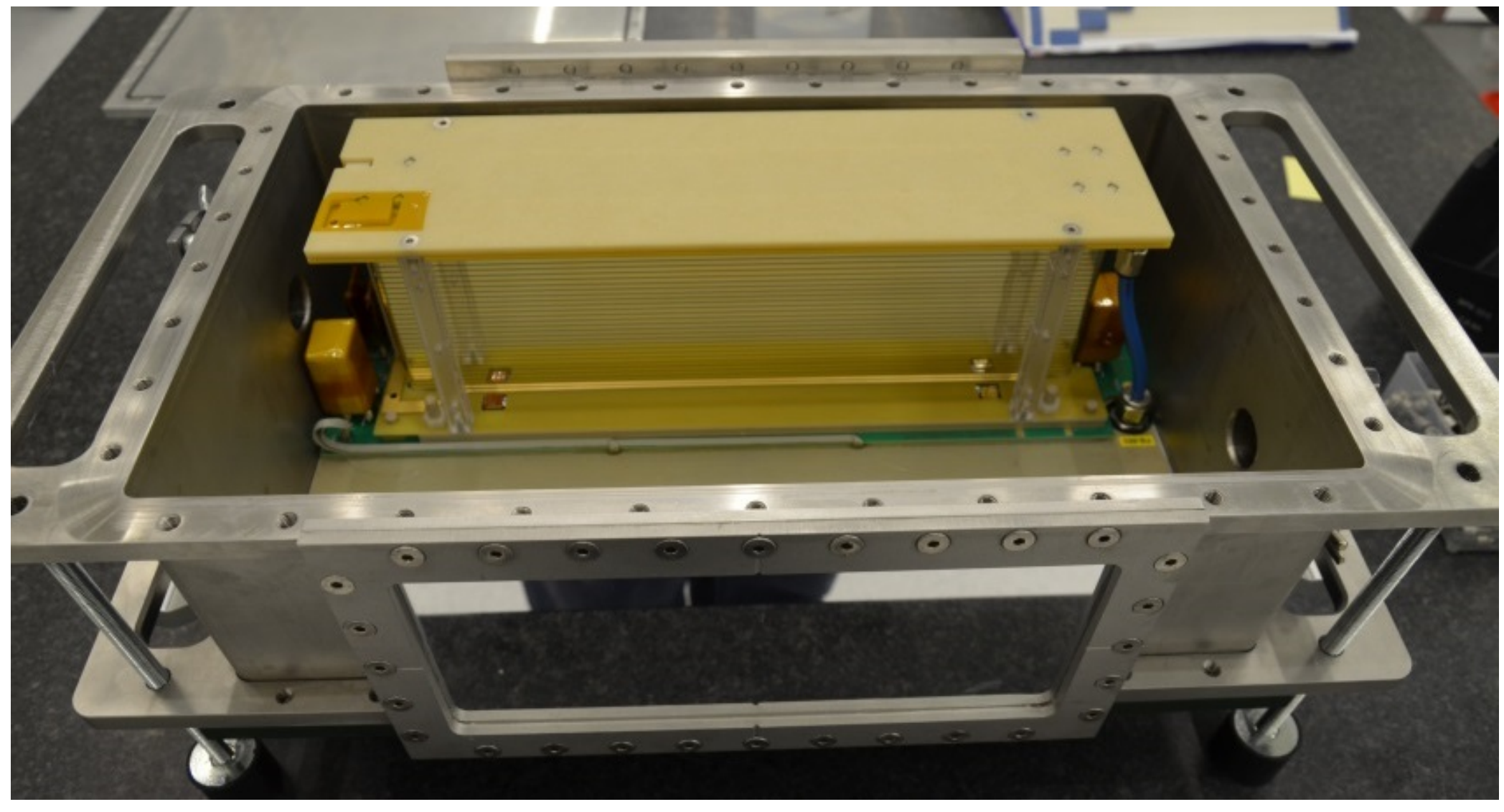

Figure 31: Picture of the first of two compartments of HGB4 installed inside the stainless steel vessel.

## 5.6 Field cage quality assurance

In order not to deteriorate/vary the equipotential levels along the strips by voltage drops, a reasonably high conductivity of the structures of the field-defining system have to be assured. This is subject to QA processes.

Moreover a stable voltage drop across the structures over several years of operation requires long-term stability of the resistors of the voltage dividers employed as well as of the interface contacts.

### 5.6.1 Foil inspection

The foil has been inspected visually using a digital microscope for any apparent damages or interruptions in the metallization or shorts between neighbouring strips. Faulty parts are rejected.

### 5.6.2 Electrical tests

Electrically the appropriate insulation between neighbouring strips as well as shorts (e.g. through pin holes) between back- and front-side strips have been tested prior to mounting of the resistor-divider PCBs. Faulty foils are rejected.

The whole field-cage set-up is tested for spark-free operation up to 5.5 kV in $Ar/CO_2$ (80/20 vol.-%) detector gas.

### 5.6.3 Performance stability

Performance stability will be checked in the future, especially for the parts which are potential subjects to ageing effects under high-dose irradiations. This refers e.g. to

- The mechanical and electrical properties of the polyimide base-material of the field-cage foils,
- The durability of the aluminium metallization,
- The absolute and relative value of the resistivity of the SMD-type resistors (of various types) of the voltage divider across the field cage and
- The contact resistivity between the respective parts.

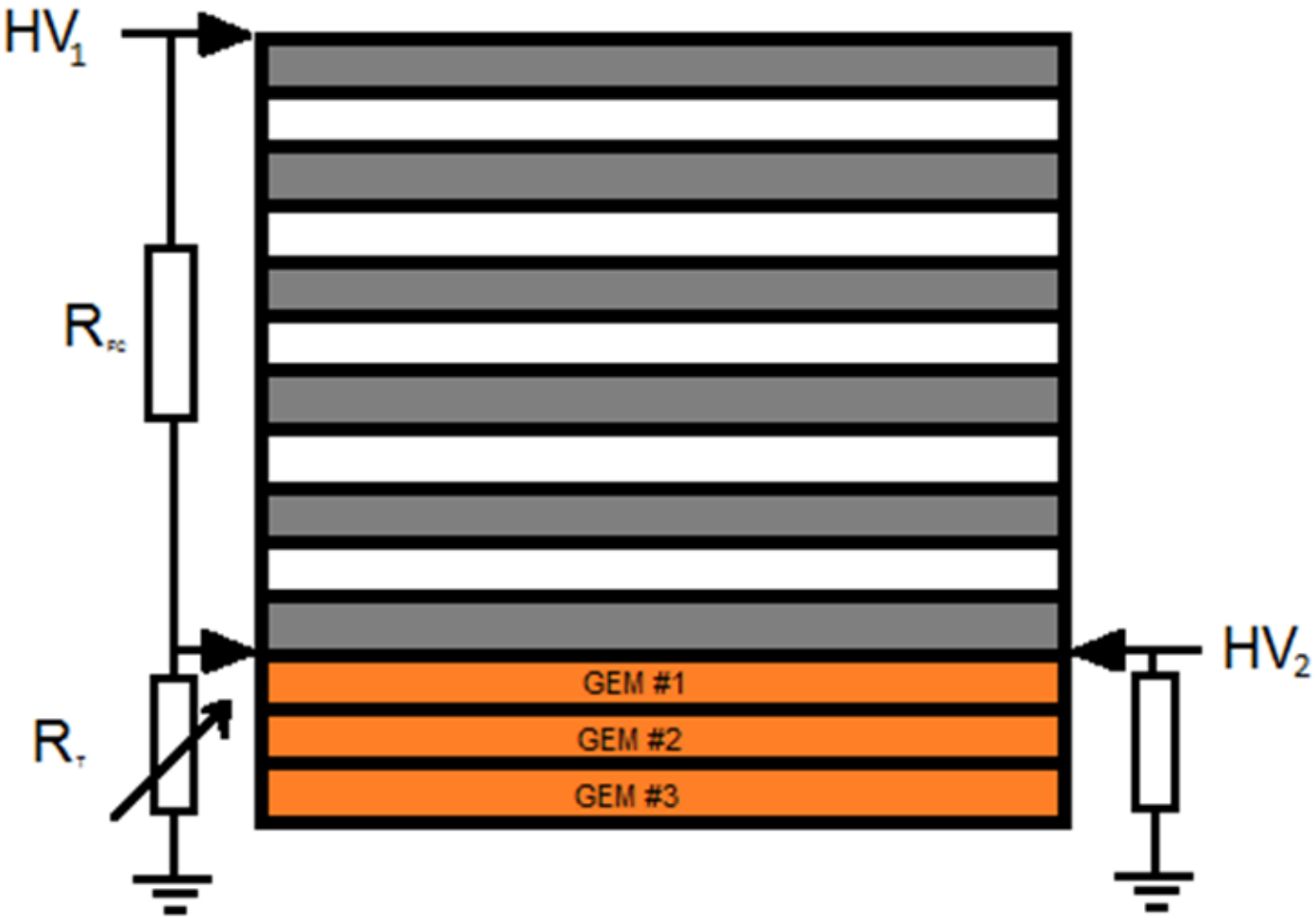

Figure 32: GEM-TPC high voltage power scheme.

# 6 HV Power Supply

The powering scheme for a GEM-based TPC needs to be adequate to accommodate the two different circuits that intrinsically are on this type of detectors. Firstly, the high voltage needed for the field cage, however a relative low current is flowing though it. Secondly, the GEM stack which will need moderate high voltage and high current, is needed for handling the high rate. Therefore in order to manage such a system, two power suppliers is needed to operate the GEM-TPC. A scheme is shown in Fig. 32. It can be seen that the $HV_1$ is connected to the cathode of the TPC and the $HV_2$ is connected to the top of the first GEM. This will allow to cope with the requirements in terms of providing different currents required by these components. One important aspect of this particular powering scheme is that one has to level out the potential on the last strip and the one of the top of the first GEM in such a way that the potential is the same.

## 6.1 Field Cage Power Supply

The main purpose of the powering for the field cage is to be able to apply a field gradient across the volume of the field cage in such a way that the electrons drift towards the GEM amplification stage. Therefore in order to have the maximum uniformity, the potential difference between strips need to be as small as possible. Results from simulations show that the field becomes very uniform in the enclosed volume, after a distance two times the strips pitch from the detector wall, therefore the charge produced by the particles crossing the sensitive volume of the chamber close to the walls will end up in a metallic plate, located on the top of the first GEM. The field cage together with a set of dividers is shown in the Figure 33. It is important to highlight that the foil used for the field cage has double-sided strips for a better field uniformity at the pitch of 3 mm on each side.

The divider itself has a set of 33 resistors, with total resistance of 100 M$\Omega$ and the current circulating across the field cage was calculated to be in the order of tens of $\mu A$ depending on the field to be applied inside the field cage. For example if one uses a field of 150 V/cm, thus having a total drop on the divider of 1.5 kV, then this will give a 15 $\mu A$ current circulating through it. In order to have the possibility to set the potential at the last strip of the field cage a resistor terminator has been developed. The main purpose of this device is to adjust the potential depending on what was the potential selected for the top of the first GEM. This was possible by the use of a high voltage potentiometer connected to ground. In the Fig. 34 the picture of the device is shown. Below it can be seen that the device has two channels one per field cage. In order to produce the total resistance required two potentiometers and one fixed resistor were cascaded (see Fig. 32). Thus, it allows us to scan fields between 150 V/cm up to 320 V/cm for a very low gain of the GEM stack.

## 6.2 GEM Stack Power Supply

The GEM stack used for this chamber consist of three foils (i.e. triple GEM). Therefore a set of 6 high voltage power lines is needed. However, the connection looks the same as a commonly used triple GEM detector, where the voltage potential for each GEM are provided via a passive resistor divider. The illustration of the powering scheme is shown in the Fig. 35.

The resistor divider used for the tests was the same as the one used for the TOTEM triple GEM [5] detector, with the main characteristics of this divider are two: the first one is that the field between the transfers and induction gaps is keep close to 3 kV/cm and the gain on the GEMs is higher at the bottom. It is important to highlight that we can achieve modulation of the gains from tens to hundred thousands, which is needed at Super-FRS since the specimens to detect varies from hydrogen up to uranium. In this way one can avoid saturation of the electronics. The divider used for HGB4 is shown in Fig.

36.

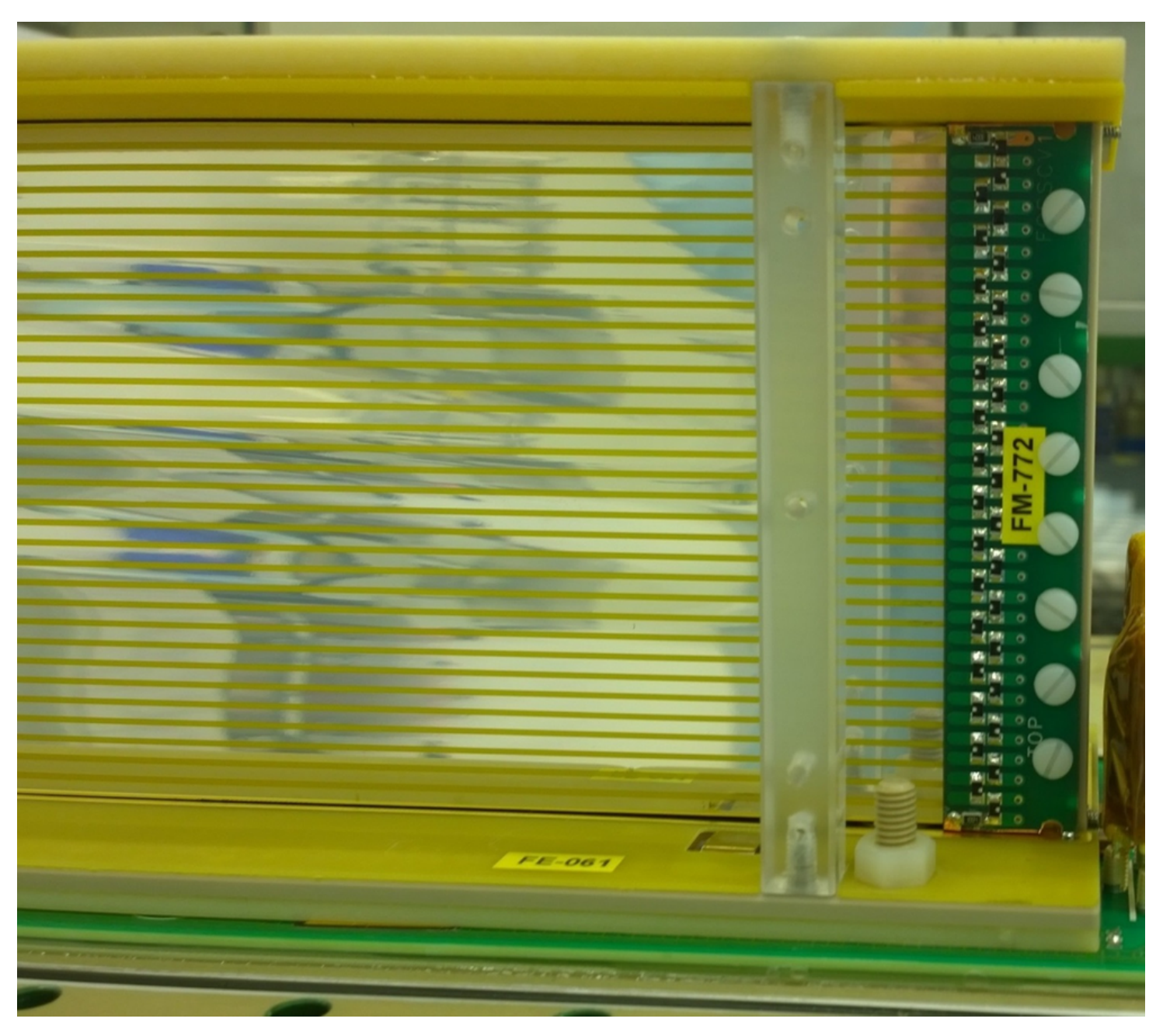

Figure 33: Field cage resistor divider.

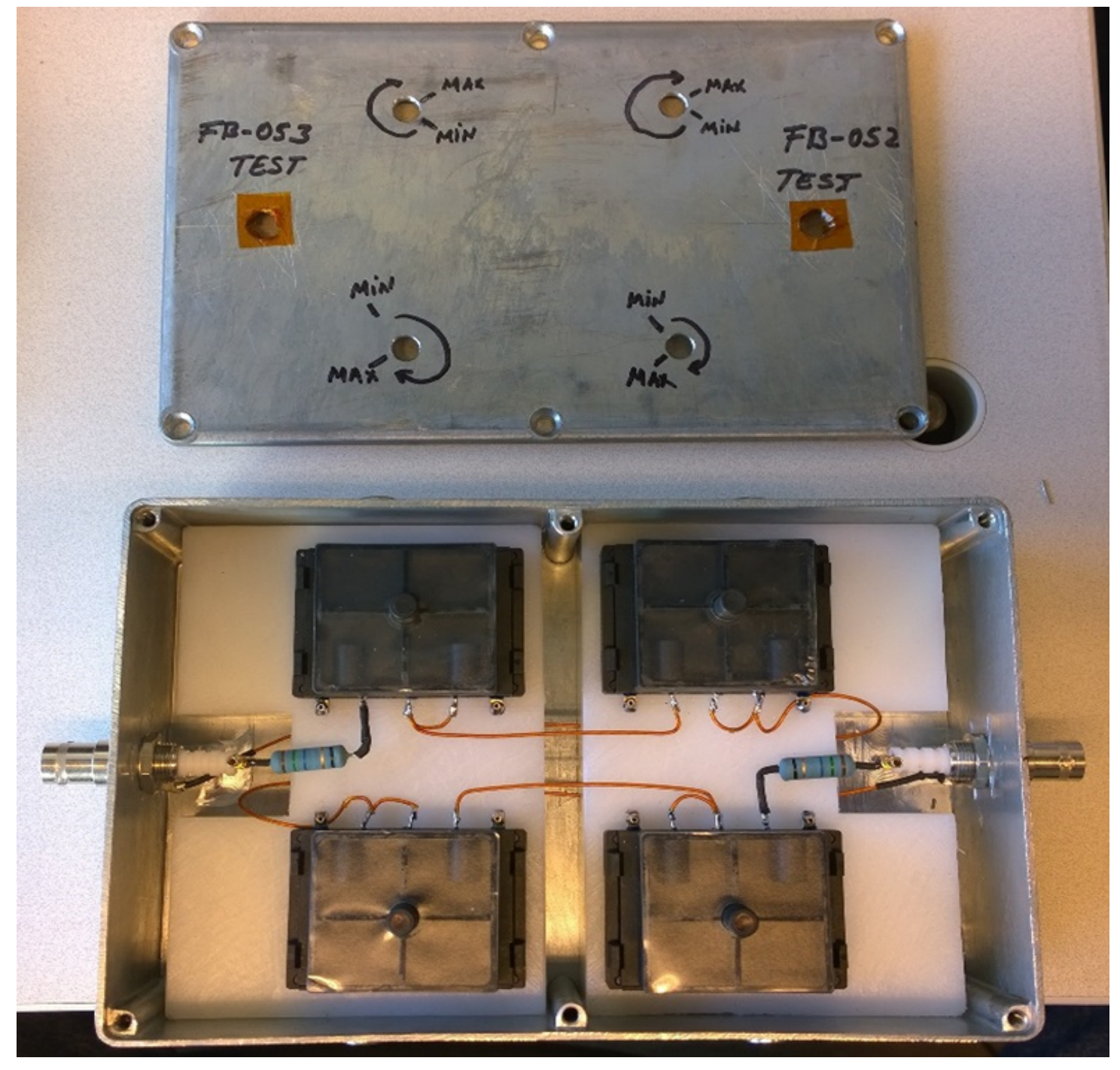

Figure 34: Field cage resistor terminator.

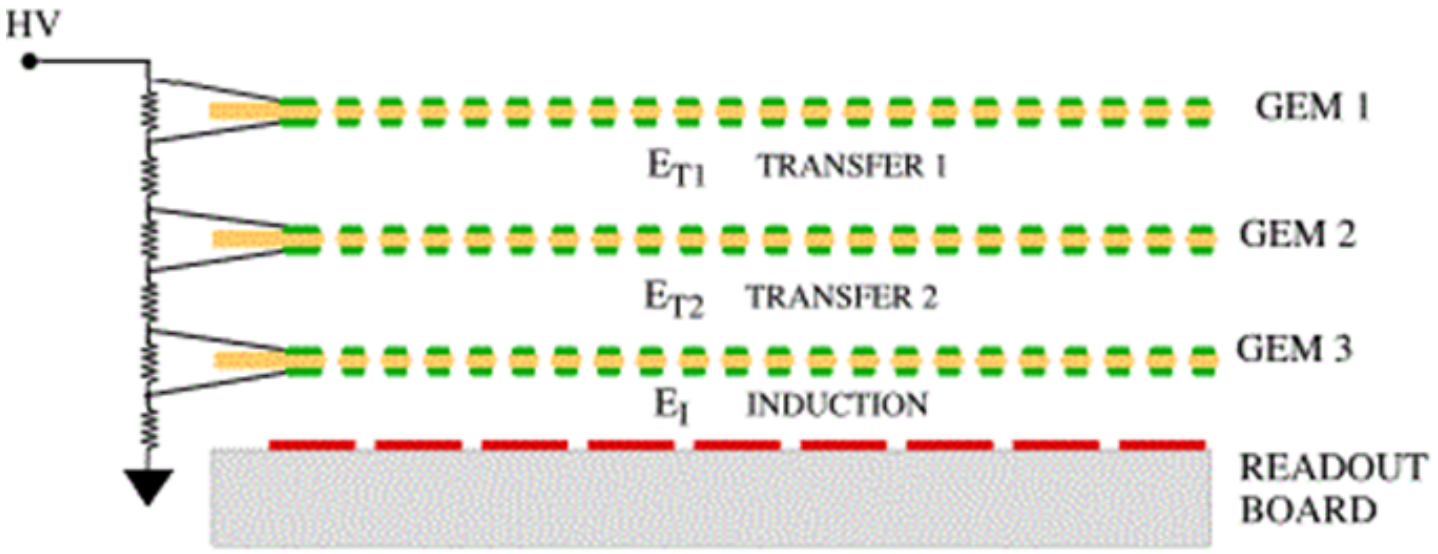

Figure 35: GEM stack power scheme.

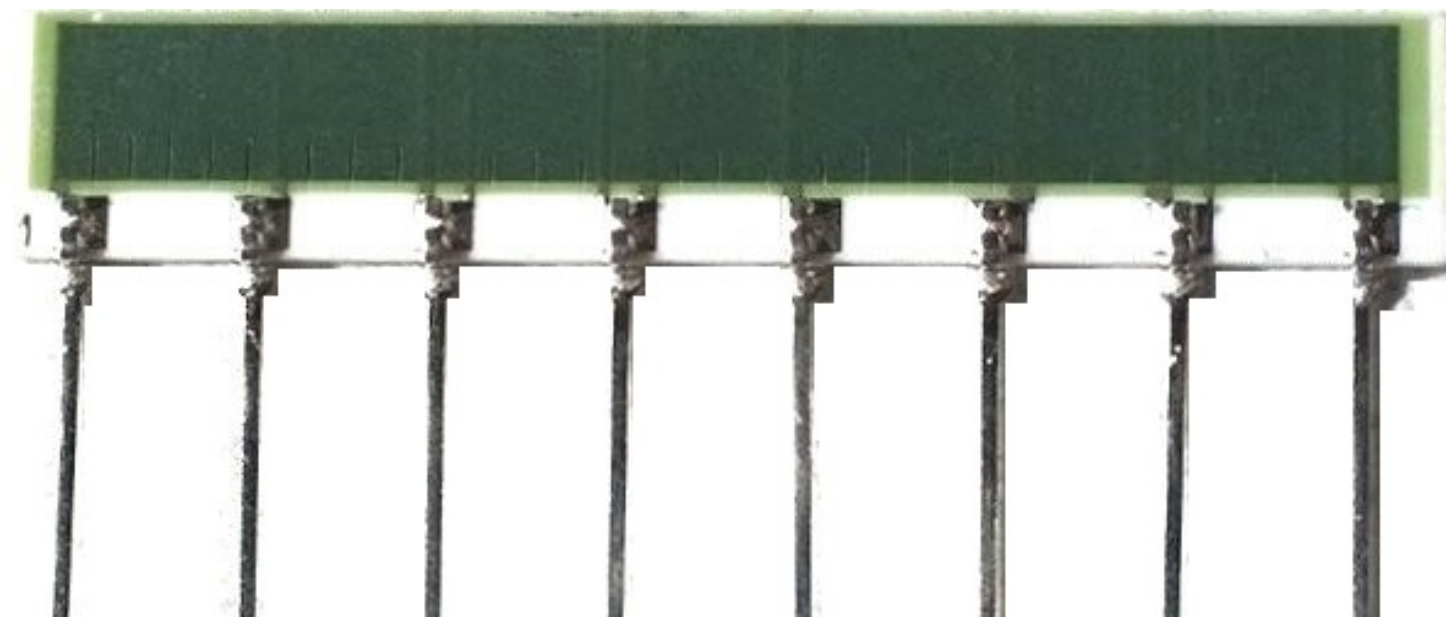

Figure 36: GEM stack passive divider.

# 7 GEM Stack Quality Assurance and Test

## 7.1 GEM Foil Quality Assurances

The proper selection of the GEM foils to be used in the GEM stack of HGB4 is of major importance [6]. Many experiments have used GEM detectors and a set of procedures has been established in order to guarantee the proper selection of the GEM foils. The first experiment to use large GEM foils was COMPASS [7], followed by many experiments such as LHCb [8], TOTEM [5] and more recently for the upgrade of the ALICE TPC [6]. This chapter will cover the requirements needed for transportation, electrical and optical checks and performance uniformity. In order to minimize the risk of failure and provide long-term operation of the chambers, a methodology has been developed.

## 7.2 Visual Inspection

As a first step upon arrival of the GEM foils a visual inspection is carried out, this procedure will serve as a first screening procedure, because it will allow us to find large side scratches, regions with defects due to under or over etching and residual deposition in the GEM foils. In Fig. 37 a single foil with five GEM foils on it is shown.

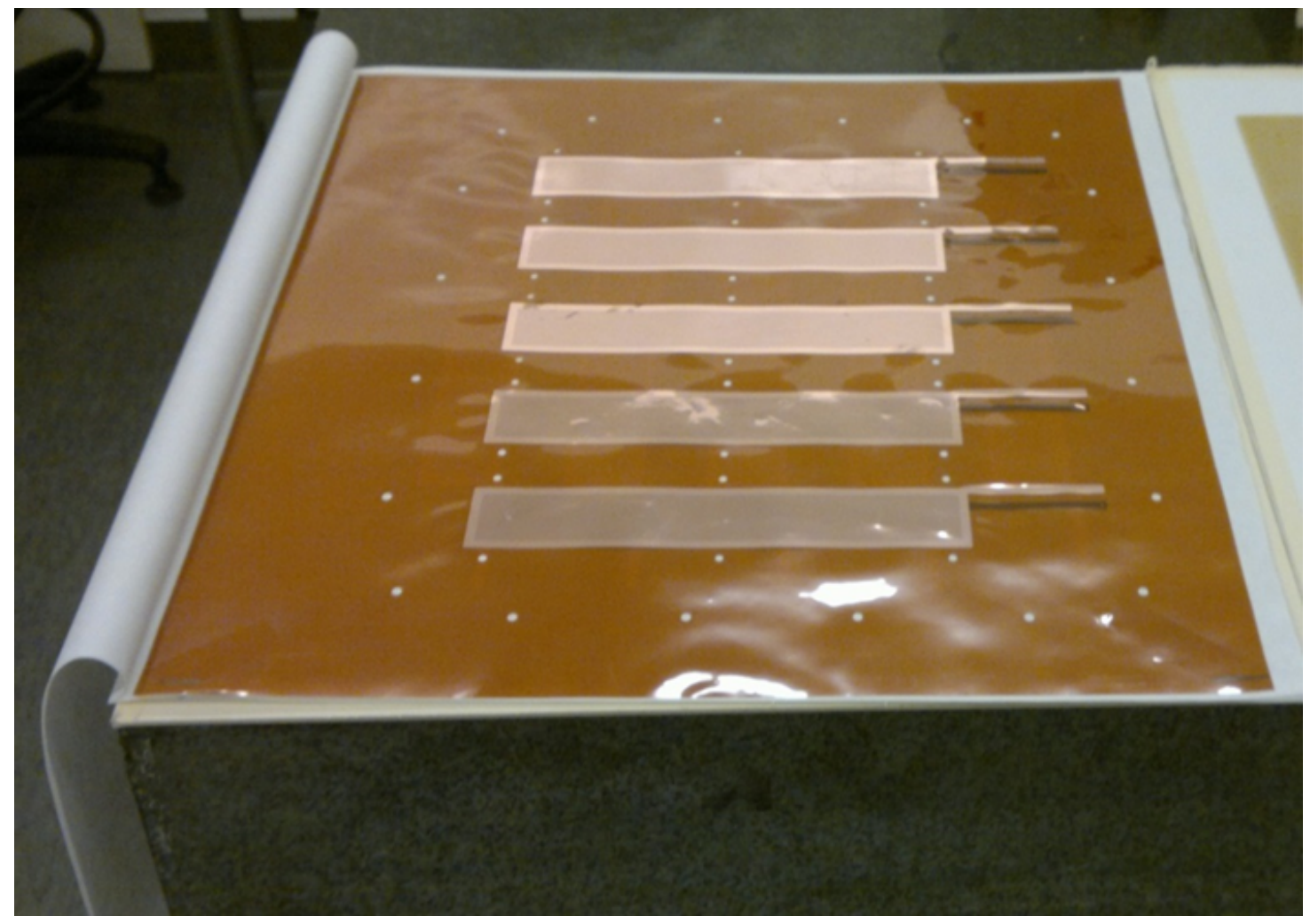

Figure 37: GEM foils for the Twin-GEM-TPC called HGB4.

One major defect that can be identified during this inspection will be the presence of wrinkles on the GEM foil. This defect is important to quantify, since depending on the dimensions one cannot expect to remove it by stretching the foil. In addition to that wrinkles could potentially point out to possible defects between copper and Kapton interface. Therefore the GEM foils with very smooth surface will preferable pass this step.

## 7.3 Electrical Characterization

The electrical characterization of the GEM foils is the next step in the process of accepting or rejecting a GEM foil. The measuring of the leakage current in particular for each sector can show how stable it can be against a potential difference across the two sides. These electrodes are located on top and bottom of the insulator Kapton insulator layer. The test is carried out in dry atmosphere in order to avoid breakdowns provoked by the water contend in air. One of the most common gases used to get dry atmosphere is nitrogen and therefore flushing the vessel prior to the test is mandatory. The set-up used for the leakage current measurement is shown in Fig. 38. The components of this set-up are a desiccator with a gas flow meter, an electrometer, a voltage source and a computer for controlling the ramping of voltages and storing the current measurements. The procedure consists of by applying a voltage difference between the top and bottom electrode from 100 V to 500 V in steps of 100 V. The current is monitored and if the value is less than 0.5 nA the next ramping step is performed. When the maximum voltage of 500 V is reached, the current is monitored for 30 min. In the case when the current is less than 0.1 nA the GEM foil is accepted for the next check. Otherwise it will undergo either thermal treatment or is send back to the producer for further cleaning.

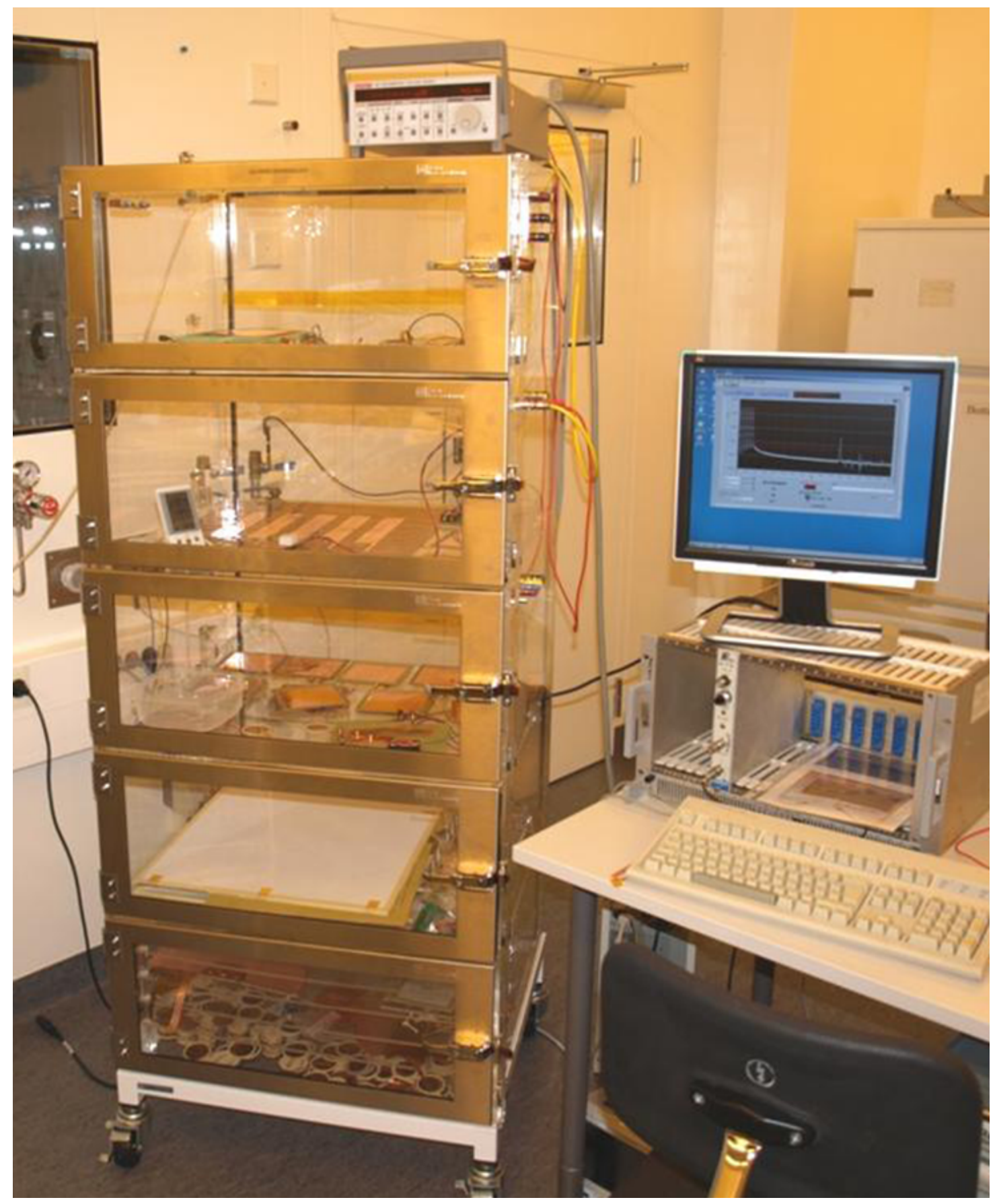

Figure 38: Leakage current measurement set-up.

A typical behaviour of the leakage current measurement is shown in Fig. 39. It can be seen that current is well below the acceptance limit, thus such a foil is accepted. However on the Fig. 40 one can see the behaviour of another type of foil when during the initial ramping of the HV the GEM foil shows typical operation, but after sometime it experienced several sparks. After flushing the vessel, the leakage current move back to normal.

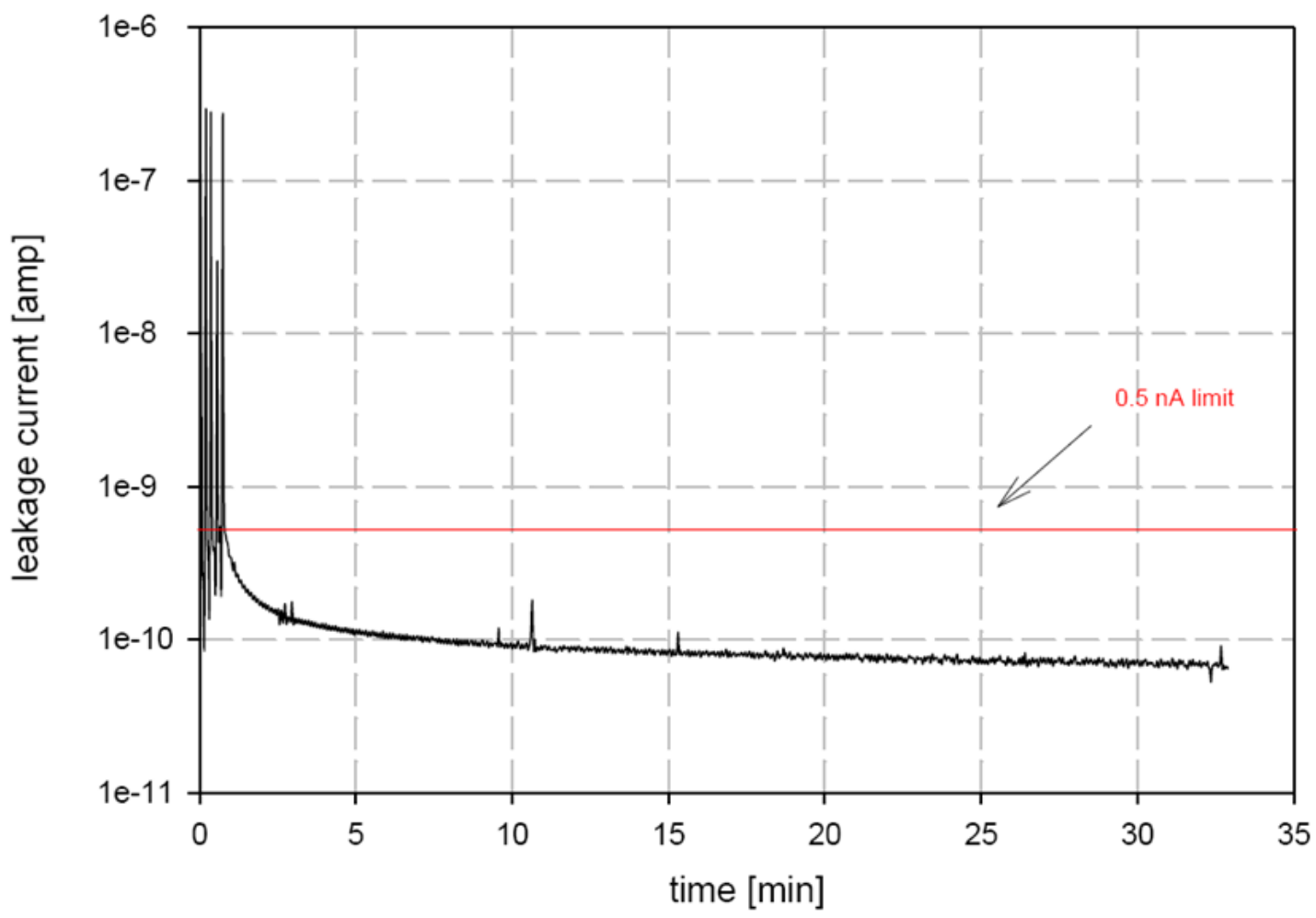

Figure 39: Leakage current resulting from a good working GEM foil. The evolution of this step in the characterization shows that the GEM foil performs well and it is accepted. The acceptance criteria is a leakage current of less than 0.5 nA for 30 min at 500 V in dry atmosphere.

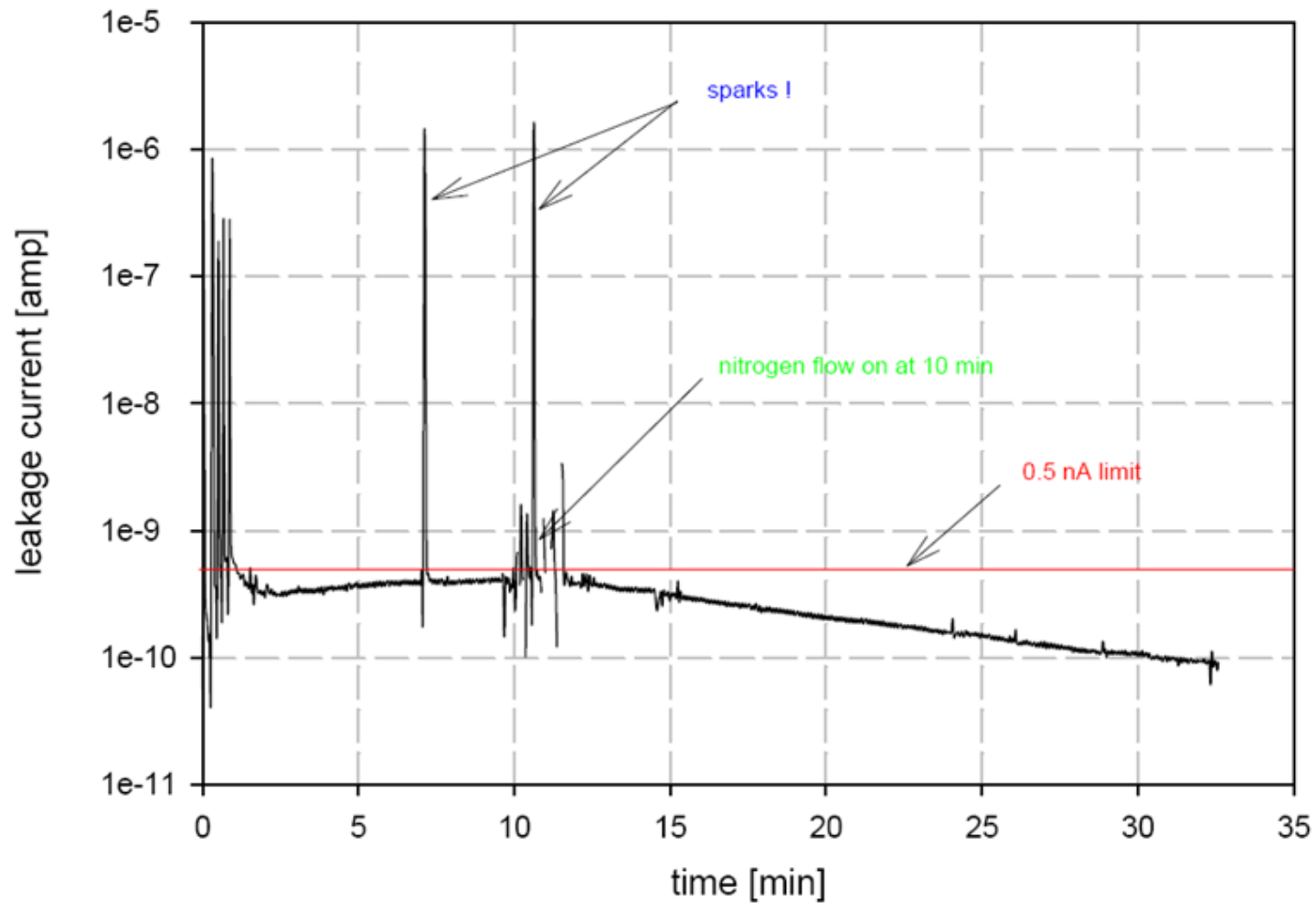

Figure 40: Leakage current resulting from a recovered GEM foil

## 7.4 Optical Characterization

After the foil has passed the coarse visual inspection and the static electrical quality assurance testing, it is subjected to precision optical inspection using a high-resolution scanning system. In this way the geometrical characteristics of the GEM foil can be studied. In particular, the distributions of inner and outer hole diameters and the hole pitch for both sides of the foil are measured. It is well known that a variation of the inner hole diameter and the ratio of the inner outer hole diameters causes a variation of the intrinsic gain of the foil. The dispersion of the distributions can be taken as a quantitative indicator of the general foil uniformity. Foils not meeting the specifications over the full surface (i.e. copper hole diameter typically $(70 \pm 5)$ $\mu m$, polyimide hole diameter typically $(55 \pm 5)$ $\mu m$) are rejected.

In addition, the method is suited to identify smaller defects that were not detected in the first visual inspection step. Defects in the form of under- or over-etched areas in the foils can occur during the manufacturing process. Other type of defects may come in the form of chemical residues from the production process, or dust attached to the foil, very large holes, missing holes, etc. All of them may cause operational instabilities. The set-up used for this purpose is based on a back-illuminated light table with area $(100 \times 100)$ cm$^2$. In Fig. 41 the scanner used for this optical characterization is shown. A nine mega-pixel camera with a single pixel size of 1.75 $\mu$mm$^2$ is mounted on an $x - y$ positioning system above the light table. The optical system has a resolution of 144 light points per mm and a field of view (single image) of $(6.1 \times 4.6)$ mm$^2$. After compression, the total image size is about 20 MB. Up to 500 individual images are required to cover the full active area on both sides.

During the scanning procedure the diameters of inner and outer holes, the pitch between holes and their shape are recorded. Distributions of the diameter, the width of the polyimide rim (the distance between the border of the outer and inner holes), and the pitch are shown in Fig. 42. Parameters describing the shape of the hole are obtained from an ellipse interpolation with a sigma of about 0.84 $\mu$m for the inner hole.

A two-dimensional map of the GEM foil characteristics is used to visualize the uniformity of measured parameters as a function of position on the foil. Fig. 43 shows an example of the spatial variation of the diameters of the inner and the outer holes. Here, the diameters of holes are averaged over an area of 2.5 mm$^2$ $\times$ 2 mm$^2$. Maps such as these can be used during assembly of the GEM stack to avoid accumulation of unwanted features in similar positions over the stack area.

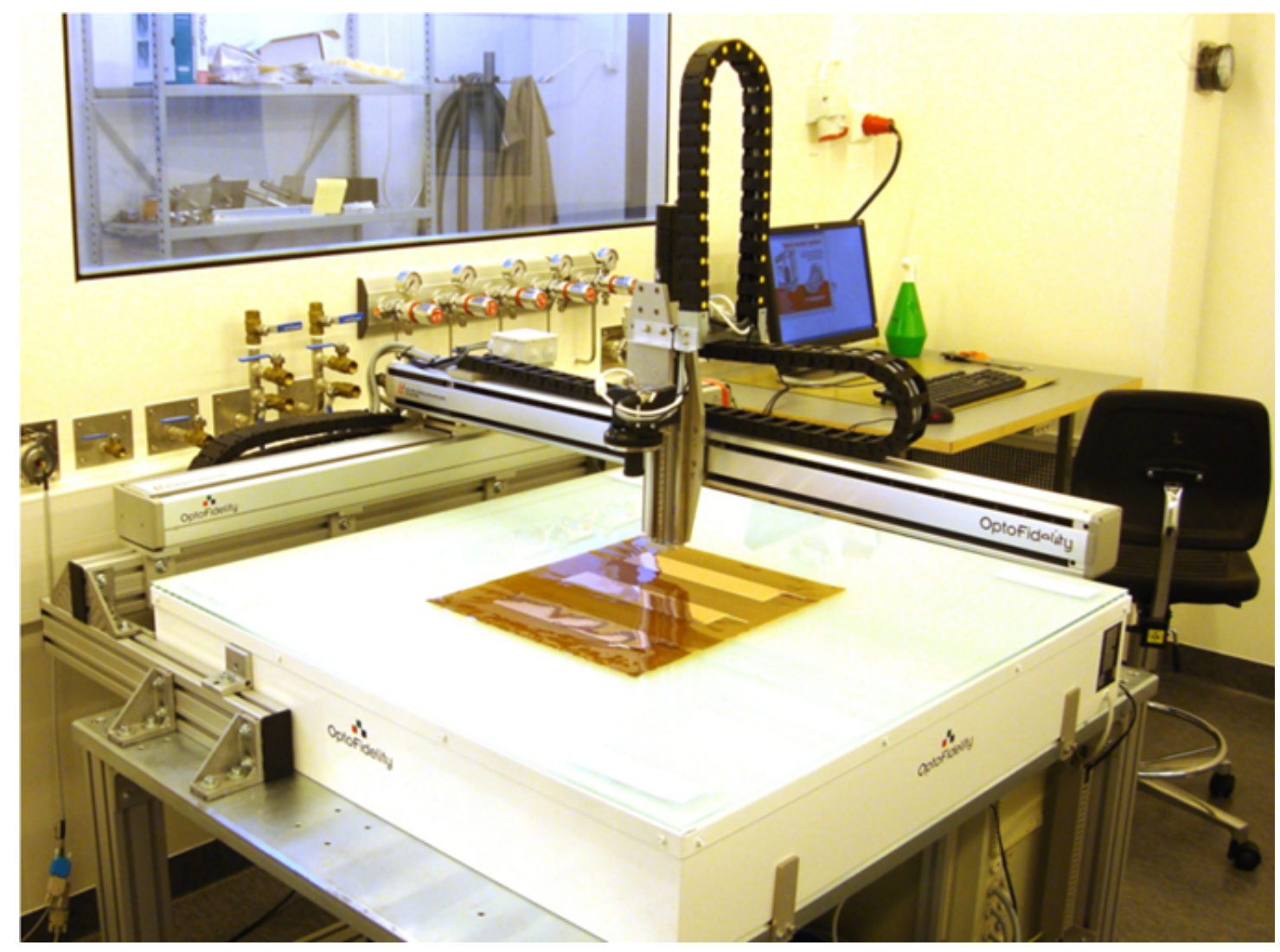

Figure 41: Set-up of the high-resolution scanning system.

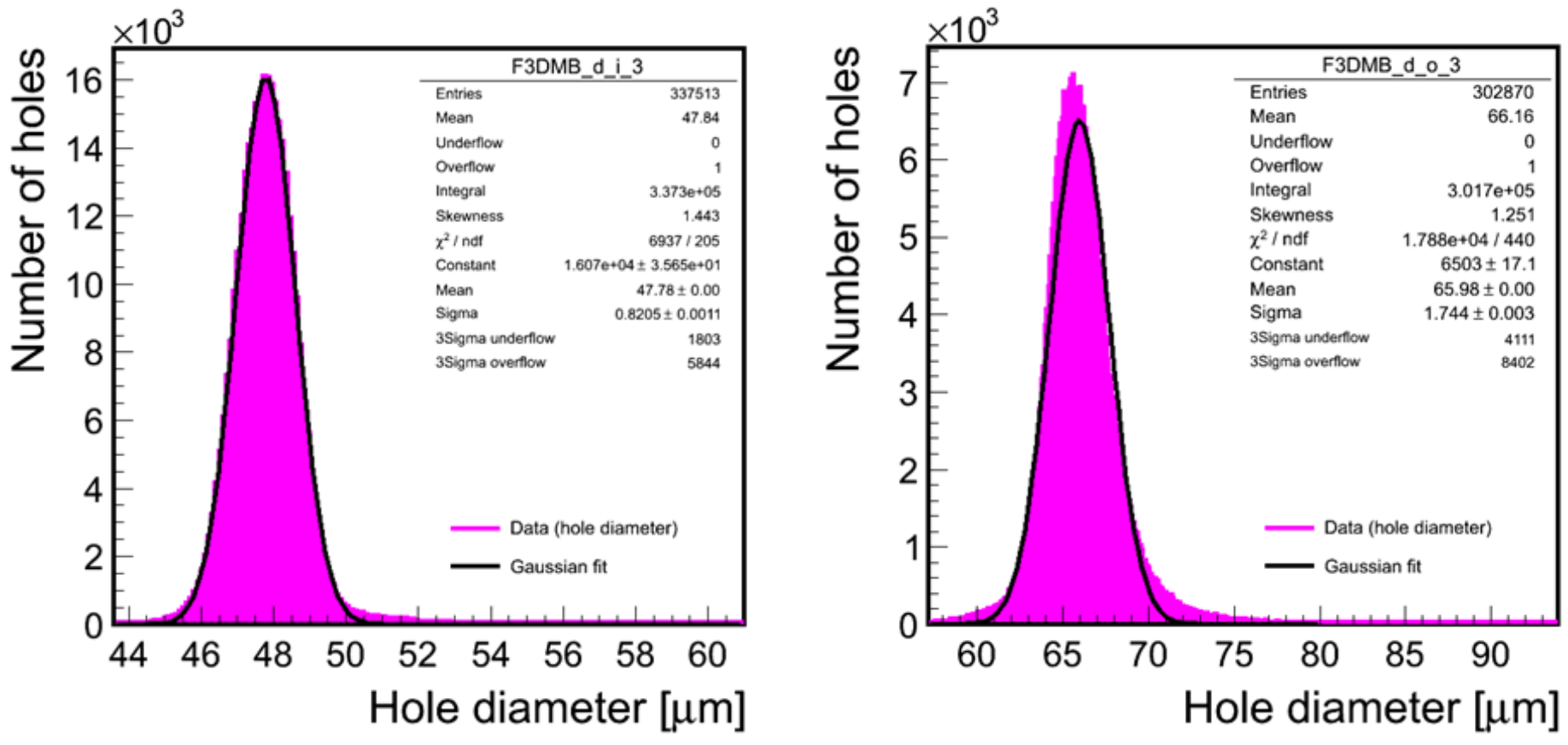

Figure 42: Distribution of the hole diameter for the HGB4 foil. The inner and outer hole diameters are shown on the left and right correspondingly.

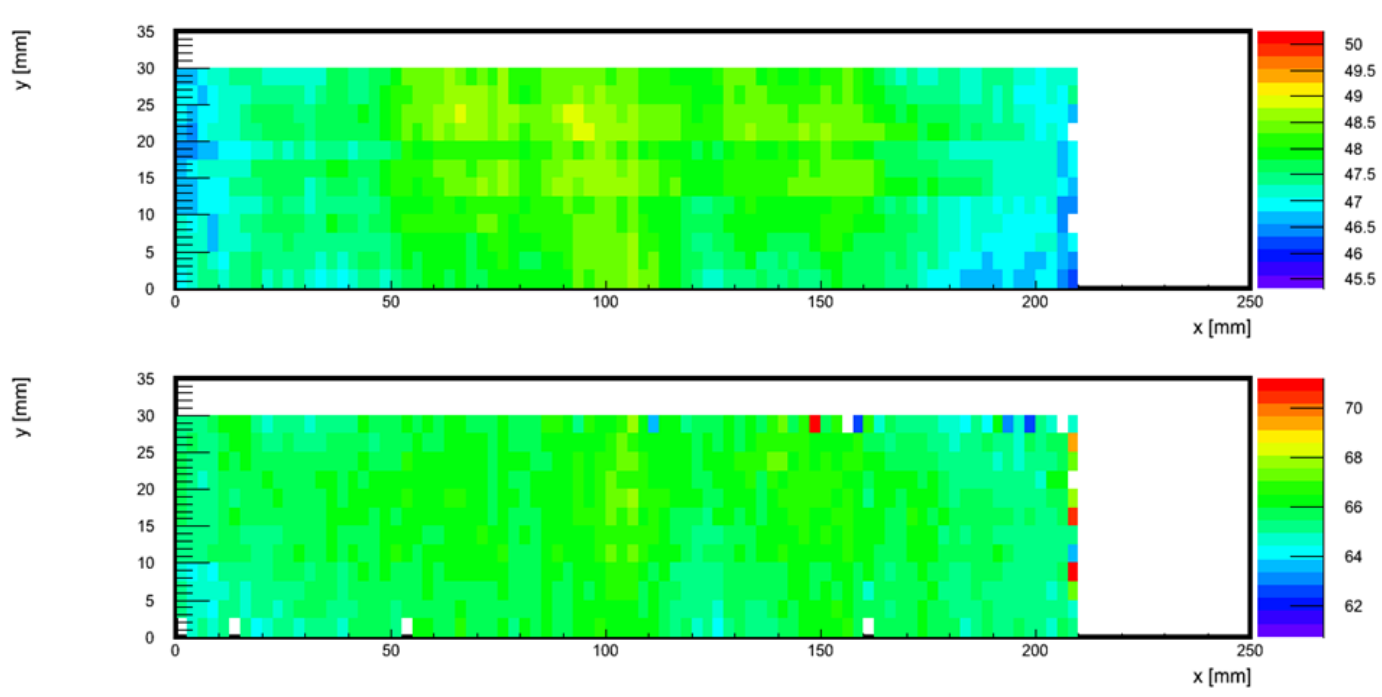

Figure 43: Map for the distribution of the hole diameter for the HGB4 foil. The inner and outer hole diameters are shown on the top and bottom panels, correspondingly.

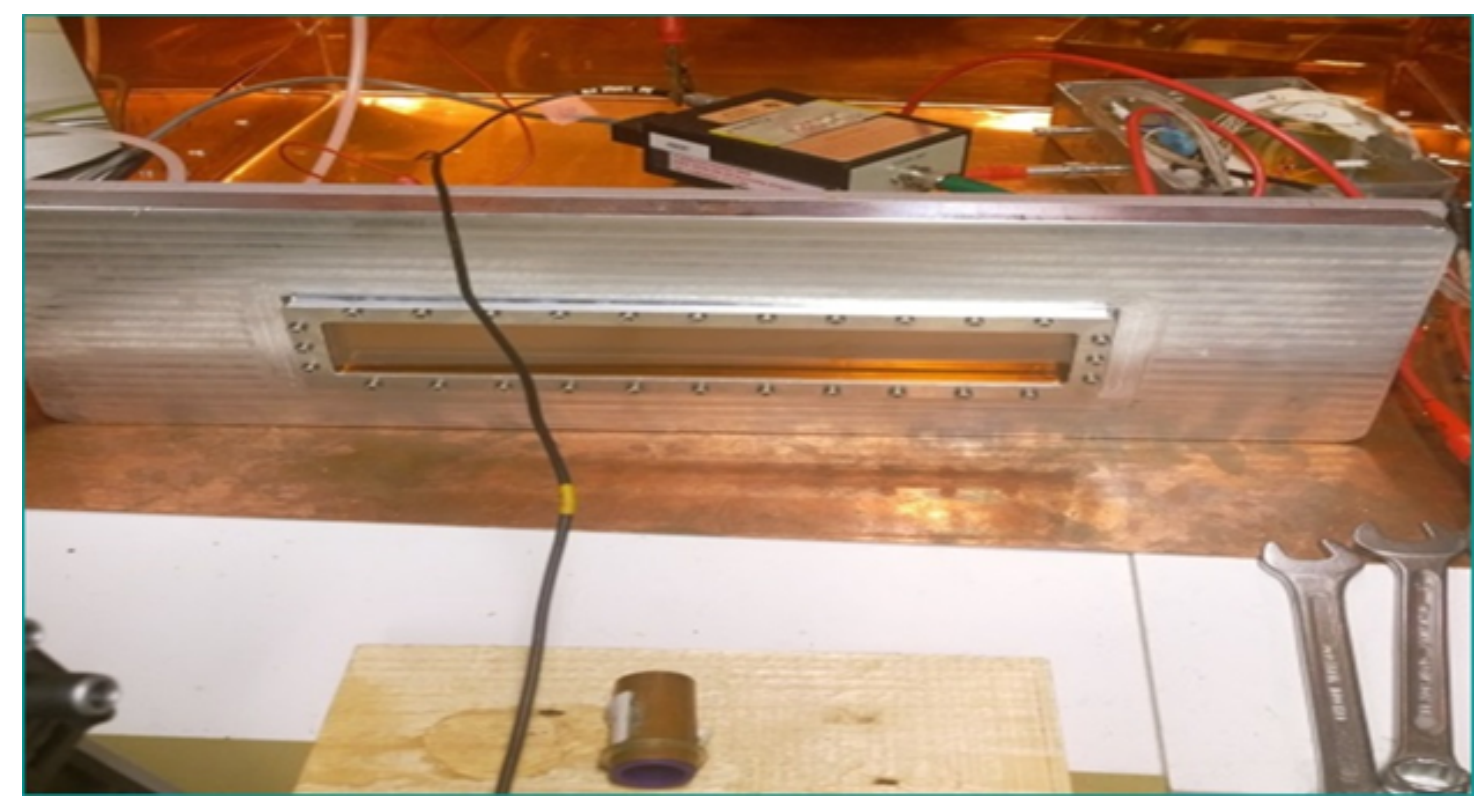

Figure 44: Schematics of the setup for the gain mapping.

## 7.5 Gain Uniformity Measurements

Acceptable performance of the GEM stack of HGB4 will be partly determined by the uniformity of the gain across the whole active area. Therefore, mapping of the gain of each GEM stack needs to be done. This mapping will allow one to find non-uniformities of gain across the whole active area and provide all the necessary information either for the final acceptance or for rejection. In the case that the gain variation of the foil will be inside the acceptance limit of about 7% then the this will be ready to be mounted inside the detector. The test bench for the measurement of the gain is shown in Fig. 44.

The gain calibration for the GEM stack in gas P10 is shown in the Fig. 45, it can be seen that the gain is linear. There are plans to provide the gain mapping in one coordinate along the beam axis and a set-up is now in

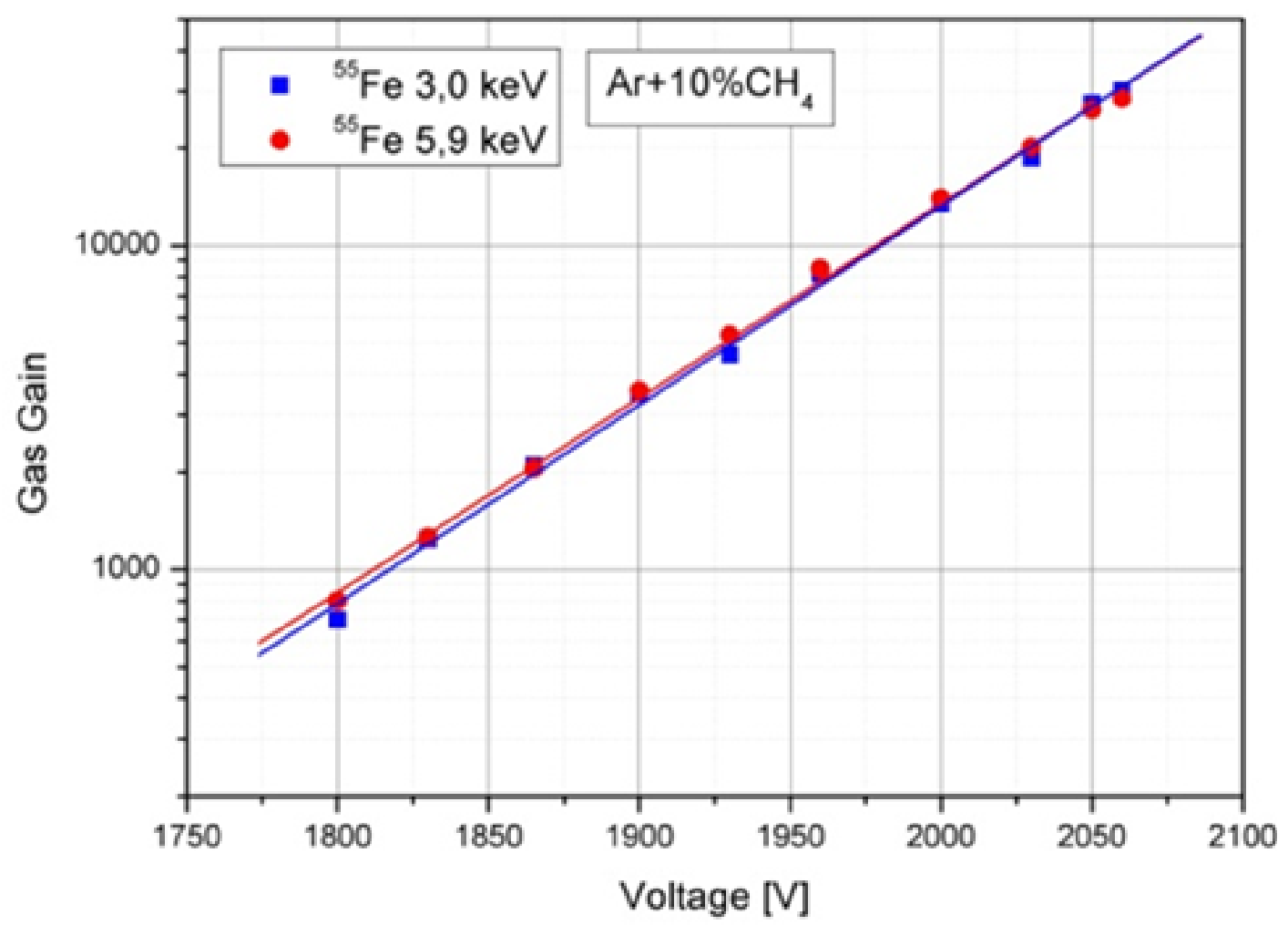

Figure 45: The gain versus the high voltage at the divider is shown for P10 construction using the scalable readout system and the APV25 hybrids.

# 8 Readout Pad-Planes

Each field-cage compartment of HGB4 is equipped with an identical pad-plane that can be operated independently. This optimizes mass-production and eases maintenance by minimizing the amount of different parts to be held on stock as well as by reducing the number of processes for their fabrication and assembly.

The PCB realizes all routes required for the high-voltage distribution for all electrical supplies such as the GEM-foils, HV-filters, low-voltage for the read-out boards and various temperature sensors. Moreover it provides sockets for the high-voltage filters as well as pick-up capacities and the outlets of the gas-distribution system. It directly connects all electrical connectors mounted on the respective panel, except the high-voltage cathode, replacing all cables inside the detector volume which would be required otherwise.

A picture of the actual pad-plane PCB mounted on one of the panel-flanges of HGB4 is shown in Fig. 46.

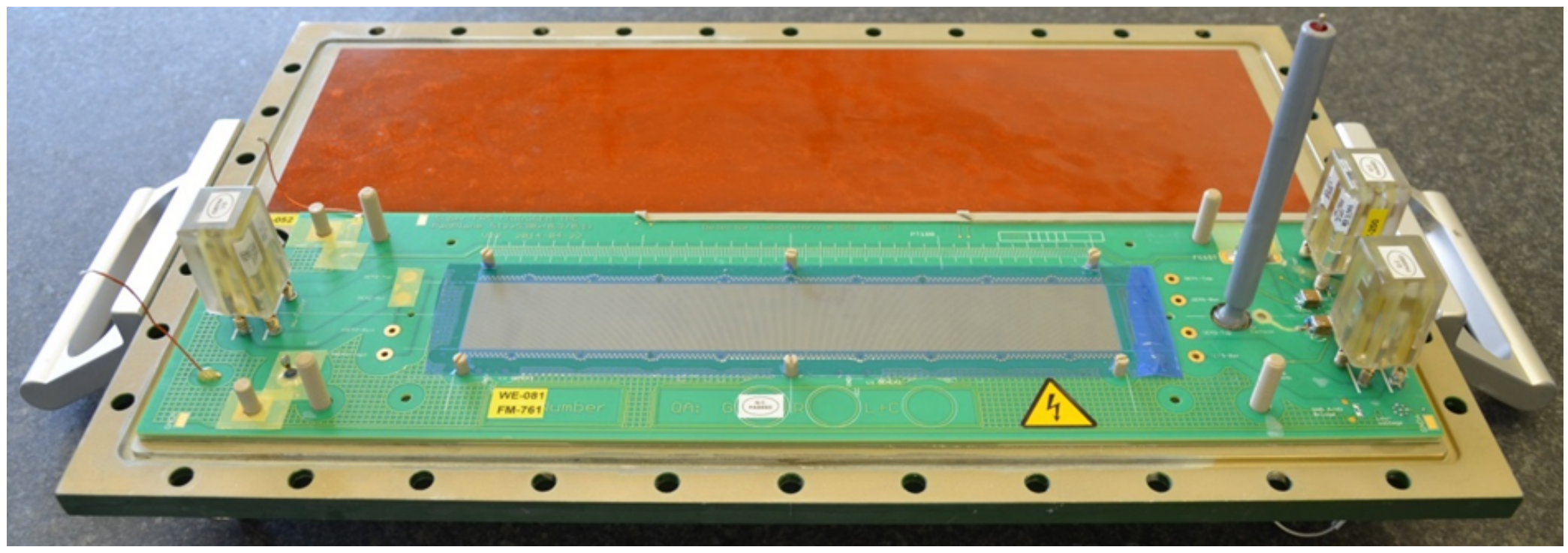

Figure 46: Picture of the actual pad-plane mounted on one of the panel flanges of HGB4.

The pad-plane PCB-sandwich itself consists of two separately produced parts that are merged/glued together during assembly at GSI:

1. A backing provides insulation and screening as well as sealing of the various fluids fed to the detector via the panels. Through its thickness it also helps to minimize the capacity of the signal lines with respect to ground and it provides additional stability and flatness for the sandwich.

2. The actual signal PCB is distributing low- and high-voltage of the supplies and realizes the routing of the electrical signals to be picked up inside the active volume and read out via the front-end boards of the read-out system, which are flanged to the panels.

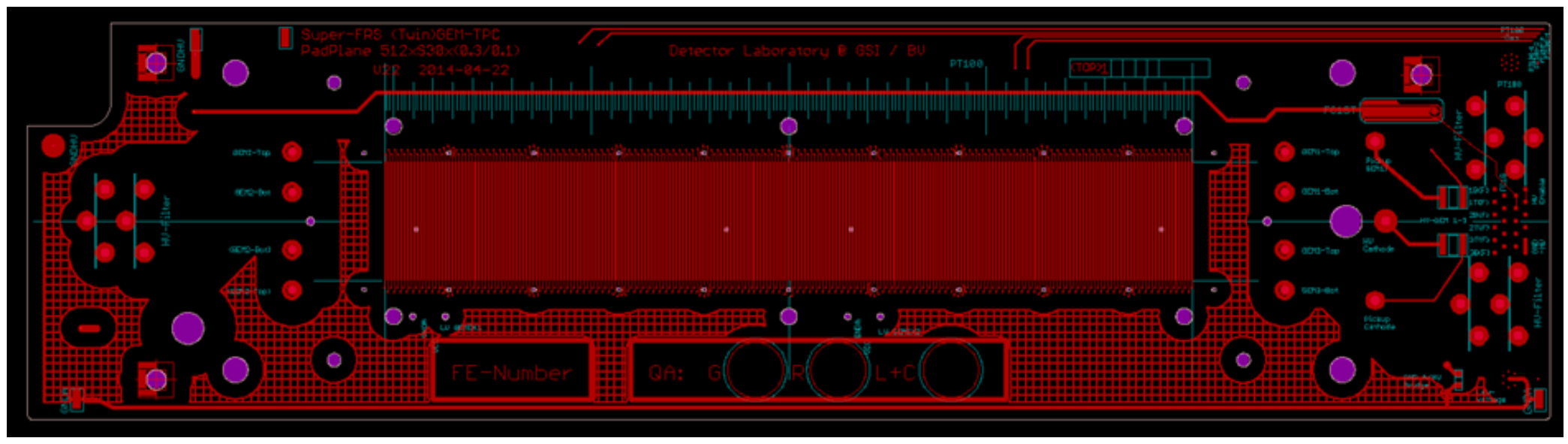

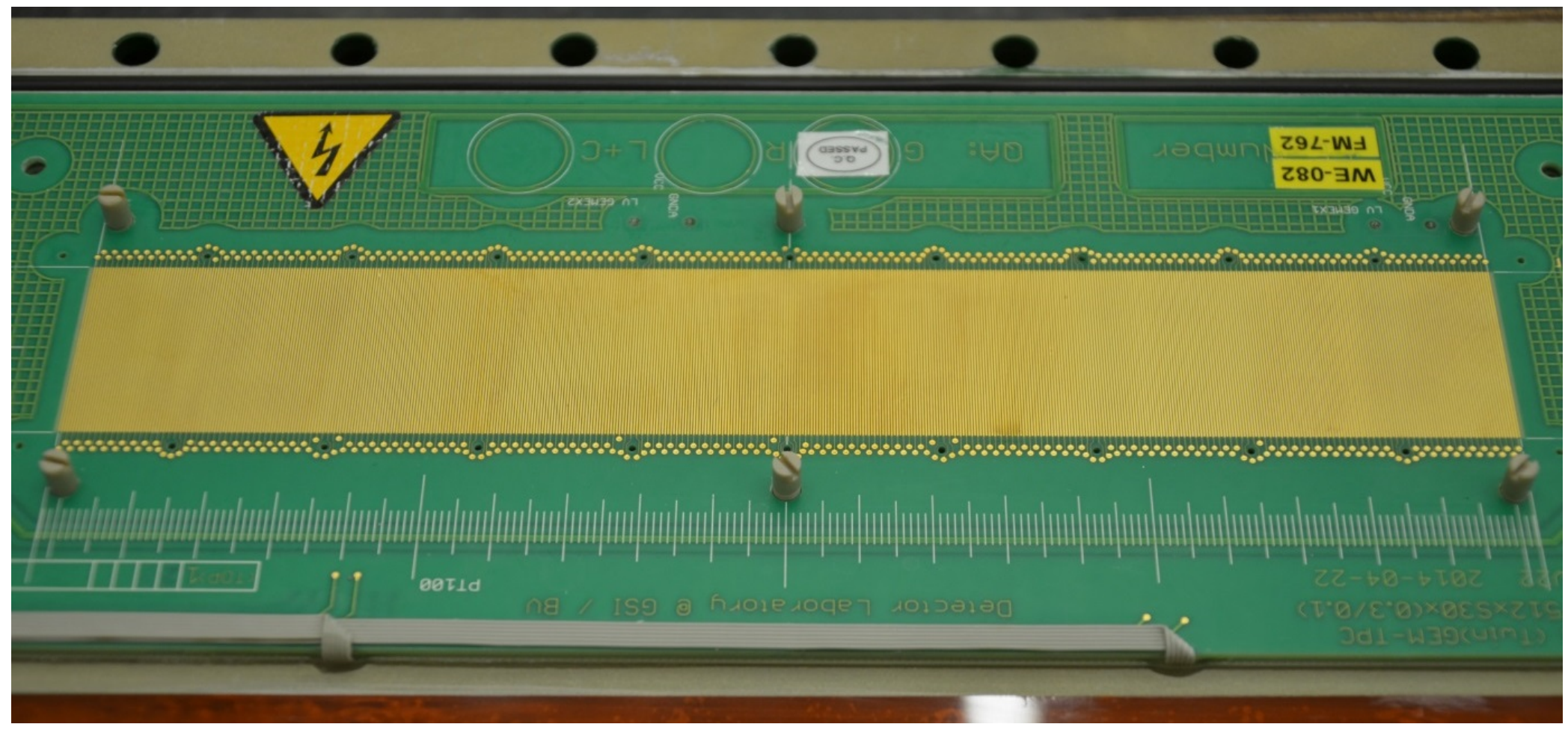

Figure 47: CAD design (upper part) and snap shot (lower part) of the surface of pad-plane of HGB4 facing the gas volume. The inner active area is patterned with 512 strip-type electrodes.

## 8.1 Geometry

Figure 47 shows a snap shot of the surface of the pad-plane of HGB4 facing the gas volume. The inner active area is patterned with 512 strip-type electrodes forming the projection layer of the detector.

Figure 48 shows a snap-shot of the strip-type electrodes forming the read-out structure on the pad-plane of HGB4. The inner active parts of the electrodes are 30 mm in length and 0.25 mm in width, the gap between neighbouring pads is 0.15 mm and the pitch is 0.4 mm.

Figure 49 shows a snap shot of the inner routing structures of pad-plane of HGB4. Gas-tightness of the PCB is assured using a 6-layer multi-layer design and by plugging the respective (blind and buried) through-holes. Realizing an appropriate high density (100 $\mu m$) routing, an optimum (minimal) spread of routing length for all 512 signals has been achieved, minimizing the spread of the respective capacities.

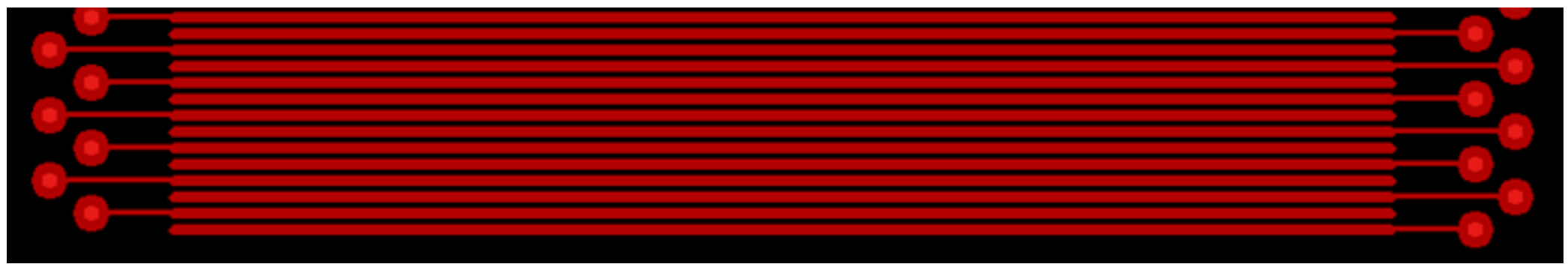

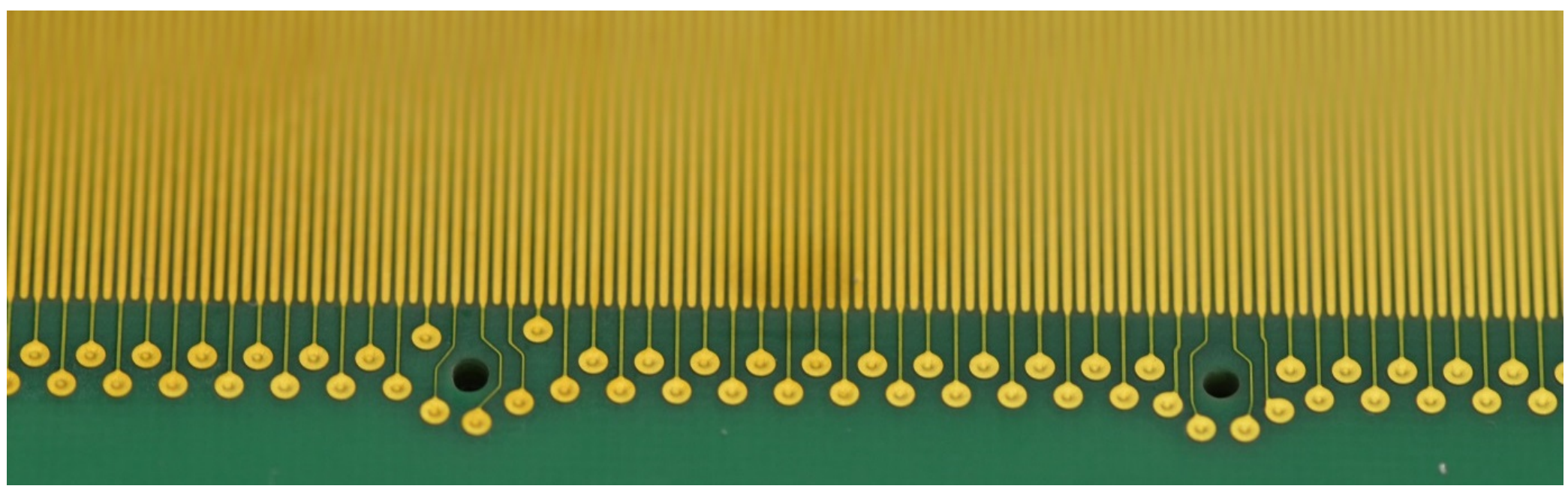

Figure 48: CAD design (upper part) and snap shot (lower part) of the surface of pad-plane of HGB4 facing the gas volume. The inner active area is patterned with 512 strip-type electrodes of 30 mm in length and 0.25 mm in width, the gap between neighbouring pads is 0.15 mm and the pitch is 0.4 mm.

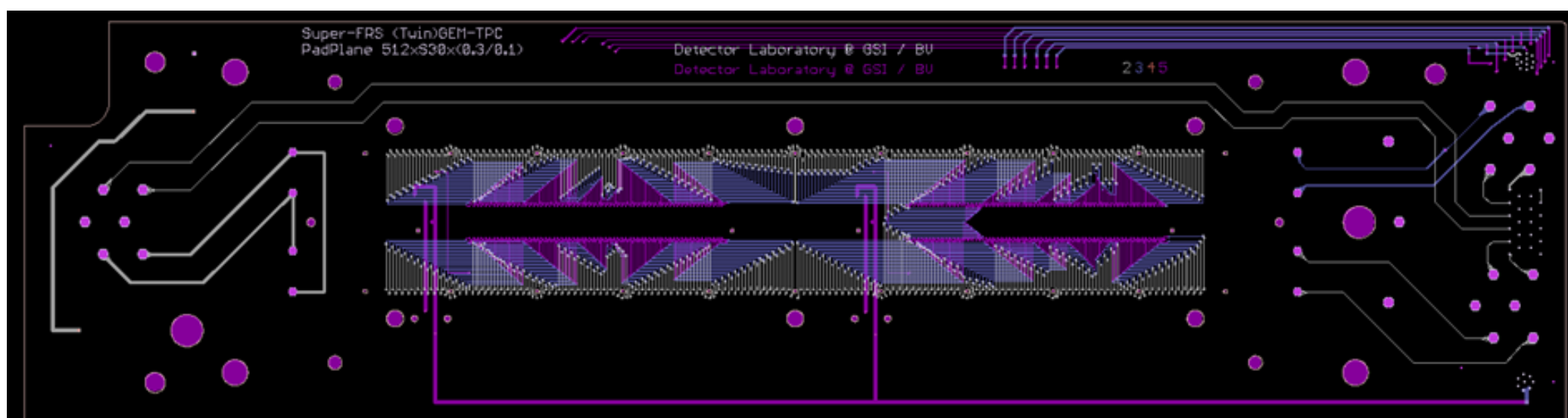

Figure 49: Snap shot of the inner routing structures of the pad-plane of HGB4.

Great care has been taken to allow high-voltage routing through the pad-plane PCB by selecting the appropriate materials and layer stacking and designing sufficient insulation distances. An acceptance test applying 6 kV to all HV routes is part of the quality assurance procedure of the pad-plane

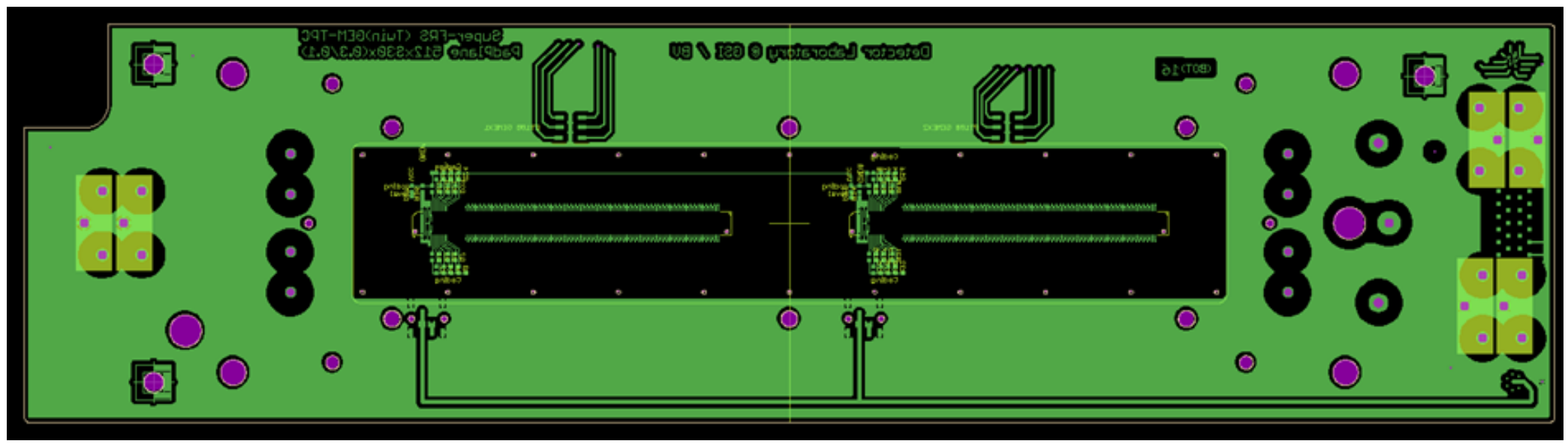

Figure 50: Snap shot of the back-side of the pad-plane of HGB4. The 512 electrodes on the front-surface are internally routed to two SAMTEC 300-pin high-density connectors each connecting 256 pads.

PCB.

Connection to the HV-supply hardware outside of the detector is provided via a 22-pin REDEL connector allowing a 1:1 connection to e.g. a HVGEM module. Separate SHV connectors are used for the biasing of the first strip of the field cage as well as for the cathode high-voltage.

The top/bottom contact-flaps of the 3 GEMs in the GEM-stack (see Figs. 56 and 57) are screwed directly to respective contact points on the pad-plane PCB.

Figure 50 displays a snap shot of the surface of the pad-plane of HGB4 facing the ambient surrounding. The 512 electrodes on the front-surface are internally routed to two SAMTEC 300-pin high-density connectors (0.5 mm pitch), each connecting 256 pads.

Several options for the powering of the read-out electronics sourced by a single input line have been realized on the pad-plane PCB for test purpose:

1. Direct powering of each front-end board via the respective SAMTEC BTHtype connector,

2. Split powering of the respective front-end boards via ERNI mini-bridge type connectors,

3. External powering via the GEMCON PCB externally interconnecting the two front-end boards (not shown here).

## 8.2 HV Filtering

High-voltage supply to the detectors usually is provided over large distances via long cables. In some cases it is advisable to implement filters directly in front of the device in order to get rid of pick-up noise and to stabilize the electrical voltage supplied to the field-cage strips. For this purpose sets of appropriate filters have been produced. The schematic applied is shown in Fig. 51 exemplary for the supplies of GEM-foil no. 1.

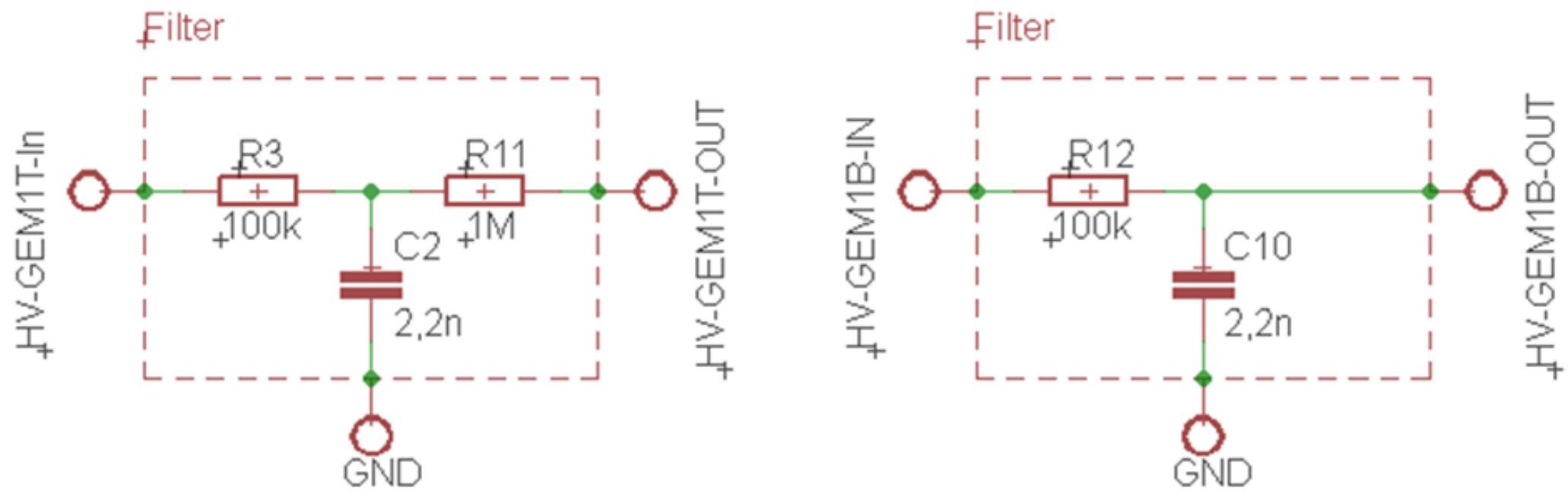

Figure 51: Electrical schematics used for the HV filters here exemplary showing connections to the top (T) and bottom (B) surfaces of GEM no. 1.

Figure 52 shows a picture of two sets of high-voltage filters employed to filter the bias voltages for the GEM-foils. Before and after potting the filters have been tested for their high-voltage stability up to 10 kV. Nevertheless, the nominal operation voltage should not exceed 5.5 kV on long term due to the ratings of some of its components.

Their position inside the detectors vessel has been chosen as close as possible to the GEM-stacks in order to provide best possible filtering; an overview on the placement is given in Fig. 53

## 8.3 Interface to Readout Electronics

Interfacing to the read-out electronics flanged to the panel and the cooler set-up, respectively, is provided via high-density fine-pitch connectors of type SAMTEC BTH-150-01-L-D-A-X which offer up to 300 pins, 256 being used for signal transfer. The rest of the pins have been reserved for powering as well as the implementation of additional functionalities like e.g. unambiguous coding, steering and control of analogue pre-attenuators etc. The actual pin assignment can be found in Table 5.

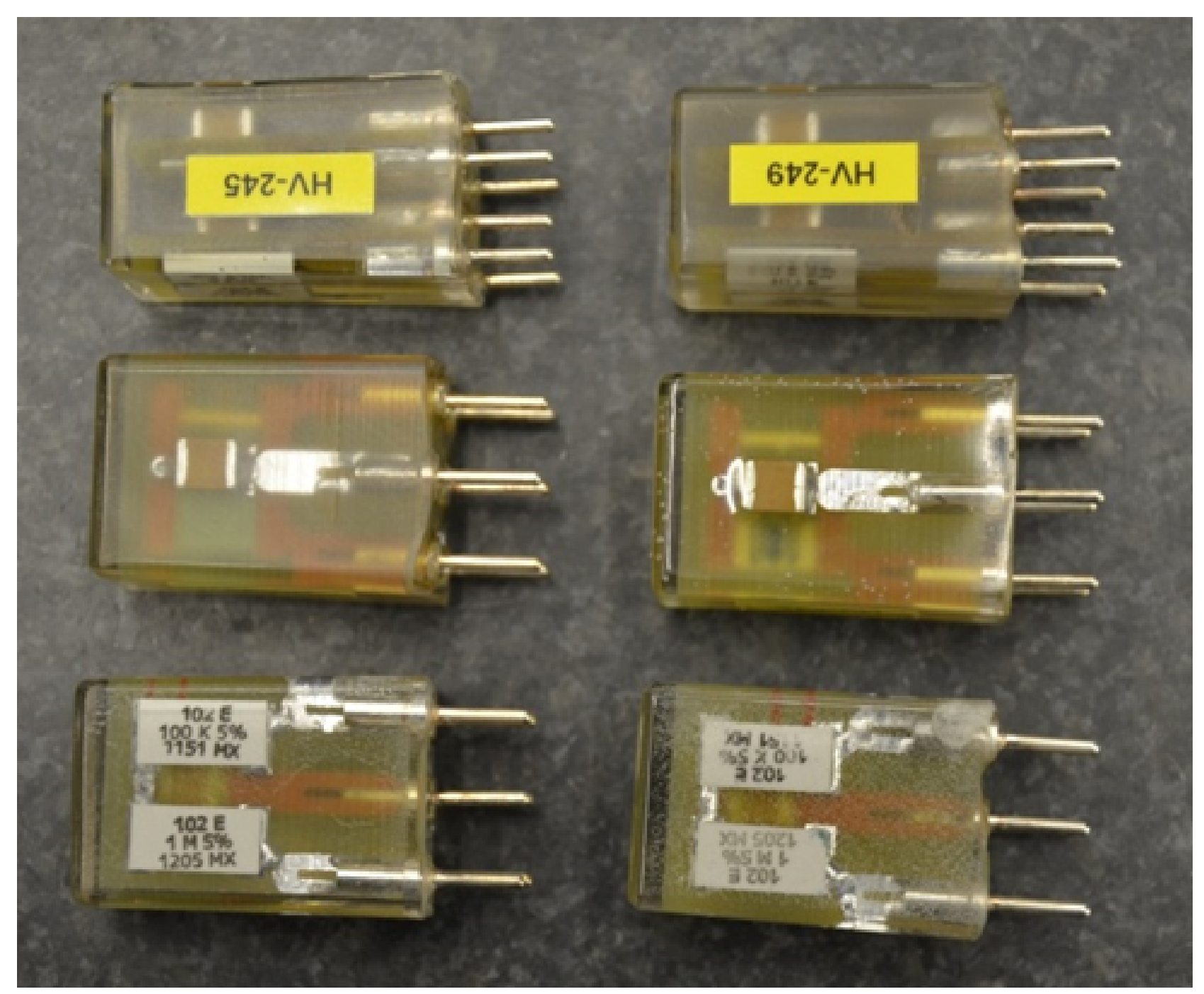

Figure 52: Picture of two complete sets of high-voltage filters employed to filter the bias voltages for the GEM-foils.

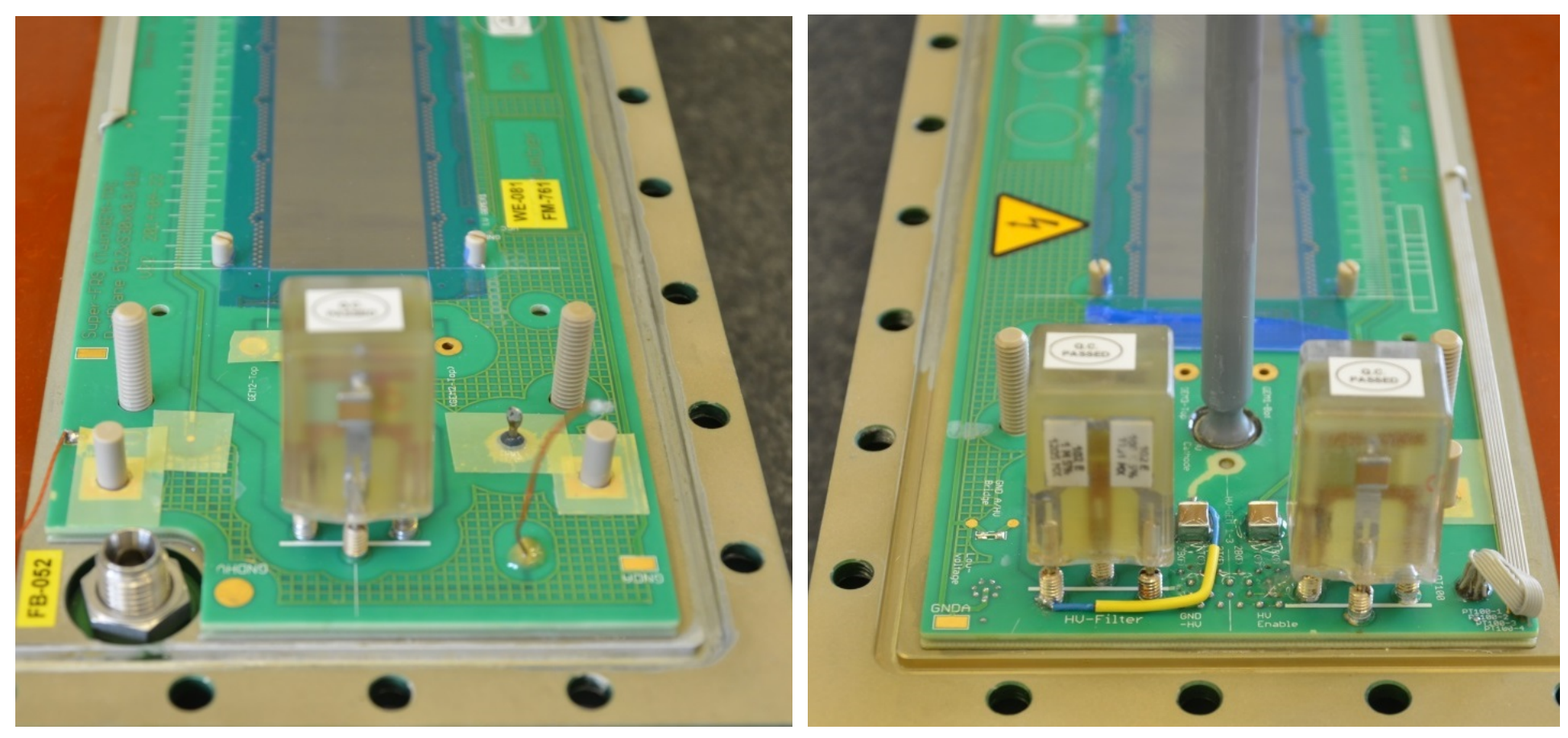

Figure 53: Picture of two complete sets of high-voltage filters employed in order to filter the bias voltages for the GEM-foils.

<table>
<tr><th colspan="6">Assignment for Super- FRS (Twin) GEM-TPC</th></tr>
<tr><th>Pin-Index</th><th colspan="2">Connector-Pin</th><th>Signal</th><th>Direction</th><th>Description</th></tr>
<tr><td>1</td><td>1</td><td>2</td><td>GND-D</td><td rowspan="5">Supply</td><td rowspan="2">(Digital) Ground</td></tr>
<tr><td>2</td><td>3</td><td>4</td><td>VCC</td></tr>
<tr><td>3</td><td>5</td><td>6</td><td>GND-D</td><td rowspan="3">+3,3V / +5V</td></tr>
<tr><td>4</td><td>7</td><td>8</td><td>VCC</td></tr>
<tr><td>5</td><td>9</td><td>10</td><td>GND-D</td></tr>
<tr><td>6</td><td>11</td><td>12</td><td>SlotNo Bit0</td><td rowspan="8">Out</td><td rowspan="8">Coded on PadPlane</td></tr>
<tr><td>7</td><td>13</td><td>14</td><td>SlotNo Bit1</td></tr>
<tr><td>8</td><td>15</td><td>16</td><td>SlotNo Bit2</td></tr>
<tr><td>9</td><td>17</td><td>18</td><td>SlotNo Bit3</td></tr>
<tr><td>10</td><td>19</td><td>20</td><td>SlotNo Bit4</td></tr>
<tr><td>11</td><td>21</td><td>22</td><td>SlotNo Bit5</td></tr>
<tr><td>12</td><td>23</td><td>24</td><td>SlotNo Bit6</td></tr>
<tr><td>13</td><td>25</td><td>26</td><td>SlotNo Bit7</td></tr>
<tr><td>14</td><td>27</td><td>28</td><td>ChipID Bit0 / VCC</td><td rowspan="7">Reserved (e.g. Pre-Divider - Absorber ...)</td><td rowspan="7">Coded on FE-Board</td></tr>
<tr><td>15</td><td>29</td><td>30</td><td>ChipID Bit1 / VCC</td></tr>
<tr><td>16</td><td>31</td><td>32</td><td>ChipID Bit2 / Out C1</td></tr>
<tr><td>17</td><td>33</td><td>34</td><td>ChipID Bit3 / TDB LF</td></tr>
<tr><td>18</td><td>35</td><td>36</td><td>ChipID Bit4 / TCK CLK</td></tr>
<tr><td>19</td><td>37</td><td>38</td><td>ChipID Bit5 / TMS Data</td></tr>
<tr><td>20</td><td>39</td><td>40</td><td>ChipID Bit6 / TDA C16</td></tr>
<tr><td>21</td><td>41</td><td>42</td><td>GND-A</td><td rowspan="2">Supply</td><td rowspan="2">(Analog) Ground</td></tr>
<tr><td>22</td><td>43</td><td>44</td><td>GND-A</td></tr>
<tr><td>23</td><td>45</td><td>46</td><td>XYTER-Channel 1</td><td>IN</td><td rowspan="3">Pads on PadPlane</td></tr>
<tr><td>...</td><td>...</td><td></td><td>...</td><td></td></tr>
<tr><td>150</td><td>299</td><td>300</td><td>XYTER-Channel 256</td><td>IN</td></tr>
</table>

Table 5: Pin assignment of the SAMTEC BTH-150 connector.

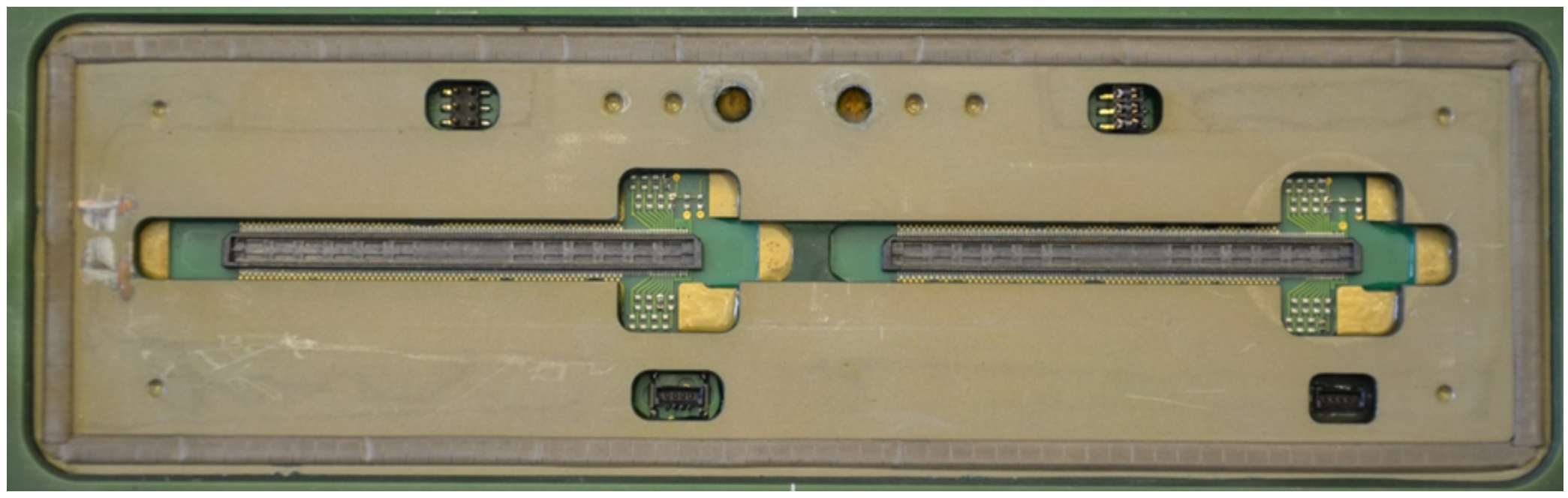

Figure 54: Picture of two adjacent high-density read-out connectors mounted on the back-side of the Pad-plane PCB (the copper cooler has been removed for that purpose).

A picture of the two adjacent SAMTEC connects mounted on the pad-plane PCB glued to the read-out flange is shown in Fig. 54. Channel counting is continuously 1-256, 257-512 following the unique scheme shown in Table 5.

## 8.4 GEM Stack

The GEM-foils employed in this project so far were produced by CERN. Their active area is of rectangular shape: $210 \times 28$ mm$^2$. A set of three framed GEM-foils are stacked in a 'typical' 2-2-2 (mm) arrangement. A picture of such a stack arrangement with so far untrimmed connector 'flaps' is given in Fig. 55.

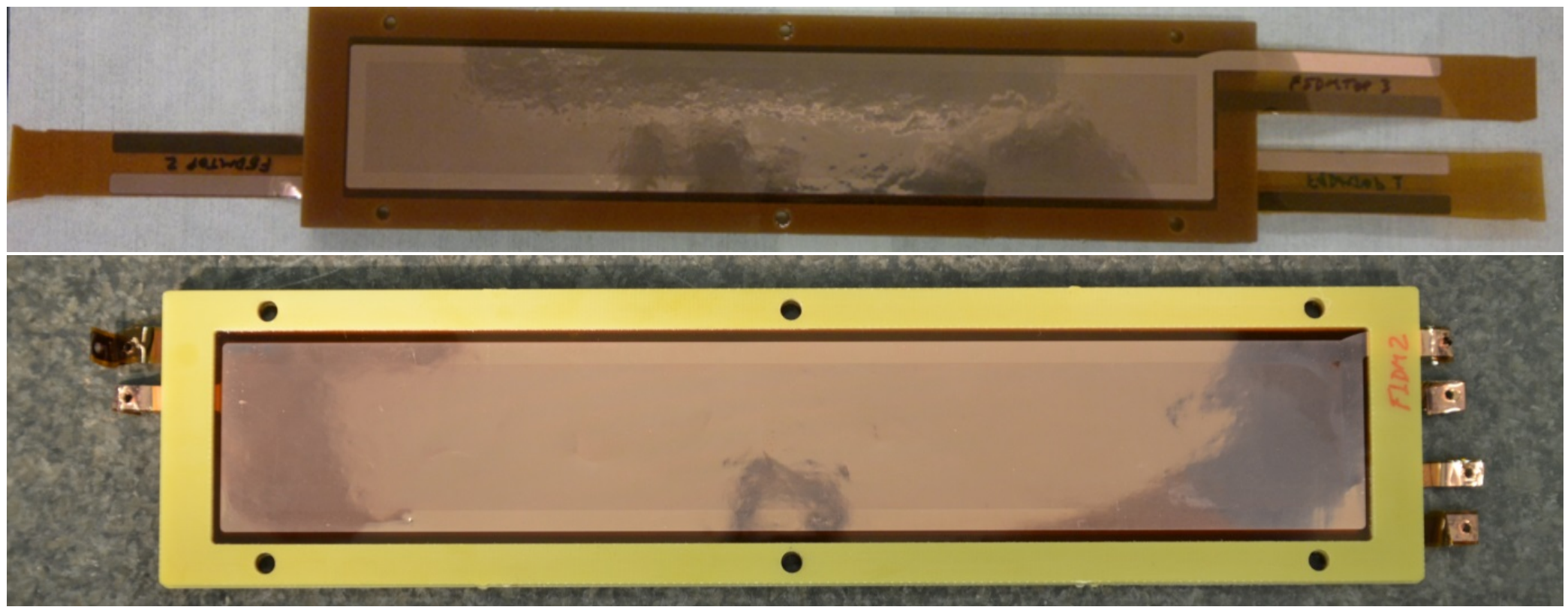

Figure 55: Picture of a set of three framed GEM-foils forming the amplification stage of a single compartment of HGB4 from the top side with uncut flaps (upper part) and from the bottom side with cut flaps 'ready-to-mount' (lower part).

The 'flaps' are used to lead out the electrical signal of each surface of the various GEM-foils. They are directly screwed to corresponding pads on the pad-plane as can be seen in Figs. 56 and 57.

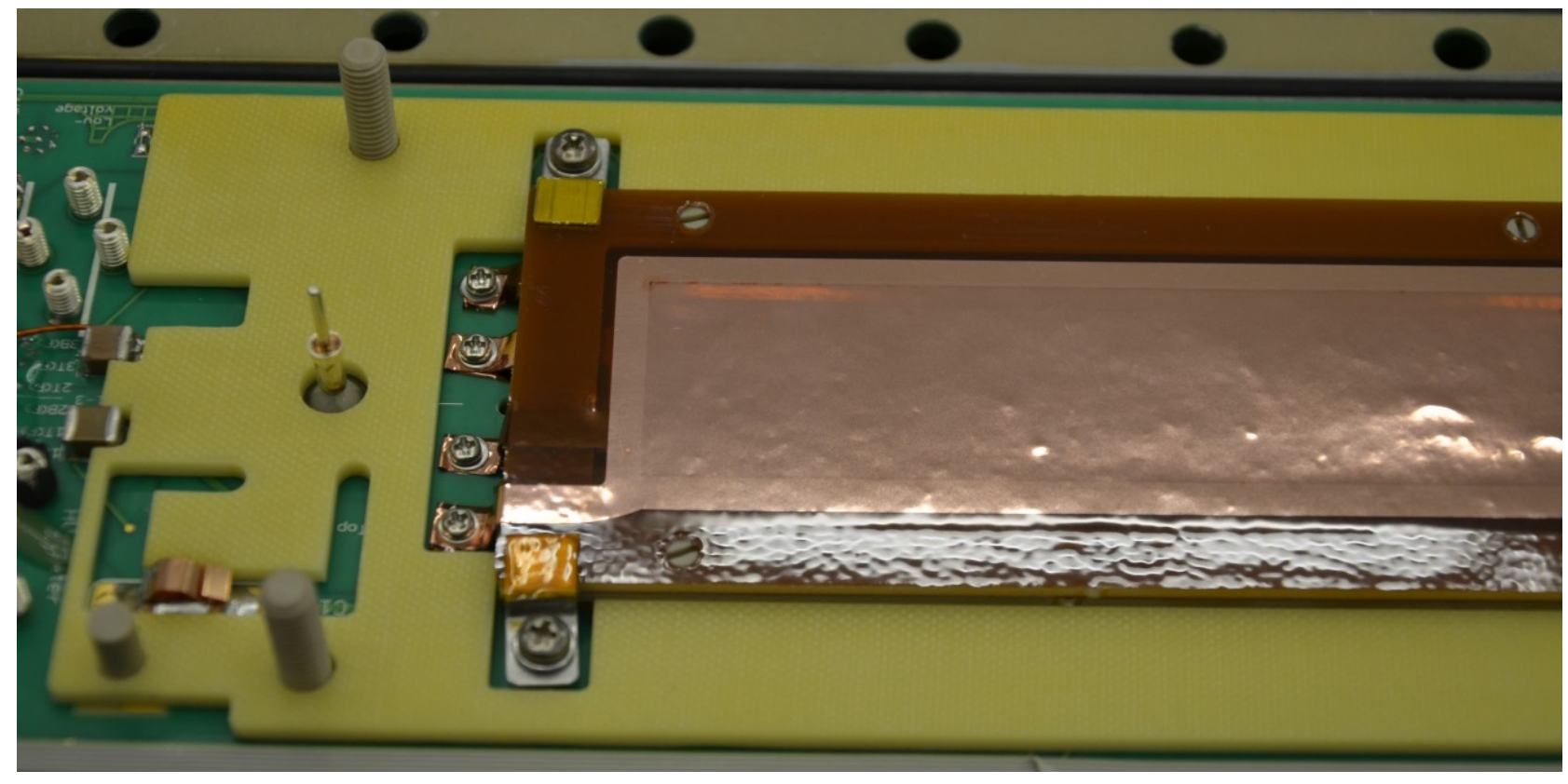

Figure 56: Picture of a GEM-stack screwed to the pad-plane.

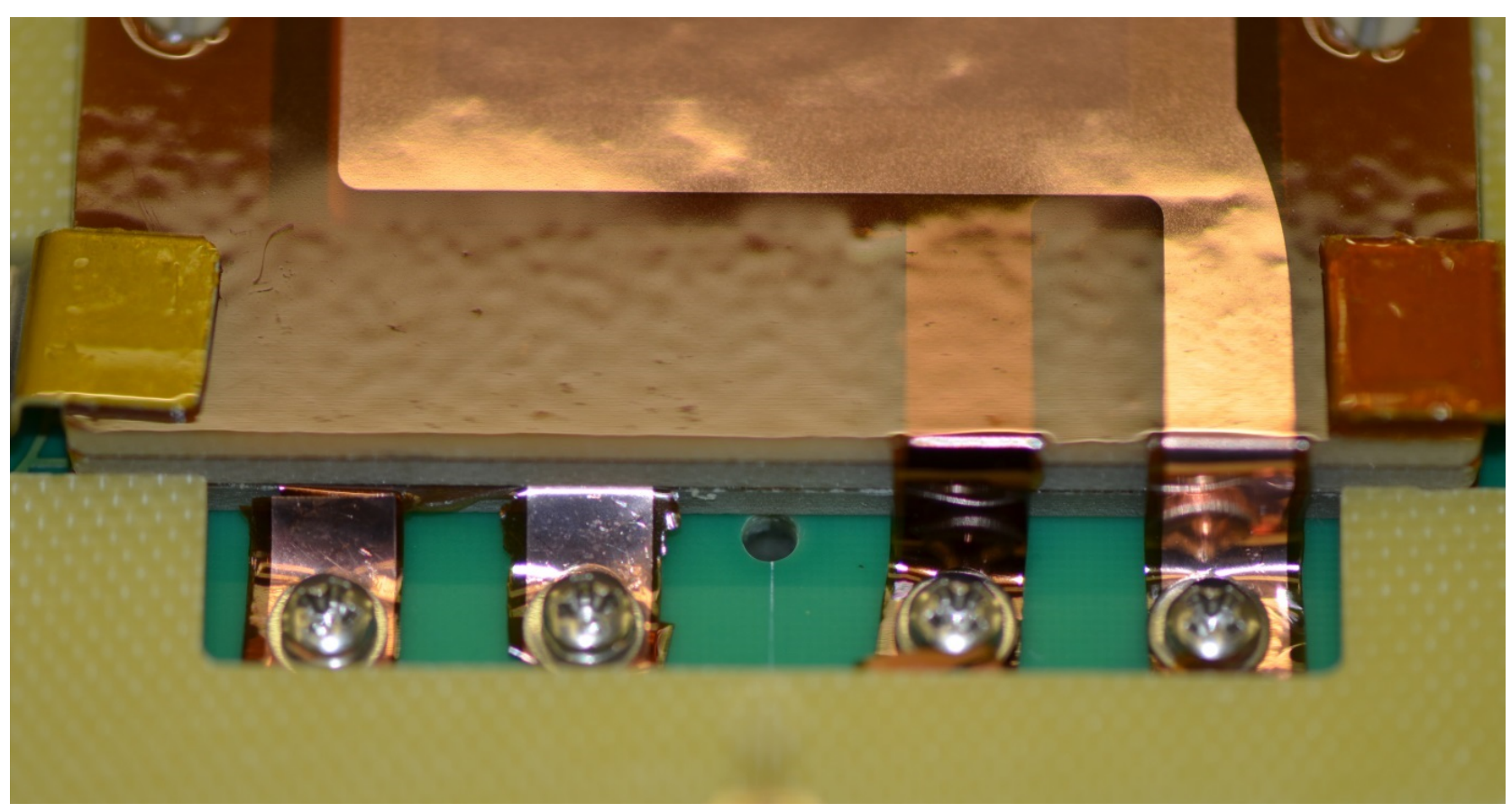

Figure 57: Picture of the flaps of the GEM-stack screwed to the pad-plane.

Special care is taken after the assembly of the whole detector to assure a leak-less operation. For that purpose a slight overpressure (100 mbar) is applied and its stability over several hours is checked. In addition, the gas purity is monitored. Usually the oxygen level is below 6 ppm and the water content in the order of 200 ppm for a detector in operation conditions.

The isolation stability of the high-voltage lines is tested by measuring the resistance between the routes leading to the GEMs as well as the respective value with respect to Ground/GND of the panel and/or housing vessel.

Accepted levels are in the order of 10 TΩ for $R_{\text{GEM-GEM}}$ and 100 TΩ for $R_{\text{GEM-GND}}$.

# 9 Front-End Electronics

The design of the Front-End readout Electronics (FEE) is driven by the following requirements [9]:

- Number of channels per single detector: 500 (200 mm wide detector), 950 (380 mm wide detector)
- Installation close to the detector in order to avoid signal transmission over large distances that would deteriorate the obtainable resolution
- MBS [10] Data Acquisition (DAQ), which is the standard data acquisition system of GSI.
- Maximum count rate: 10 MHz (ion rate in the whole detector).
- The timing reference to be used in the long term is White Rabbit [11].
- Vacuum operation.
- Radiation hardness.

Furthermore, simulations have been performed to determine the dynamic range needed for the charge measurement (see chapter 10, Fig 73), which is approx. 0 to 100 fC.

The spatial resolution in the $x$ coordinate is directly given by the design of the pad-plane (pitch of 0.4 mm) of HGB4. Since the standard deviation ($\sigma$) of a uniform distribution is $\frac{w}{\sqrt{12}}$ (with $w$ being the width) one obtains $\sigma_x = 0.12$ mm or $\text{FWHM}_x = 2.355 \times \sigma_x = 0.27$ mm. The $y$ coordinate is determined from the drift-time measurement ($y = v_{\text{drift}} \times t_{\text{drift}}$). In order to obtain a similar spacial resolution as for $x$ a time resolution below 3 ns is required (see Table 2).

## 9.1 Developments of pre-series

So far the N-XYTER [12] based electronic board called GEMEX [13] was used for the tests of HGB4. The N-XYTER chip was selected as ASIC since it offers 128 independent readout channels on a very compact chip (the die size is 6.5 mm × 3.1 mm). Furthermore, there is in-house experience with this ASIC at GSI, giving us the unique opportunity to design the full board together with the ASIC developers.

The GEMEX comprises two N-XYTER chips, an ADC and a FPGA readout chain based on the EXPLODER [14] card. It is the 256 channel front-end board hosting a trigger, a time-stamp logic, an external clock input to the FPGA and a high-precision PLL (Phase Locked Loop) synthesizer. The N-XYTER chip operates at a sampling rate of 32 MHz with 12 bit resolution and with the dynamic range up to $1.2 \times 10^5$ electrons. The readout of the buffered data is synchronized to the 32 MHz clock in a token ring. The token Read-Out Controller (ROC) register is set if there are data in the buffers of one N-XYTER channel. For each clock cycle, the data of the channel are sent via a data bus to the FPGA. In that way, a readout rate of 250 kHz per channel is reached (32 MHz/128 = 250 kHz) [15].

Each of the boards comprises a 300 pin connector, where up to 256 detector signals can be connected. The remaining free connections are used e.g. to supply low-voltage to the electronics (see table 5).

Due to the temperature drift of the baseline of the version 1.1 of the chip, cooling is required. The heat dissipation of the GEMEX card was realized by a partial solid backing of copper directly mounted to the ASICs on the heat spreader. Due to stringent spatial restrictions heat is dissipated effectively via the edges, where an active fluid cooling system is flanged to the surfaces [9, 16].

The main issue with the applied electronics is the dynamic range. Here, as mentioned, it is $1.2 \times 10^5$ $e^-$ - which expressed in charge units leads to the range up to about 20 fC (see Fig. 58). This is too small compared to the requirements (see Fig. 73). However, only a part of the shown full range calculated for the ions from protons up to uranium, is required in real experiments at Super-FRS that concentrate on a given mass range. The issue of insufficient dynamic range will also be solved with a controllable add-on pre-attenuator card.

The general requirements given in the previous section become more specific if one considers the specification of the chosen ASIC and readout design:

- 185 GEMEX cards are required in total (256 channels each).
- A reference clock for FPGA and ASIC is required.
- The N-XYTER version 2.0 will be used with the next revision of the GEMEX card.
- (Fluid) Cooling is needed.
- As for protection of the ASIC, glob-topping is envisaged.
- The powering scheme and sequence starts at N-XYTER and proceeds then to the FPGA.
- N-XYTER and ADC emulation will be realized as software features.
- Internal and external pulsers are required for probing the N-XYTER/ADC range.

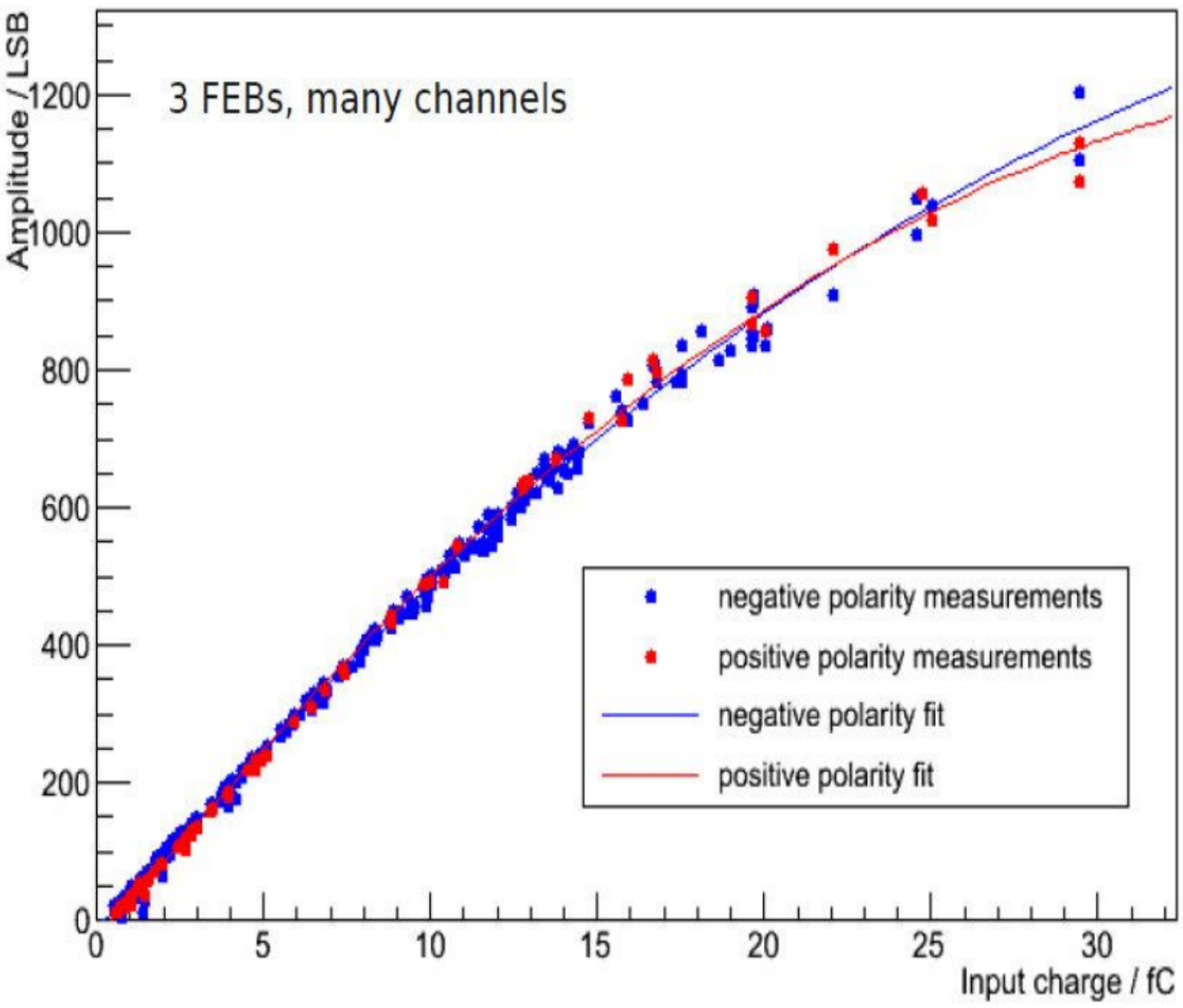

Figure 58: N-XYTER 2.0 calibration result [17]

## 9.2 Outlook on the development of the readout electronics

The last version of the readout electronics (GEMEX-1c) showed several deficiencies and is currently subject to revision. The production of the board is very challenging due to the following facts:

- Very compact design (PCB form-factor): Combination of the FPGA (multi-layer PCB needed) and ASIC (50 $\mu$m fine structures) on the same PCB.
- ASIC mounting/bonding pitch adapter is needed (N-XYTER has a 50 $\mu$m inter-channel pitch).

The integration of the N-XYTER and FPGA in such a compact design increased the probability that one of the components caused the full board not function correctly. Some of the samples produced in the last revision worked in an erroneous state, due to the fact that the powering sequence (N-XYTER first, then FPGA) and scheme (on board) was not correctly executed.

As a consequence, a revision of the board has already been started. In the next version of the readout electronics some components will be replaced:

- Instead of the previously mentioned FPGA [1] a XILINX Spartan 6 (same as used for EXPLODER) will be used. This FPGA is larger and makes the transformation of the EXPLODER software easier.
- The N-XYTER 1.1 will be exchanged with the N-XYTER 2.0, most probably offering more stability with temperature.
- Integration of the ASIC will be realized using a pitch adapter.

In order to be able to reduce the overall complexity and also to go back to a space-wise more relaxed design enabling us to disentangle problematic issues, three single boards will be built (see Fig. 59).

The new design of the read-out chain offers the following advantages:

- Components can be tested individually.
- Standalone tests of new the FPGA are possible.
- Standalone tests of (new) N-XYTER 2.0 are possible.
- The connector in between the cards (e.g. between N-XYTER and the FPGA) gives access to all signals .
- More space to realize additional test pads is available.
- Production of individual PCBs is easier. Separation of the difficult production processes (multi-layer/fine structures) to different boards minimizes the total risk for production failures

[1] this device was discontinued by the manufacturer

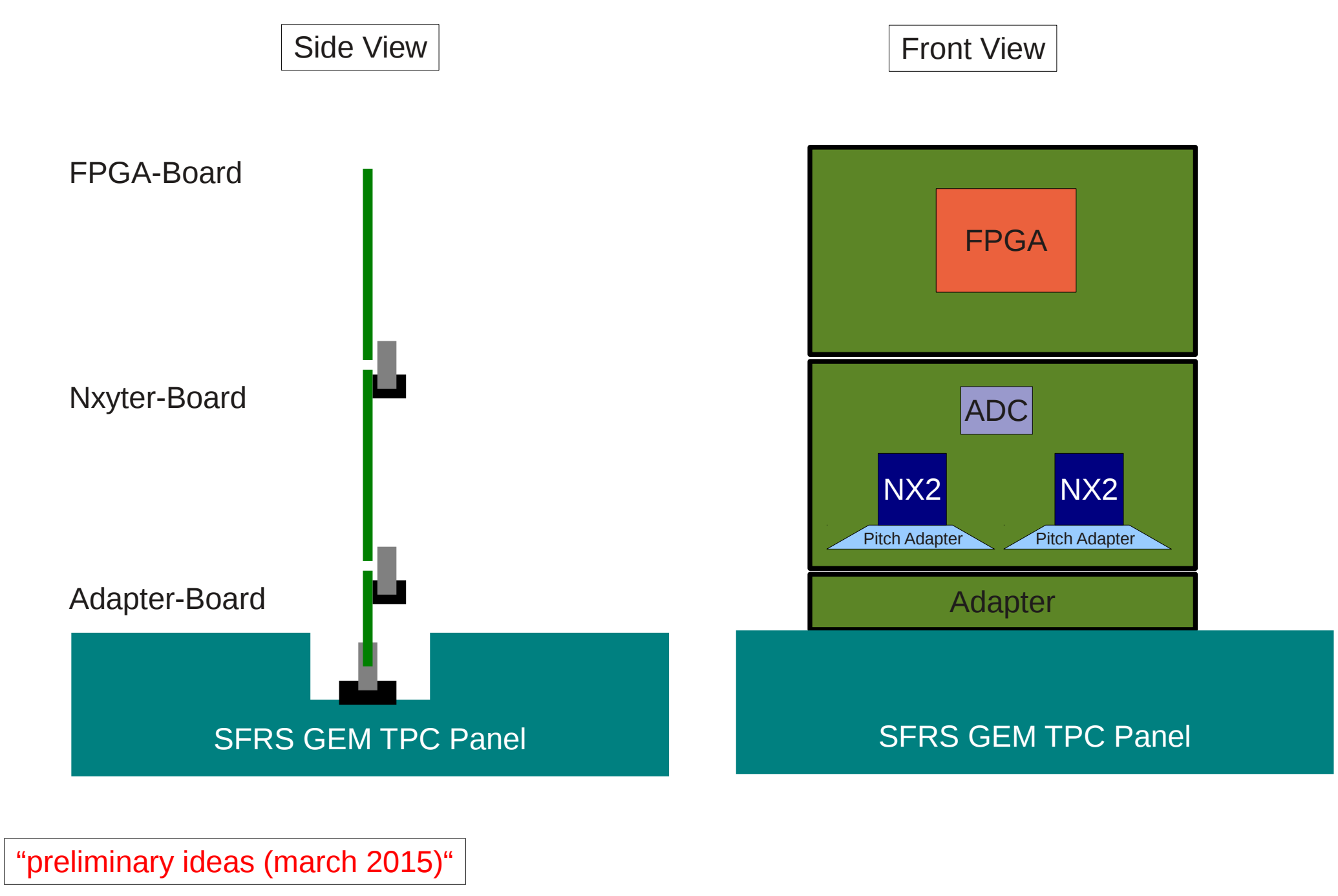

Figure 59: Design scheme of the next version of the readout electronics chain.

## 9.3 DAQ integration/Data transmission

A scheme of how the GEMEX board is integrated into a the MBS DAQ is shown in Fig. 60. The triggers and the external clock are provided via a LevCon and are sent as LVDS signals to the GEMEX. A 32 MHz clock is integrated in the LevCon. Up to four different trigger types can be forwarded from the DAQ to the GEMEX via the LevCon. The trigger types could e.g. be beam-On, beam-Off, calibration, etc.

The data from the GEMEX are sent to a PC which contains two special modules developed by the GSI electronics department–the PEXOR [19] and the TRIXOR [18]. The PEXOR has four SFP connectors for optical links. All the data transfer between the GEMEX and the PEXOR is done via the optical link. Also the settings of the GEMEX are set via this optical link, typically the trigger window, the baselines and the thresholds. The TRIXOR is a triggering module and can be connected to the trigger bus of the main DAQ. If the trigger is sent from the DAQ, data from the GEMEX will be accepted by the PEXOR. The received data are verified by checking the data structure and by comparing the trigger type and the trigger counter with the information received from the TRIXOR. Validated data are then sent via network to the event builder [15].

The above presented chain leads to the following parameters for the data transmission:

- 200 MB/s per SFP
- transmission from the PEXOR to the DDR3 (PC): 600 MB/s
- ethernet connection, e.g. 10 Gigabit: approx. MB/s

To estimate the real data through-put of the full chain (detector + FEE + DAQ) the information on the data structure is mandatory. For each N-XYTER chip one two-stage double RAM Buffer is implemented in the core of the FPGA. The first stage receives all data, coming from N-XYTER and the ADC during the trigger window. At the end of the post-trigger window:

- the second arm of the double RAM buffer is activated
- the transfer of the valid data from the first into the second stage is started.

During the transfer the data are packed. The data packets have the structure shown in Table 6.

The length of the packet depends on the registered hits. The minimum length is 5 words of 32 bit (no registered hits or a packet of synchronization trigger). The maximum data packet length is approx. 4k words of 32 bit.

Using all the above information it is clear that the possible rate of data taking (event rate) depends on various parameters, like e.g.:

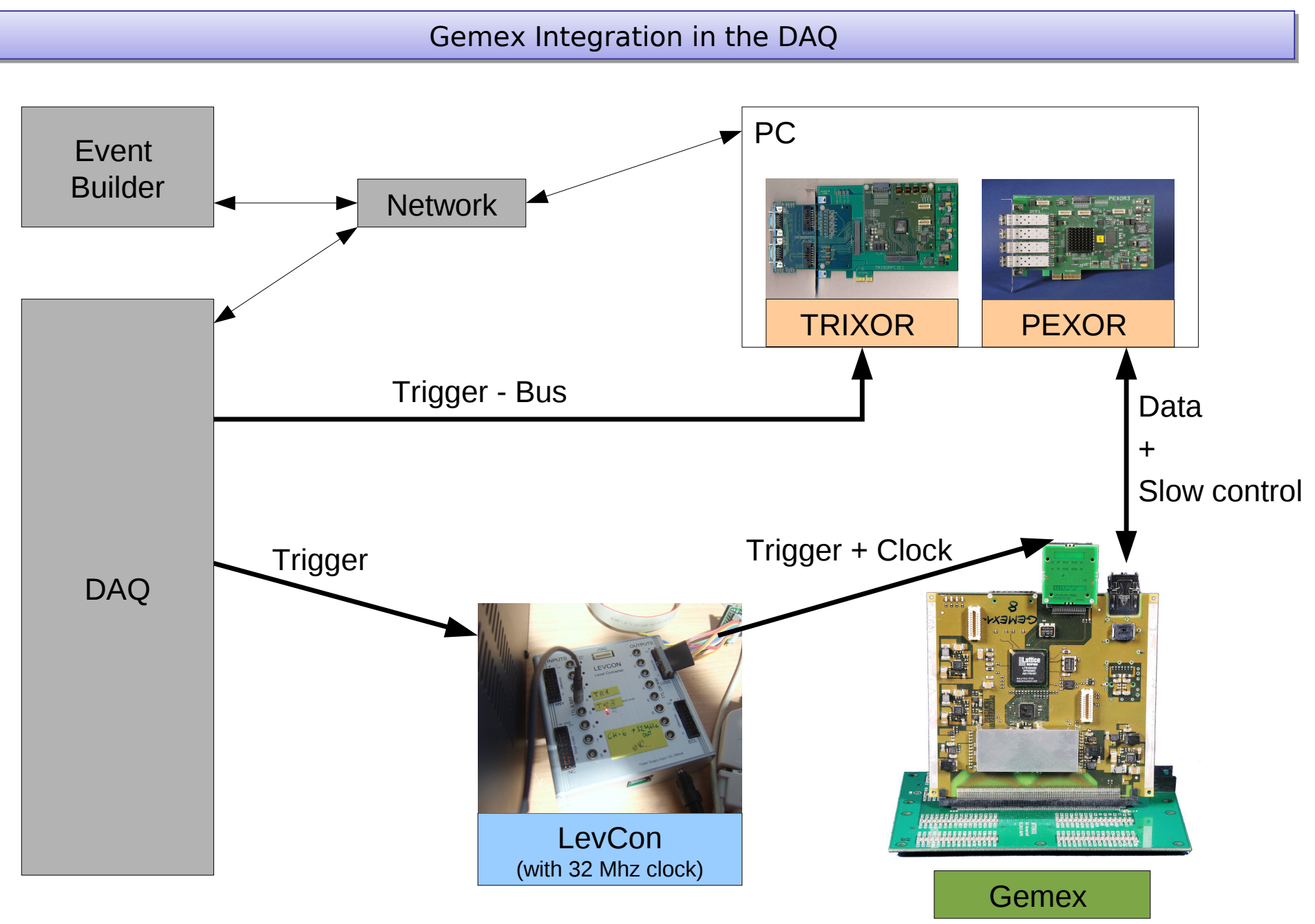

Figure 60: Integration of the GEMEX card into the MBS DAQ [15] .

- number of the GEMEX cards per SFP
- ethernet connection
- the trigger-window setting

One simple example will be discussed in order to give an idea on the rate capability/data throughput. For 1 event approx 15 strips will fire (see Fig. 72). Since the detector consists of the twin field-cage configuration, 15 channels will fire in two N-XYTERs. Thus one would have 2 data packets containing 35 words of 32 bit each, leading to 280 B/event. In this example let us consider a set-up in which one GEMEX card has been connected via one

| Packet Header | 2 words of 32 bits |
|---|---|
| Data Header | 1 word of 32 bits |
| Epoch Time Stamp | 2 words of 32-bits |
| Packet's Data Body | 2 words of 32-bits per hit |
| Error | 1 word of 32 bits |
| Data Folder | 1 word of 32 bits |

Table 6: Structure of data packet, see also Fig. 61

SFP. If we now take the 200 MB/s from the SFP as the limit this leads to the maximum event rate of 0.7 MHz for which the data could be transferred. Note this is a very naive calculation, data padding has e.g. not been considered at all. Furthermore the SFP speed (MB/s) depends on the size of the data packets.

However, the above given number is not the 'real' rate. In Super-FRS-type experiments a reaction trigger is constructed which drastically decreases the rate of the events that have to be stored. The maximum rate for accepted triggers in the MBS DAQ with small event size is 50 kHz [21].

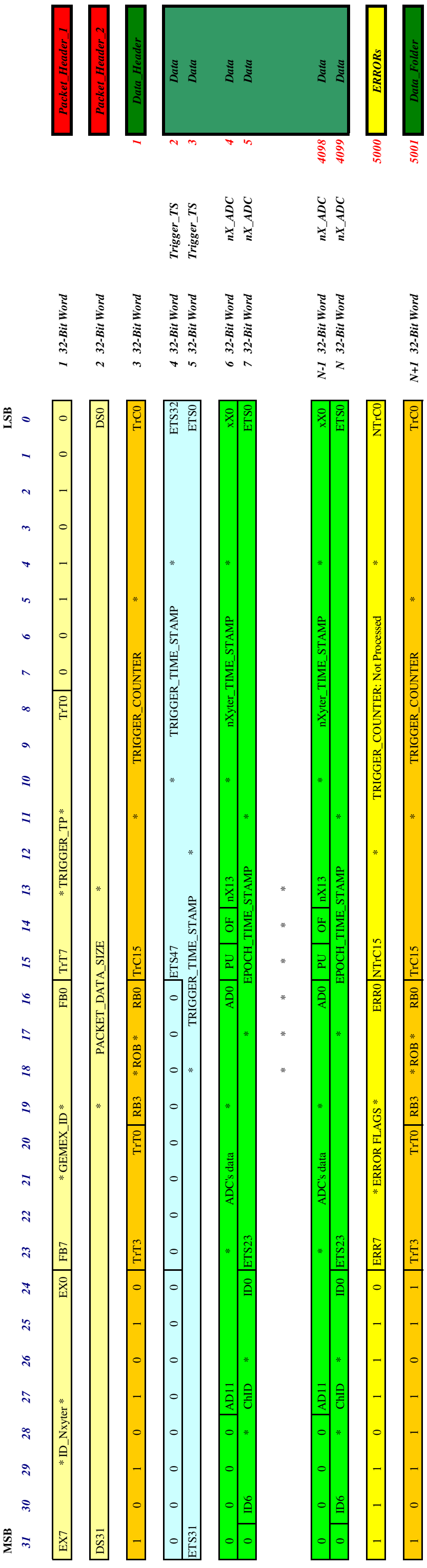

Figure 61: The N-XYTER data format [20].

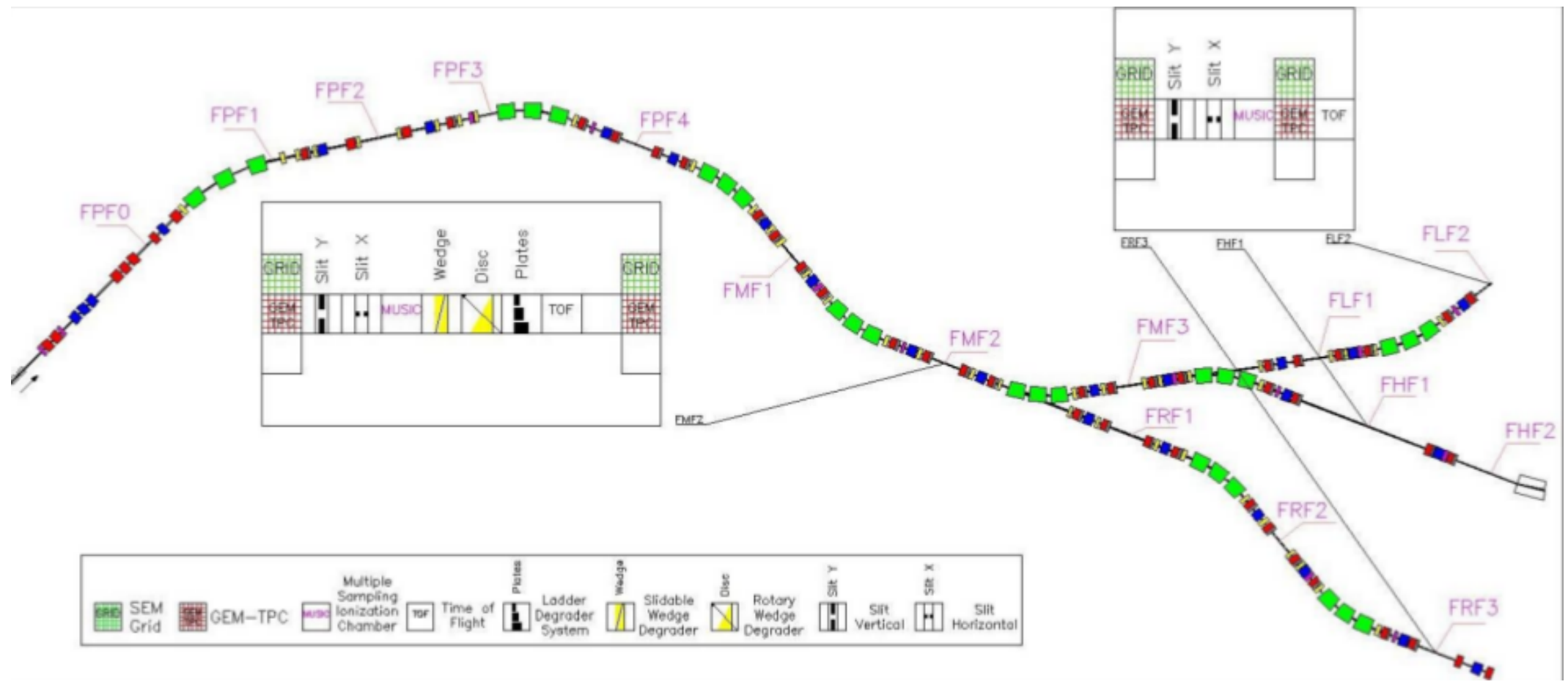

Figure 62: Layout of Super-FRS and the standard detector configuration at FMF2, FLF2, FHF1 and FRF3.

# 10 Simulations

In the first part of the following Section we describe the effects of the tracking resolution on the mass identification of the produced fragments. In Fig. 62 the layout of the separator and the standard detector configuration are shown for the mid-focal plane (FMF2) and the last focal planes of the three branches (FLF2,FHF1 and FRF3). Results of Monte Carlo simulations performed for the HEB (from FPF0 up to FHF1) are discussed. The simulations were done using the MOCADI code[22]. In the second part the charge deposition in the HGB4 has been simulated. In addition to MOCADI, the GARFIELD++ [26] and GEANT4 [25] codes were used. A third part describes the results of the simulated efficiency of HGB4.

## 10.1 Super-FRS tracking requirements

### 10.1.1 Momentum measurements and mass identification

A major role in the particle identification (PID) at Super-FRS is played by the tracking detectors. The identification in the mass and charge is based on TOF$-\Delta E - B\rho$ method ([31]). The charge of the fragment is determined from the energy-loss measurement in the ionisation chamber. The mass of the fragment is obtained from the velocity and momentum measurement as:

$$A/Q = \frac{B\rho}{\beta\gamma} \tag{1}$$

where $B\rho$ is the magnetic rigidity of the fragment, $\beta$ is the velocity of the fragment and $\gamma$ is the relativistic Lorentz factor.

The magnetic rigidity of the particle can be determined from the measured position at the focal planes, i.e. FMF2 and FHF1, as:

$$B\rho = B\rho_0 \left(1 - \frac{x_{\mathrm{FHF1}} - M \cdot x_{\mathrm{FMF2}}}{D}\right) \tag{2}$$

where $B\rho_0$ is the reference magnetic rigidity between FMF2-FHF1, $D$, $M$ are the dispersion and the magnification, respectively, $x_{\mathrm{FMF2}}$ and $x_{\mathrm{FHF1}}$ are the measured focal-plane positions.

The fragmentation reaction of $^{238}$U at 1.5 GeV/u on C target(1 g/cm$^2$) were simulated. The position and angular distributions of the primary beam hitting the target were Gaussian ($\sigma_x$,$\sigma_y$ = 1 mm, $\sigma_a$,$\sigma_b$ = 2 mrad). The fragments were identified between FMF2 and FHF1 using two HGB4 detectors placed at FMF2 and FHF1, respectively. The angular and energy straggling on the matter of the tracking detector were taken into account. The assumed matter in HGB4 is: 40 cm of P10 gas (1.7 mg/cm$^3$), $5\times20\mu$m Mylar windows, $2 \times 100\mu$m Fe windows. A TOF resolution of 20 ps is also assumed.

The calculated $A/Q$ resolutions for $^{236}$U fragments after fragmentation as a function of the horizontal TPC position resolution are shown in Fig. 63 for two different FHF1 distances. The dashed (red) and full (blue) curves were obtained assuming a distance equal to about 7 m and 1 m, respectively. For larger distance the mass resolution increases by about 10%. The third curve (green) represents the case with no extra matter, i.e. no tracking detectors. The calculated $A/Q$ distributions for the three fragments $^{235-237}$U for 1 m and 7 m HFH1 distances, assuming a position resolution of 0.5 mm, are shown in Fig. 64.

The TPC resolution better then 0.5 mm is needed to separate the fragments in the mass region around $A = 238$ ($Z = 92$). Less stringent position resolution is required for lower masses, as shown in Fig. 65 for fragments produced in the regions $A = 40$ ($Z = 18$).

#### 10.1.2 Angular resolution

In order to reconstruct the momentum of the fragments one needs to measure horizontal and vertical angles at the focal planes. The calculated horizontal angular distributions $a_1$ of the $^{236}$U fragments at the final focal plane is shown in Fig. 66. The width (sigma) of $a_1$ angular distribution is around 0.75 mrad.

In Fig. 67 the calculated FHF1 angular resolutions of the $^{236}$U fragments for 1 m distance are plotted as a function of the position resolution. The angular resolution is the width of the distributions obtained from the difference between $a_1$ and the angle measured by the two HGB4s. The dashed line indicates the width of the angular distribution in the case of no extra matter.

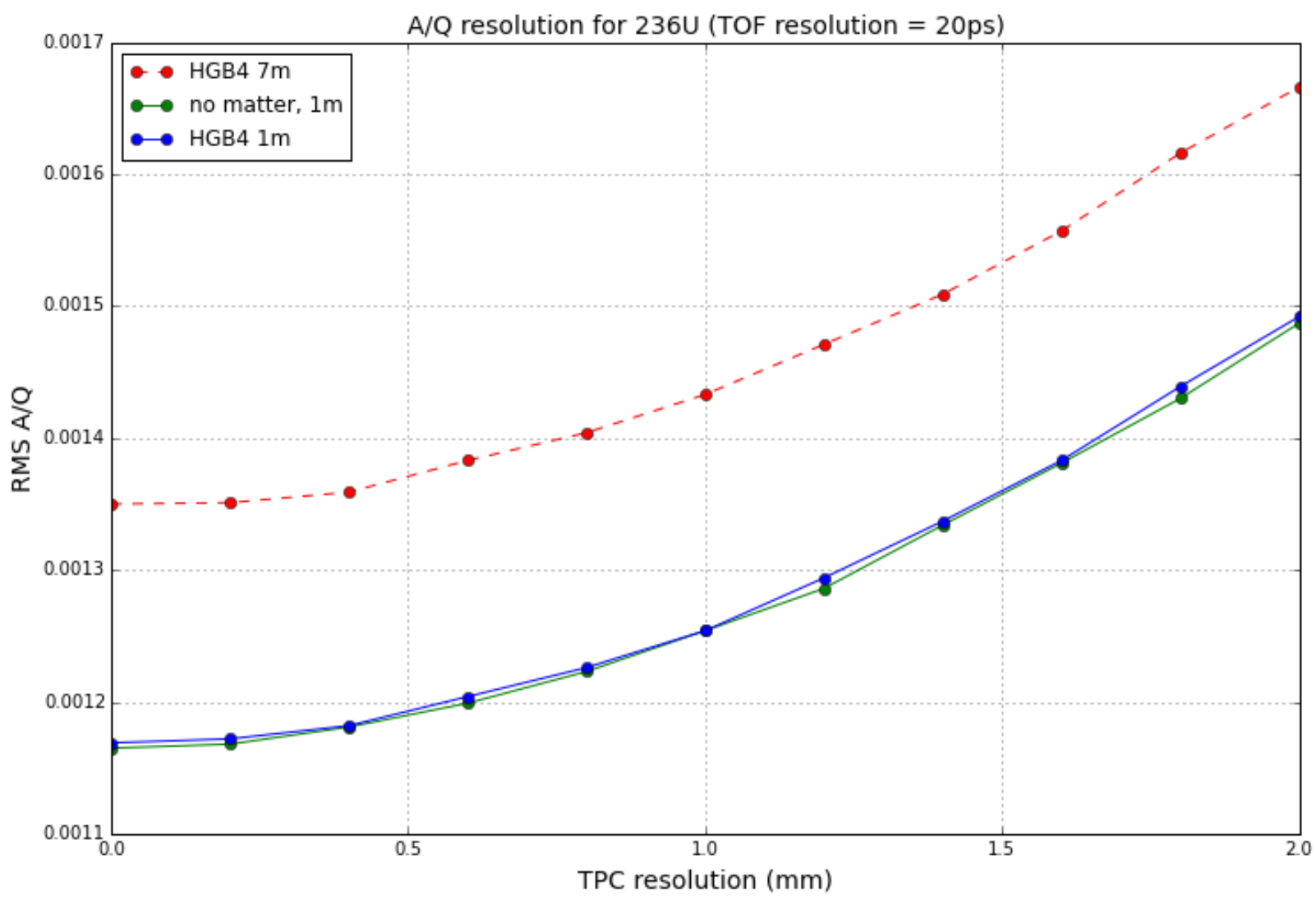

Figure 63: $A/Q$ resolution of $^{236}$U fragments at about 1.4 GeV/u as a function of TPC position resolution. The TOF resolution of 20 ps is assumed.

#### 10.1.3 Angular straggling

The simulated position and angular straggling were calculated using the ATIMA code [24]. The position straggling $\sigma_x$ in the middle of the HGB4 detector calculated for different fragments and different $B\rho$ is plotted in Figs. 68.

The calculated angular straggling $\sigma_a$ due to the whole HGB4 matter is shown in Fig. 69.

### 10.2 Charge deposition

The following parameters were used to calculate the deposited charge in HBG4:

- active volume size: 400 mm × 80 mm × 30 mm

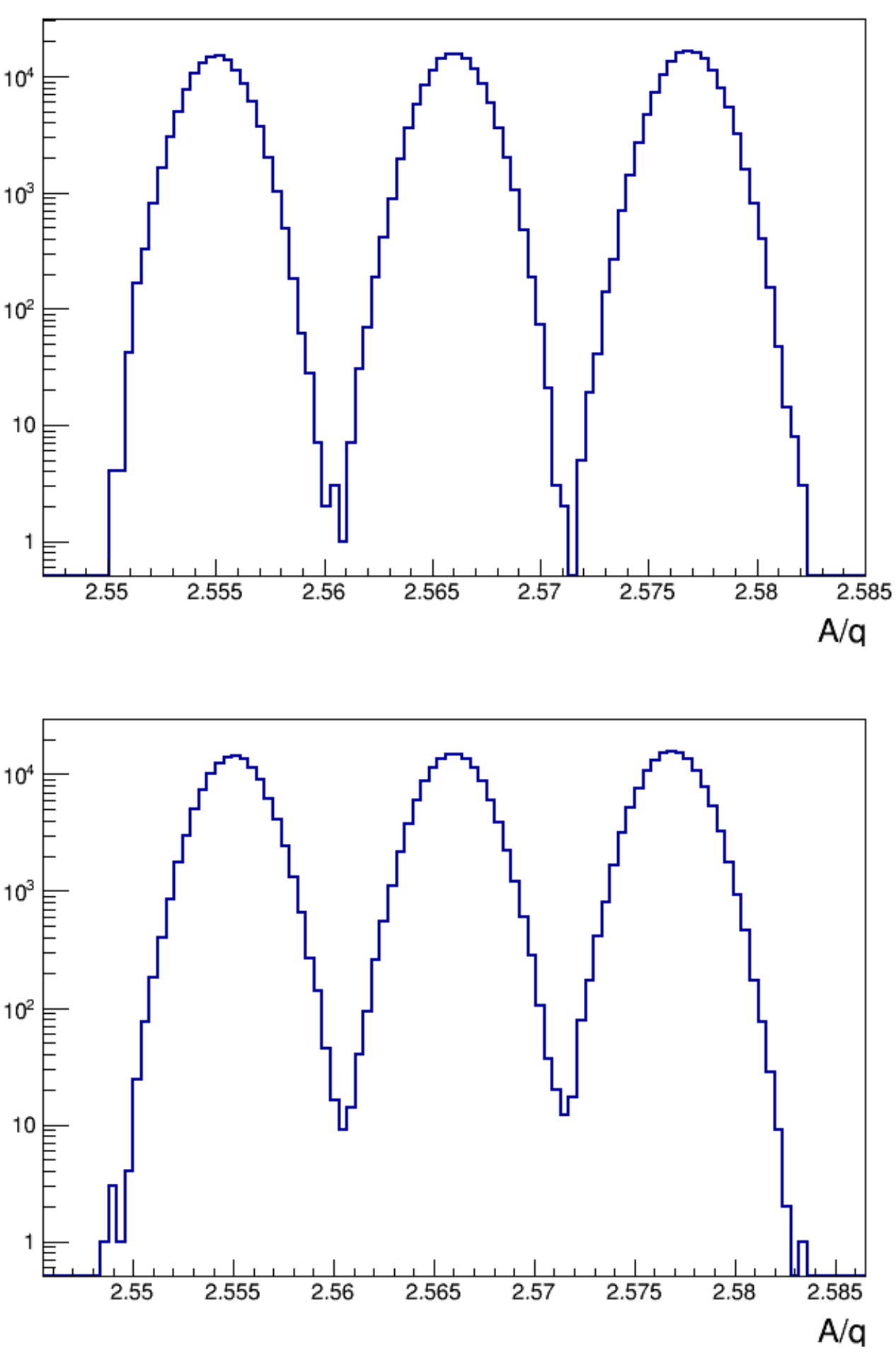

Figure 64: $A/Q$ distributions of $^{235-237}$U at about 1.4 GeV/u assuming 0.5 mm TPC resolution and 20 ps TOF resolution. Top: HFH1 distance equal to about 1 m, Bottom: HFH1 distance equal to about 7 m

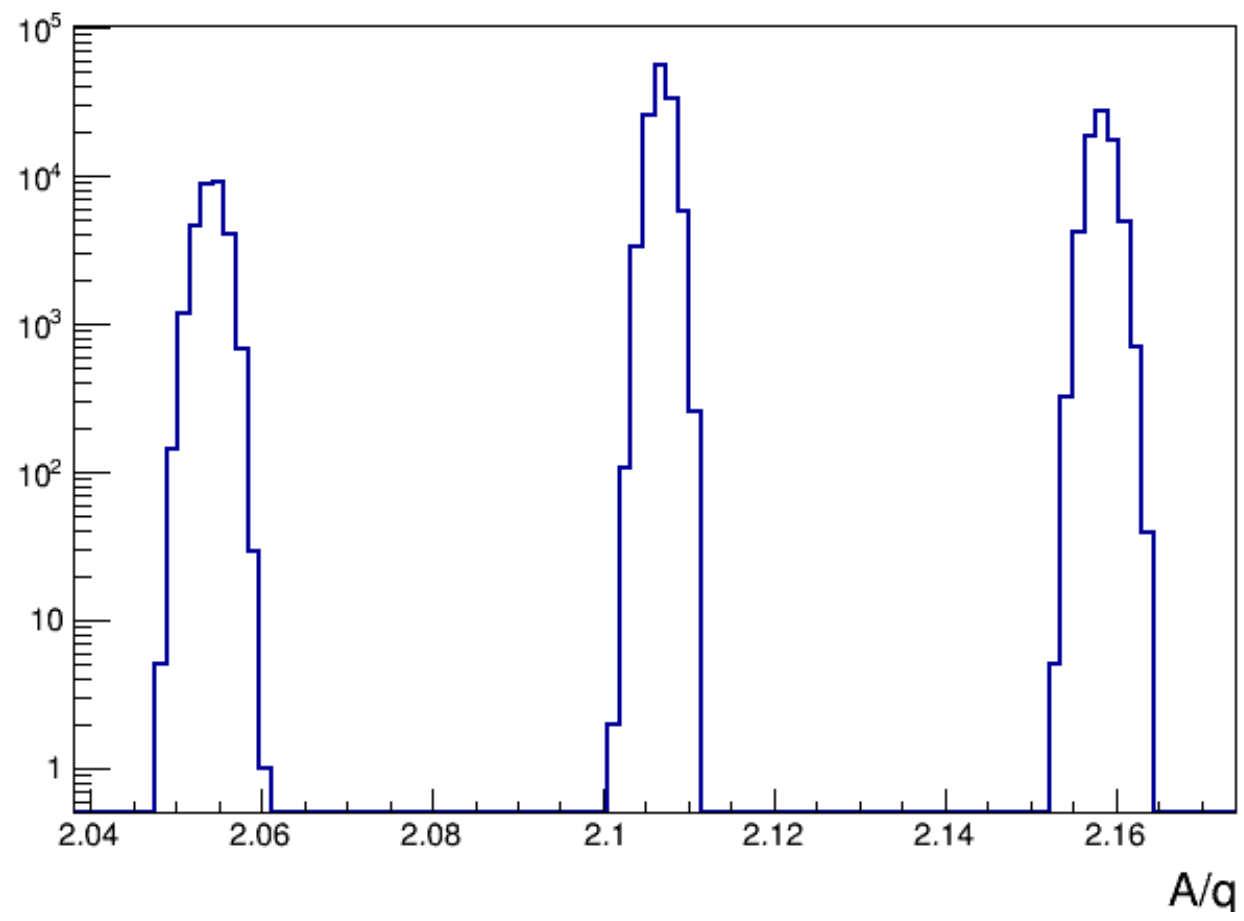

Figure 65: $A/Q$ distributions of $^{37-39}$Ar at about 1.4 GeV/u assuming 0.5 mm TPC resolution and 20 ps TOF resolution at FHF1 distance of about 1 m.

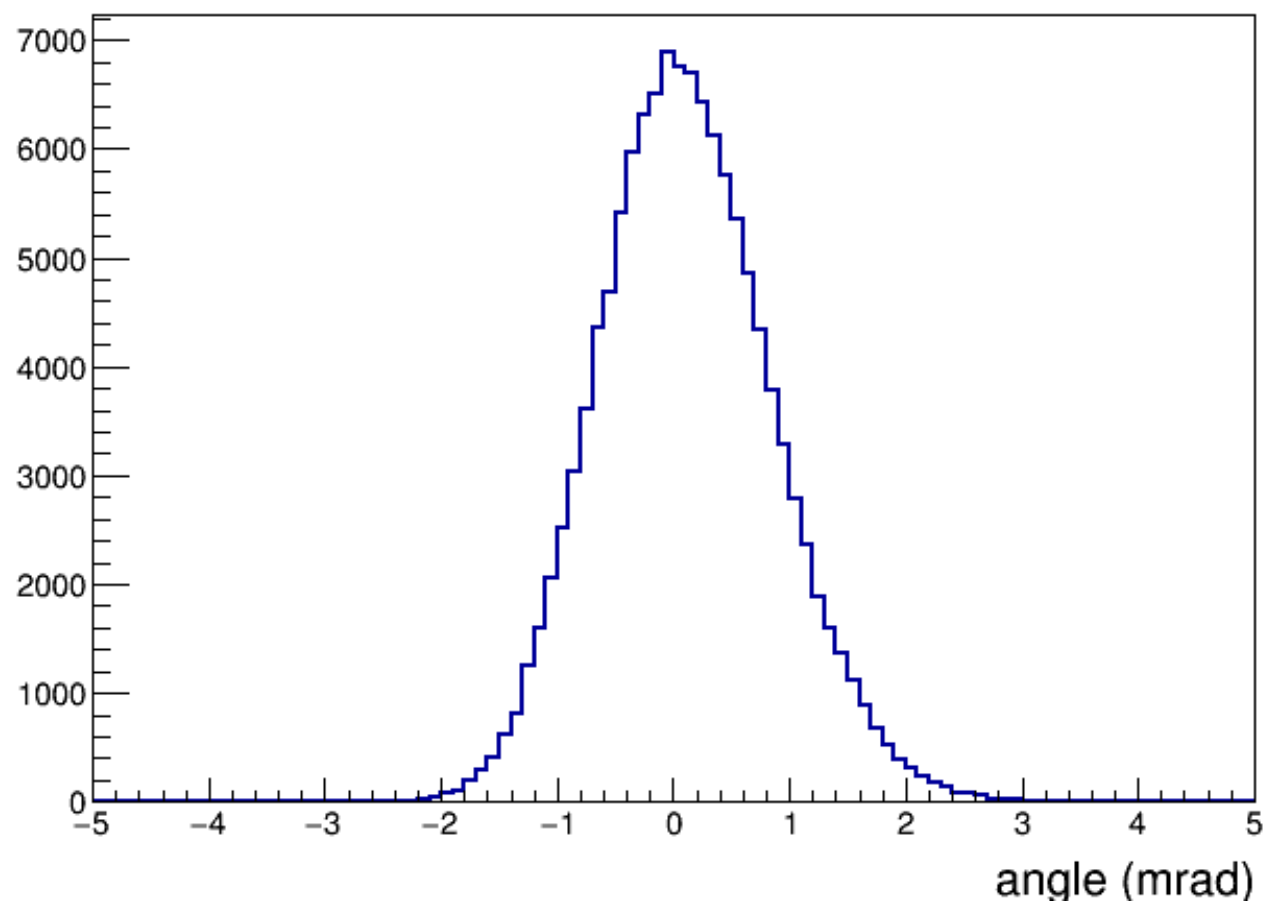

Figure 66: The horizontal angular distribution of $^{236}$U fragments at FHF1 for no extra matter case.

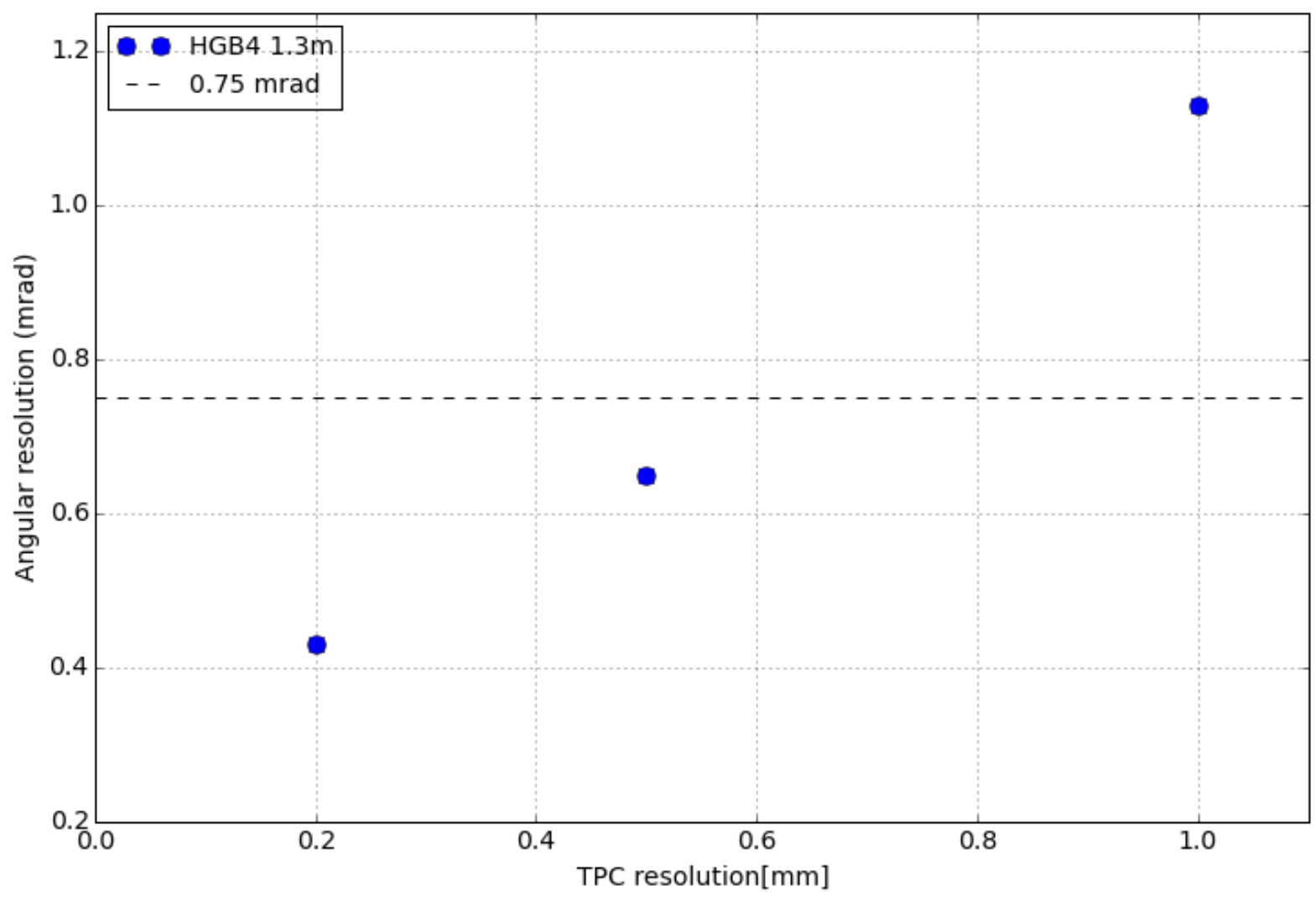

Figure 67: The calculated FHF1 angular resolutions of $^{236}$U for 1 m distance as a function of the position resolution.

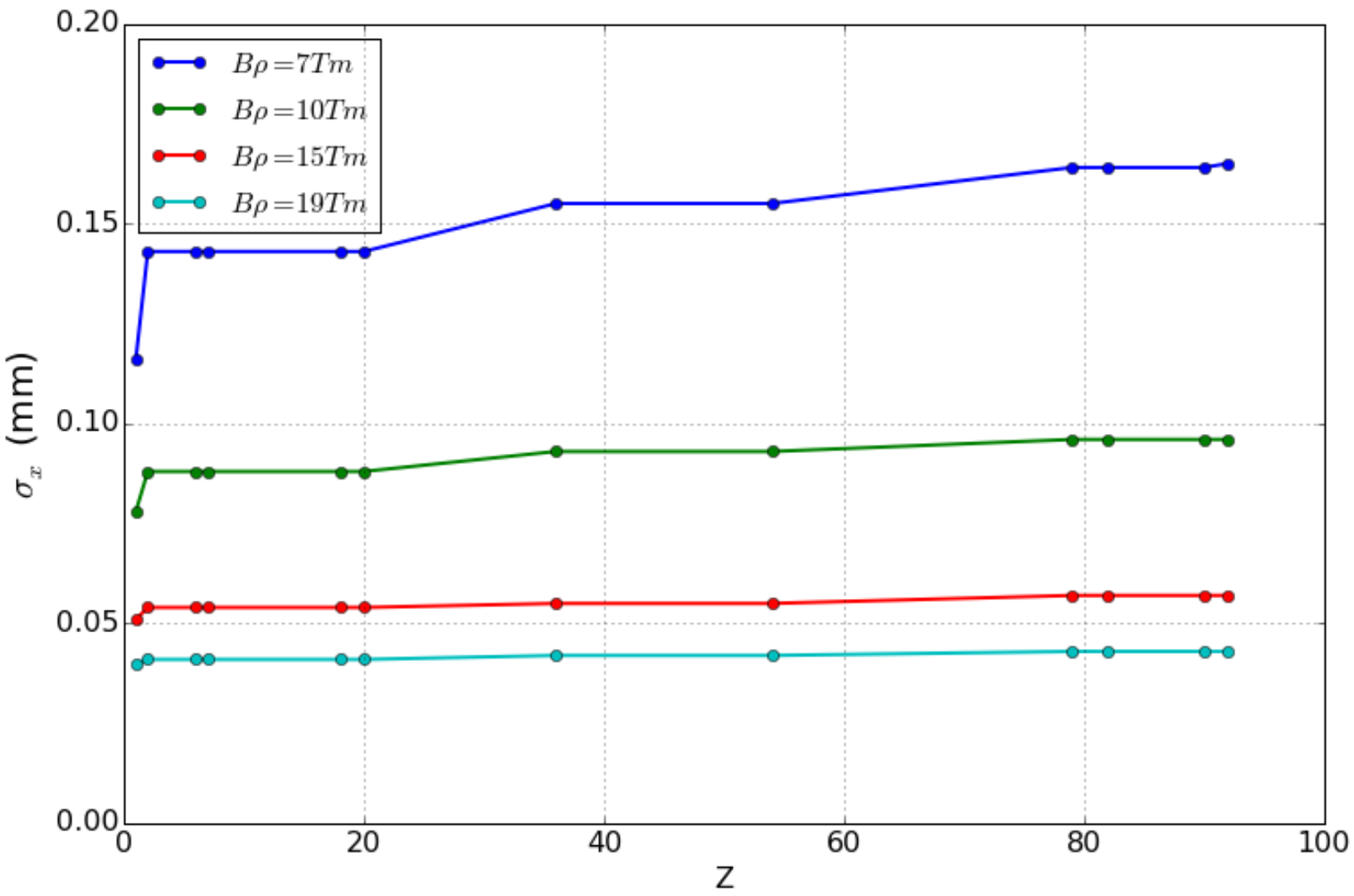

Figure 68: Calculated position straggling in the middle of the HGB4 detector.

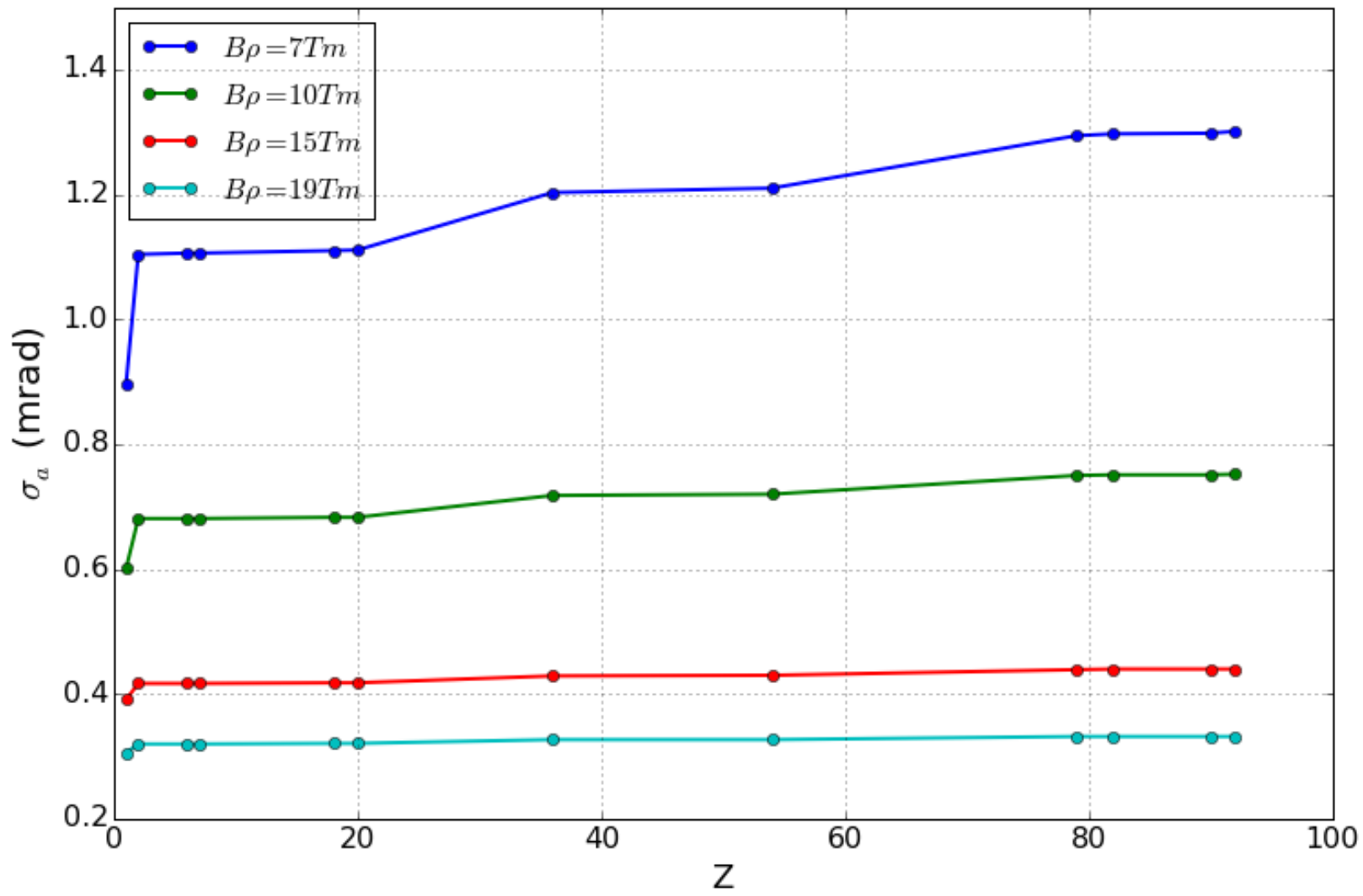

Figure 69: Calculated angular straggling due to the whole HGB4 matter.

- active volume gas: P10 at normal temperature and pressure ($W_i = 26.2$ eV, $\rho = 1.7$ mg/cm$^3$)

The charge deposition in the active volume was calculated from the projectile energy loss $dE$ using ATIMA [24]. Since The $dE$ does not fully correspond to the deposited energy in the active volume of the detector. The projectile energy loss in HGB4 does correspond to the deposited energy in the gas. It needs to be corrected for the energy taken away by delta electrons. This correction, which depends on the projectile type and energy and on the geometry of the active volume, was calculated using GEANT4 [25]. The relative amount of deposited energy calculated for several projectiles at different energies are shown in Fig. 70. The charge coming from the ionization of the projectile can be extracted from the deposited energy $E_D$ and mean ionization potential $W_i$ according to the following Eq.

$$Q_i = E_D/W_i \tag{3}$$

The calculated charge $Q_i$ for different projectiles and B$\rho$ is shown in Fig. 71.

The deposited charge in the active volume will be spatially scattered until it reaches the induction gap and the pad-plane. The main contributions to the spread are coming from the diffusion in the drift space and the spatial straggling in the GEM stack. The electric field used in the simulations was considered as constant, and put equal to 400 V/cm. This assumption is justified by FEM calculations [27], which have been performed to estimate the field homogenity. The spatial straggling produced in the GEM stack was calculated using the FEM method (electric field map calculation) and Garfield (electron transport) (for details see Ref. [29]). The total effect on the spatial straggling were approximated by the following formula

$$\sigma_{ex} = \sqrt{0.359 * Y + 0.1225} \tag{4}$$

where $\sigma_{ex}$ is the spatial straggling of the electrons at the induction gap in mm and $Y$ is the drift length in cm.

The charge after the GEM stack (3 foils) will induce a charge on the pad-plane (see Chapter Readout Pad-Planes).

The induced charge was calculated using GARFIELD++. The relative induced charge on the pad-plan is shown Fig. 72 for two different drift lengths. The $x$ axis indicates the strip number. The maximum charge induced on the central strip corresponds to 13-17% of the total one, depending on the drift length.

The maximum induced charge per strip is plotted in Fig. 73 for different projectiles at different $B\rho$. The results of the simulations suggest that the electronics readout needs to accept up to 100 fC.

### 10.2.1 Efficiency simulations

The standard FRS TPC can work up to $3 \times 10^4$ ions/s [23] with about 90% efficiency. To improve the ion capability of this detectors one can first adopt a single-strip readout. Simulations show that the rate dependence improved further by use of the twin design. The twin field-cage configuration used in HGB4 allows to calculate the control sum, defined as the sum of the drift times of the both drift volumes $t_{\mathrm{CS}} = t_1 + t_2 - 2 \cdot t_{\mathrm{t}}$, where $t_1$ and $t_2$ are the drift times from the two drift spaces and $t_{\mathrm{t}}$ is the reference time. In a real experiment $t_{\mathrm{t}}$ is the time delivered by a plastic scintillator used as a trigger.

To simulate the rate capability of HGB4 we assumed that beam particles can be described by the Poisson process. The time distribution between the ions is therefore an exponential function. The vertical position of the ions passing through the detectors were a Gaussian distribution, centred in the middle of the chamber with width equal to 5 mm. Electronic noise, delta electrons and recombinations in the gas were here neglected. The drift time resolution of 3 ns and minimum time to resolve multiple hit 50 ns is assumed. Assuming a collection time equal to 1.6 $\mu$s (8 cm drift length in P10) the probability of multiple hits within the collection time increases starting from few kHz. In Fig. 74 the mean number of hits within the collection time is plotted with green points.

Whenever two or more signals arrive in the same time window, the two arrival times, and thus the $y$ coordinates, are mixed up and signal can not be unambiguously sorted. The probability to register just 1 hit in the time window as a function of rate is shown by blue points. By using the control sum an multiple hits can be recovered and at given configuration an efficiency above 90% is reached at 2 MHz, as shown in FIg. 74 (red points).

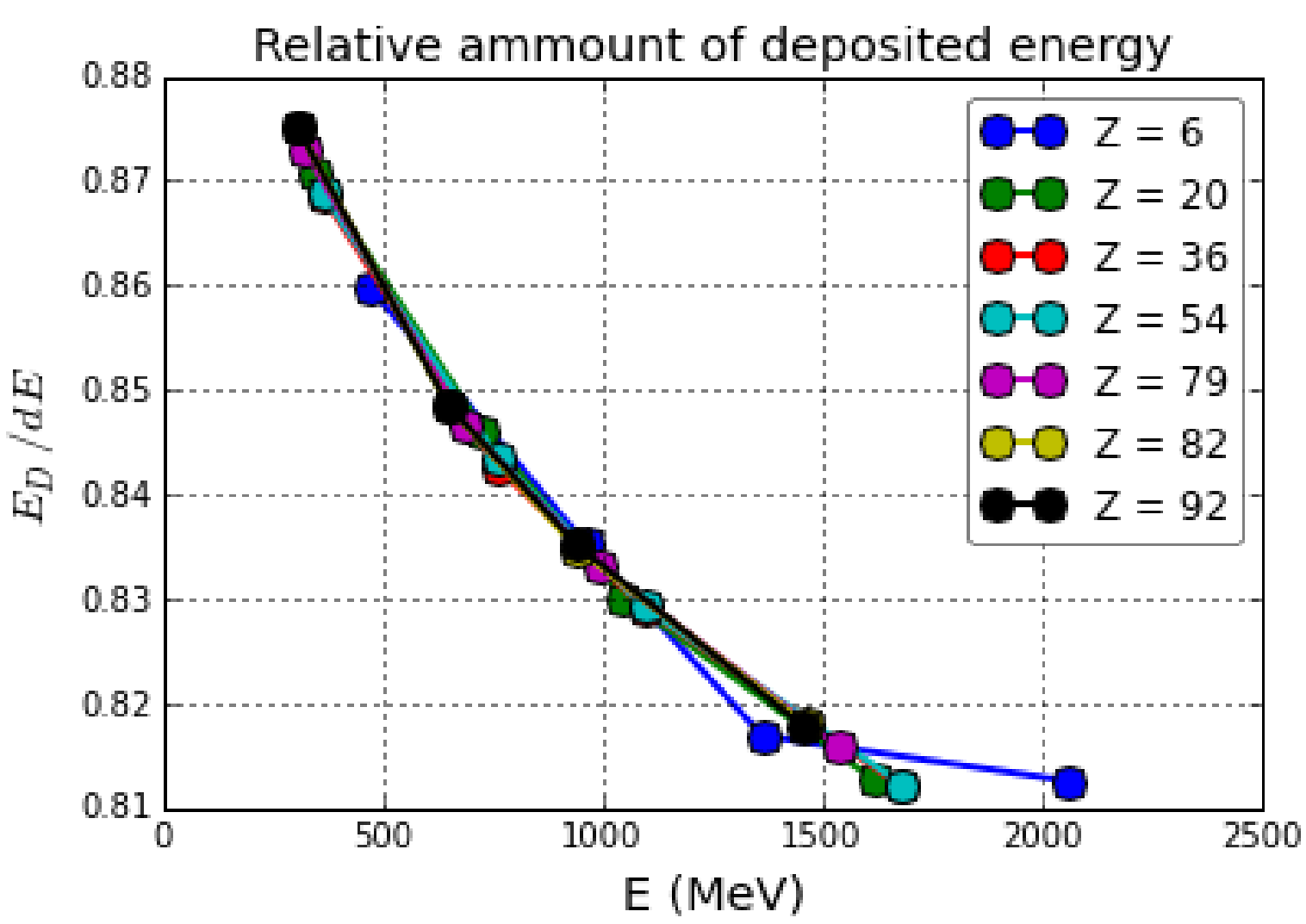

Figure 70: Calculated relative deposited energy in HGB4 as a function of energy and charge of the projectiles.

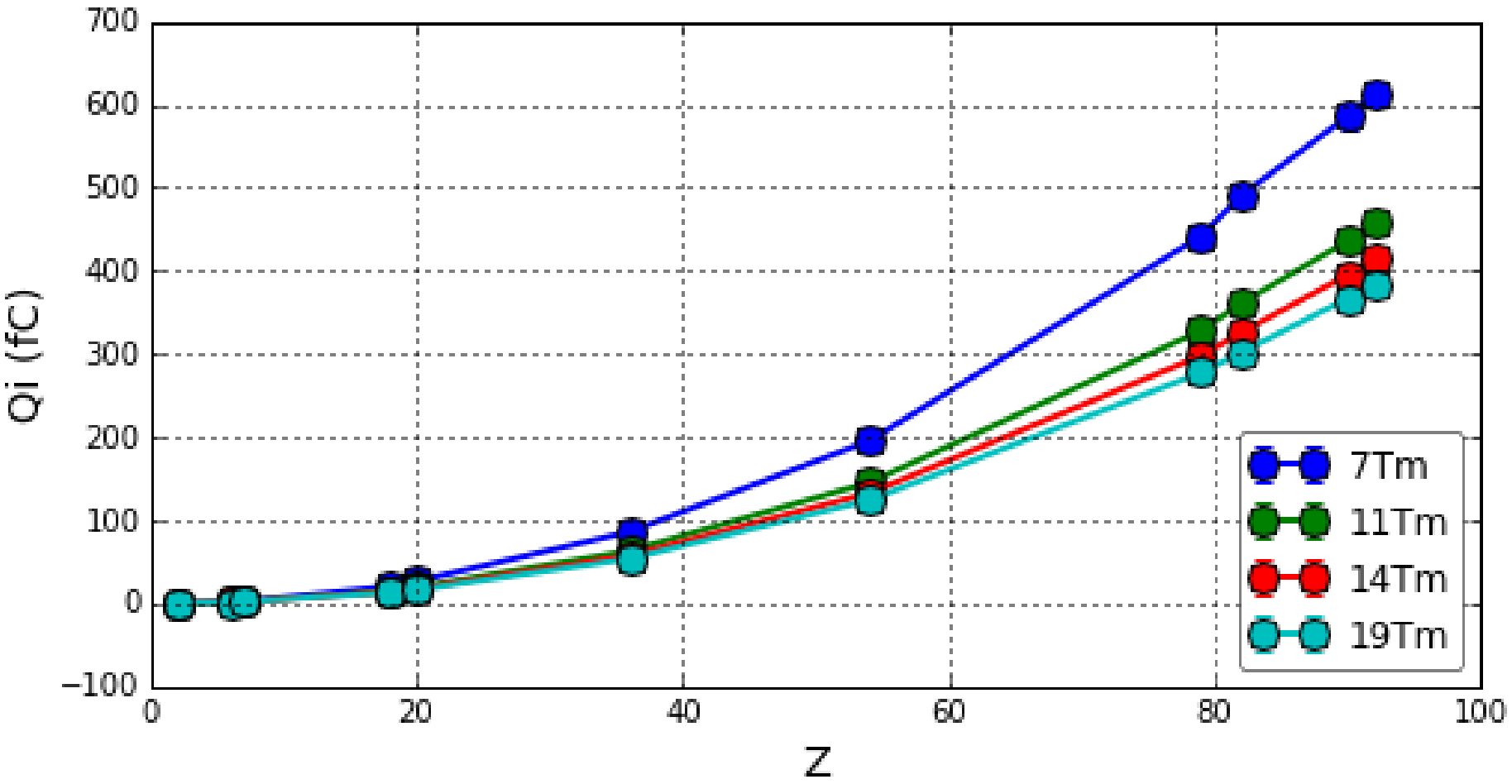

Figure 71: The charge deposited in the active volume of HGB4 for different ions and $B\rho$.

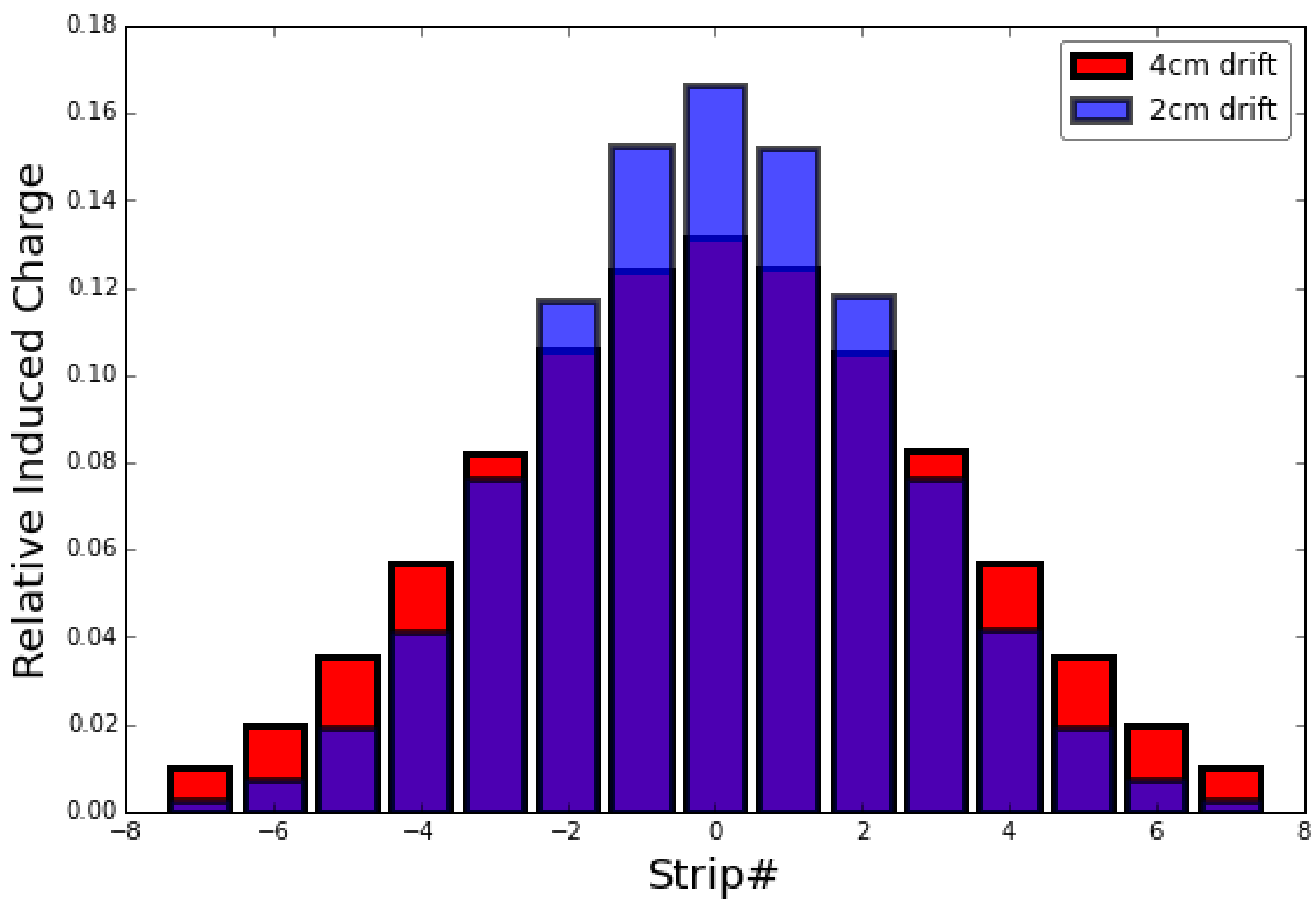

Figure 72: Relative induced charge on the pad-plane. The red and blue histograms corresponds to the 4 and 2 cm drift lengths, respectively.

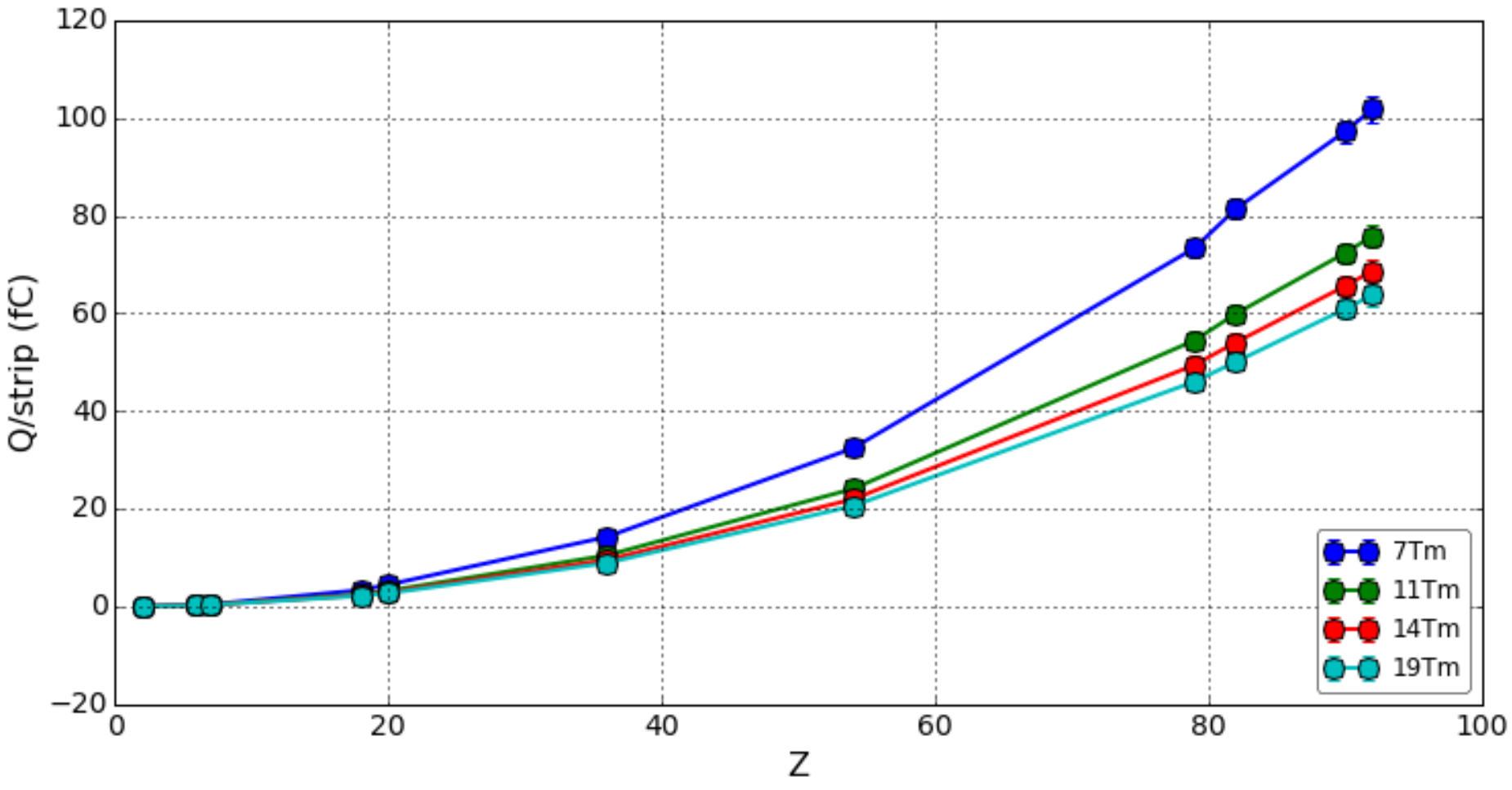

Figure 73: Maximum induced charge per strip for different fragments and $B\rho$.

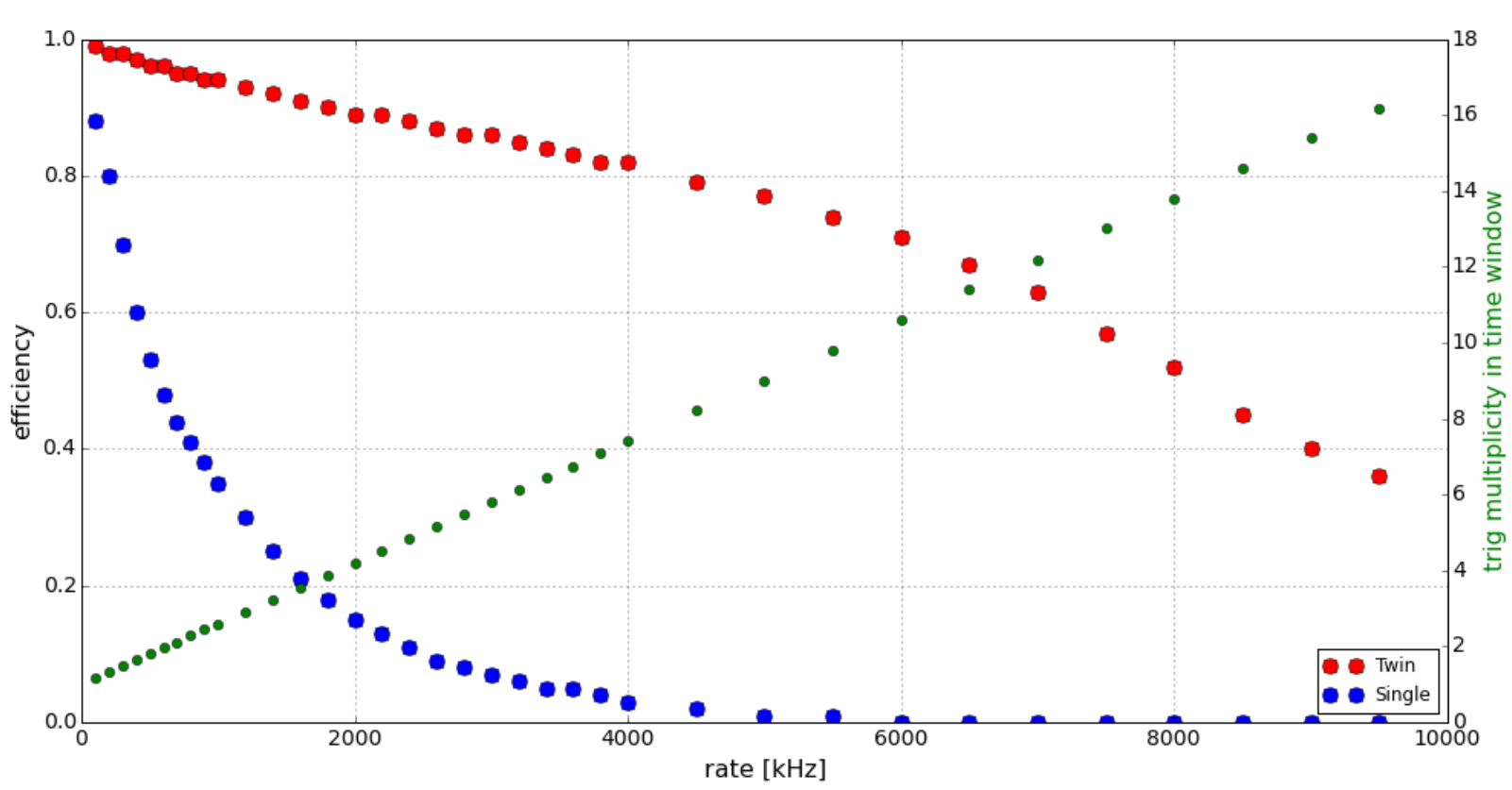

Figure 74: Simulated efficiency for HGB4 twin design (red) and single volume detector (blue) as a function of rate. The green points represent the mean number of hits in collection time at given rate (scale on the right axis).

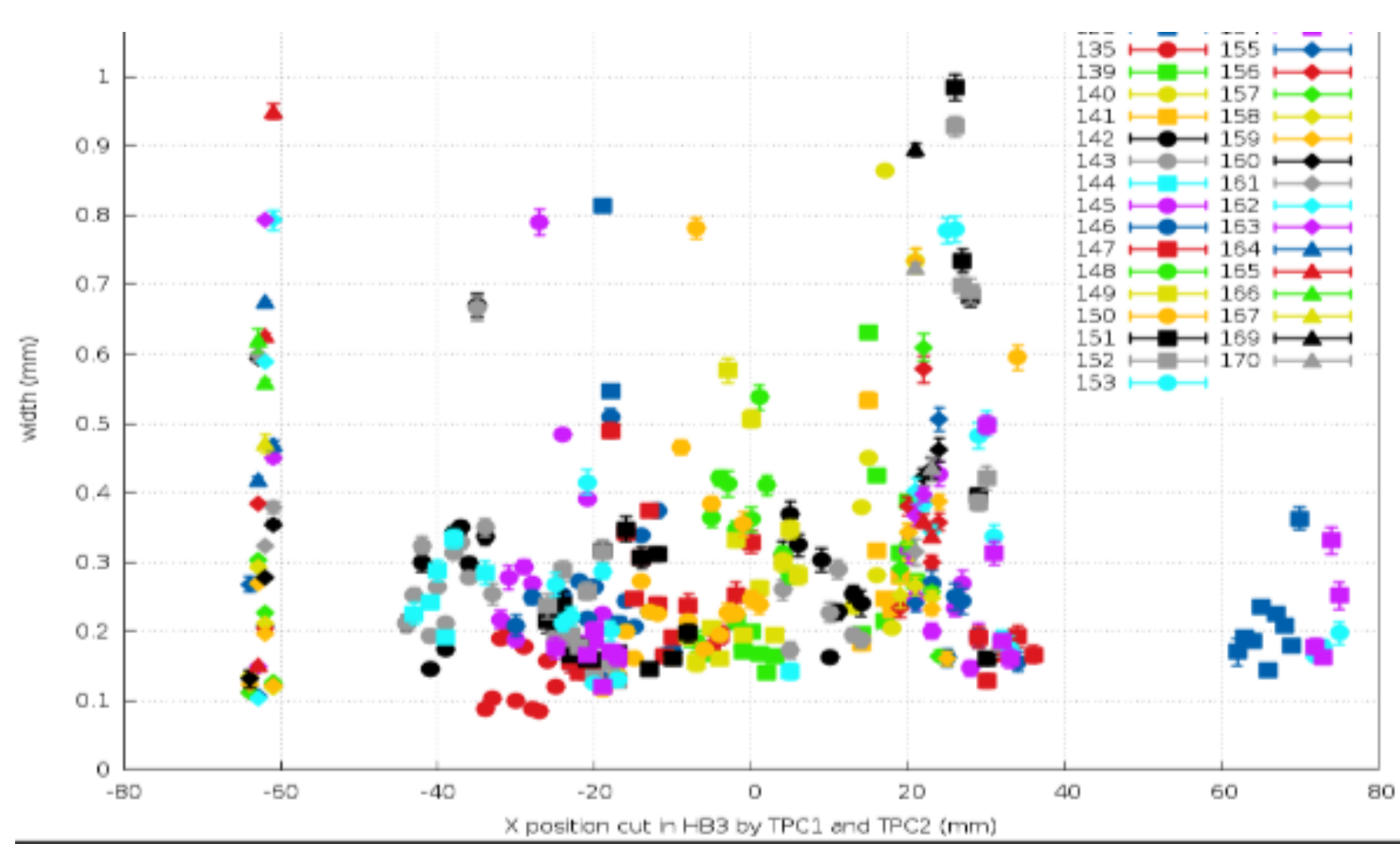

Figure 75: Measured horizontal position resolution of the GEM-TPC prototype (no twin) for different runs, corresponding to different beam positions.

# 11 Prototype results

The current design HBG4 is based on the tests and results of the TPC and previous generations of the GEM-TPC. Currently, standard TPC signals are read by delay lines. In the first GEM-TPC prototype (HGB1) the proportional part with gas amplification was substituted by GEM stacks and the signals were read by delay lines and standard electronics [23]. This was tested in-beam and reproduced all characteristics of the standard FRS TPC. As a next step delay-line readout was replaced by the pad-plane (see Chapter 8) and a single-strip digital readout front-end electronics (see Chapter 9) was adopted. This prototype HGB3 was tested at GSI in 2012 with Au primary beam. A horizontal position resolution about 0.2 mm (RMS) over whole volume of the detector was obtained (see Fig. 75. An efficiency over 99% compared to a plastic scintillator, placed in front of the detector, was measured for beam intensity up to 60 kHz (see Fig. 76).

The latest prototype, HGB4, which has a twin field-cage design was tested in 2014. The control sum measured with HGB4 during the GSI campaign is shown in Fig. 77. The U primary beam with moderate rate was used [36].

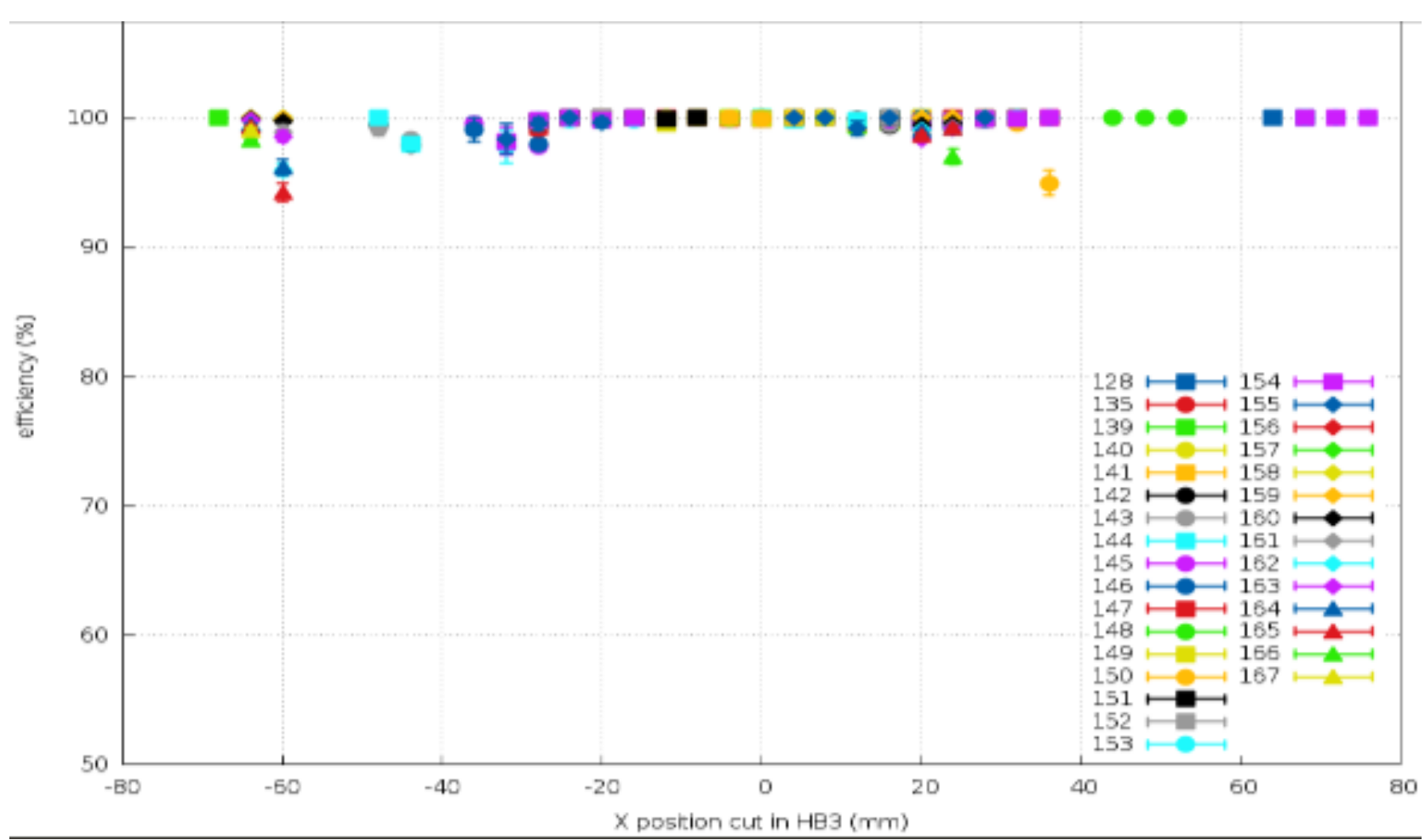

Figure 76: Measured efficiency of the GEM-TPC prototype (HB3) (no twin) for different runs, corresponding to different beam positions.

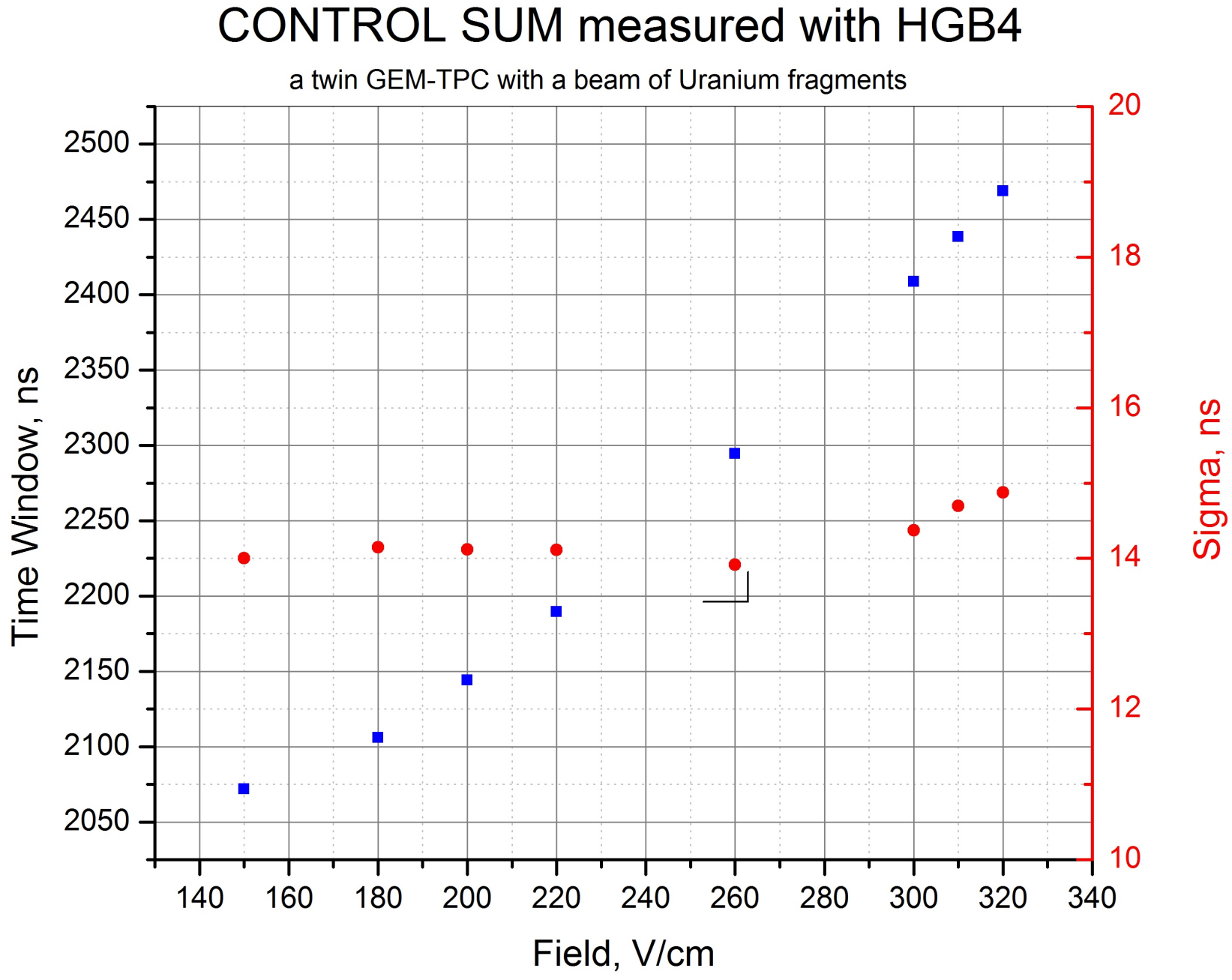

Figure 77: The control sum scan at different fields; from 150 V/cm up to 320 V/cm. In blue the total sum and in red the time resolution.

# Glossary, terms and definitions

| Term | Definition |
|---|---|
| GSI | GSI Helmholtzzentrum für Schwerionenforschung GmbH |
| FAIR | International Facility for Antiproton and Ion Research |
| Super-FRS | Superconducting Fragment Separator |
| NUSTAR | NUclear STructure, Astrophysics and Reactions |
| TPC | Time Projection Chamber |
| PS | Pre-Separator |
| MS | Main Separator |
| EB | Energy Buncher |
| HEB | High Energy Branch |
| LEB | Low Energy Branch |
| RB | Ring Branch |
| FRS | Fragment Separator |
| FEE | Front-End Electronics |
| DAQ | Data Acquisition |
| RIBs | Rare Isotopes Beams |

Table 7: Glossary

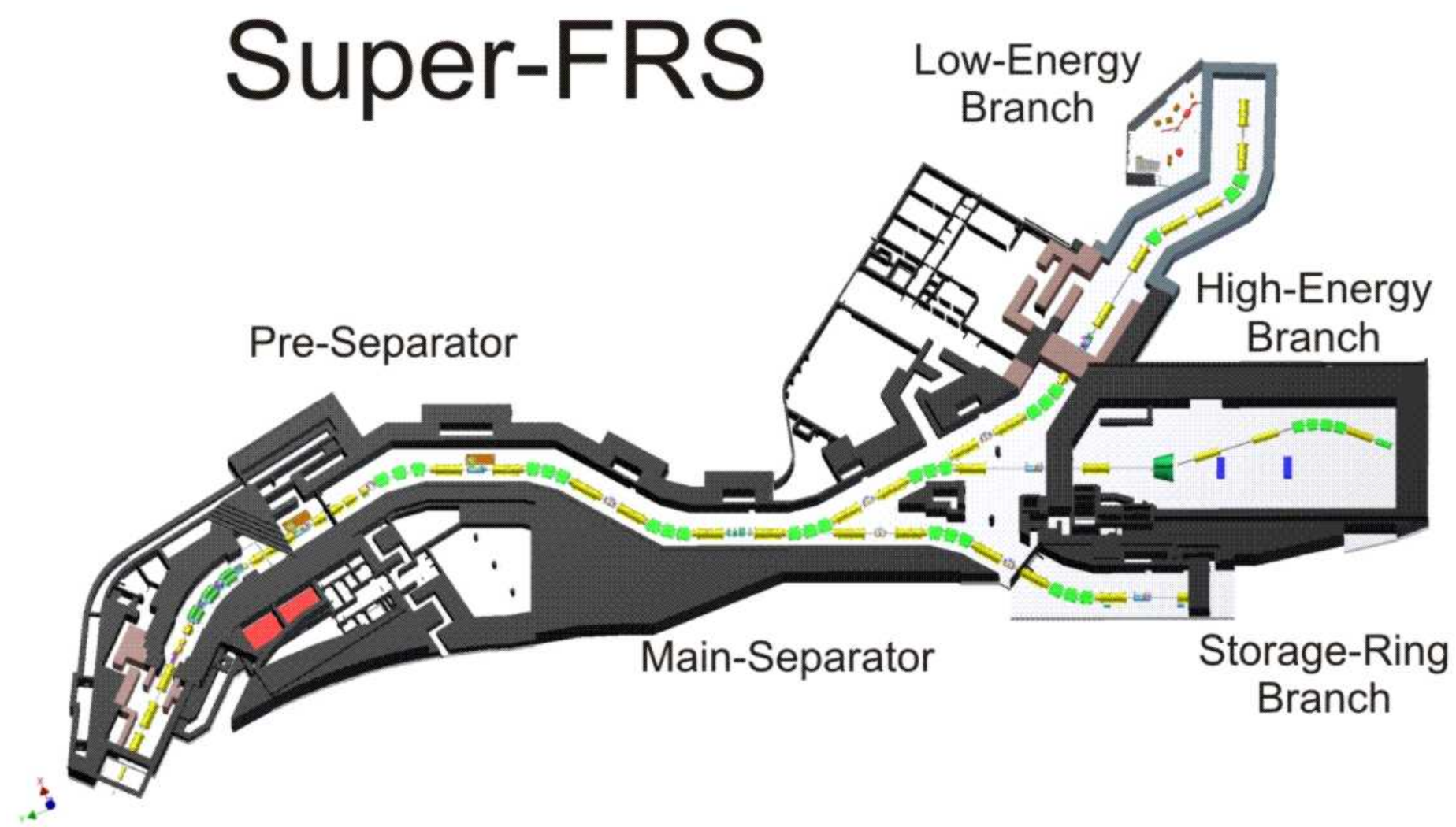

Figure 78: Layout of the Super-FRS.

# Appendix I

The Super-FRS is the central device of the NUSTAR facility at FAIR [34]. It will be used to produce rare isotope beams from projectile fragmentation and from in-flight fission of uranium with energies up to 1.5 GeV/u. It is expected to deliver exotic nuclei to three branches: the Low-Energy Branch (LEB), the High-Energy Branch (HEB) and the Storage-Ring Branch (RB). The three branches are shown in Fig. 78.

To produce in-flight rare isotope beams (RIBs), the primary beam delivered by the SIS100 (e.g. $3 \cdot 10^{11}$ $^{238}$U/s) will impinge on a graphite target of a few g/cm$^2$ located at the entrance of the Super-FRS. The high-separation power of the device will allow to deliver high-quality RIBs spatially separated within a few hundreds nanoseconds. For these dedicated experimental branches, the Super-FRS will provide spatially separated mono-isotopic or cocktail beams RIBs depending on the goals of the experiments. Challenging requirements from experiments at those branches are achieved with a multi-stage magnetic system, comprising intermediate degrader stations. The main parameters of Super-FRS are listed in Table 8.

Specialized detector systems for full particle identification will verify the separation performance. The fragments are uniquely identified in charge $Z$ and in mass numbers $A$ by using the TOF$-\Delta E - B\rho$ method. According to this method one can distinguish three classes of detectors. Particle identification at the Super-FRS will be achieved by using precise tracking detectors to measure positions and angles at focus, Multi-Sampling Ionization Chamber (MUSIC) for $\Delta E$ measurements and Time-of-Flight (TOF) detectors to measure the fragment velocity. Their combined information allows particle identification on event-by-event basis as well as the measurement of the particle momentum. Details about the particle identification are given in [33]. The physics goals of the NUSTAR collaboration are given in Appendix II [34, 35].

| | |
|---|---|
| Magnetic rigidity range | $2-20$ Tm |
| Horizontal emittance | $40\pi$ mm mrad |
| Vertical emittance | $40\pi$ mm mrad |
| Momentum acceptance | $\pm 2.5\%$ |
| Horizontal angular acceptance | $\pm 40$ mrad |
| Vertical acceptance | $\pm 20$ mrad |
| First order resolution first stage (for 2 $mm^2$ target ) | 750 |
| First order resolution second stage (for 2 $mm^2$ target ) | 1500 |
| Momentum dispersion (FMF2-FHF1) | 7.6 cm/% |
| Horizontal magnification (FMF2-FHF1) | 1.45 |
| Length of Pre-Separator | $\approx 71$ m |
| Length of the Main-Separator (HEB) | $\approx 105$ m |

Table 8: Main parameters of Super-FRS.

# Appendix II

## NUSTAR Physics Cases

- **Super-FRS collaboration:** The selected experiments can be pursued exclusively at FAIR with Super-FRS, due to the combination of high energies, high momentum-resolution and the characteristics of the multiple-stage magnetic spectrometer. Examples are the study of exotic atoms (mesonic nuclei), exotic hypernuclei, high-momentum components of the tensor force, delta resonances in asymmetric nuclear matter, resonant coherent excitation in crystals, and in-flight decays in extremely short-lived nuclei and resonance states. Important are production of new isotopes and atomic collision studies (stopping, straggling) with heavy fragment beams at the highest energies of FAIR.

- **HISPEC/DESPEC** (High-Resolution Spectroscopy / Decay Spectroscopy): The aim is to comprehensively benchmark nuclear models and the $r$-process along and beyond $N = 126$, and along neutron-rich $Z \sim 82$ isotopes.

- **MATS/LaSpec** (Precision Measurements of very short-lived nuclei with Advanced Trapping System)/Laser Spectroscopy): It is expected that LaSpec and MATS will be competitive especially in the regions of neutron-rich refractory elements, which cannot be delivered from ISOL-type facilities in large amounts. This is an interesting region to study because for example, around $^{108}$Zr, new shape evolutions (from very deformed nuclei to spherical ones) are theoretically predicted. The region east and northeast of lead – especially beyond $N = 126$ - is of high interest concerning the $r$-process path.

- **R$^3$B** (Reactions with Relativistic Radioactive Beams): Specific examples of the R$^3$B physics program are quasi-free scattering reactions, dipole response and dipole polarizability of heavy neutron-rich nuclei, the study of the dipole response of halo nuclei up to high excitation energies of 30 MeV, electromagnetically-induced fission, and the measurement of the fission barriers.

- **ILIMA** (Isomeric beams, LIfetimes and MAsses): ILIMA will be unique worldwide for the study of short-lived nuclei enabling masses and half-lives to be measured for the first time. In particular, this is true for nuclei with $A \sim 200$ that are key to understanding the heaviest ($A \sim 195$) $r$-process abundance peak.

- **EXL/ELISe** (Exotic nuclei studied in light-ion induced reactions at the NESR storage ring/ Electron-Ion Scattering in a Storage Ring): low momentum transfer measurements with recoil particles with energies below 500 keV can be performed in a storage ring to investigate properties such as the nuclear compressibility of medium-heavy and heavy isotopes.